\documentclass[%
 10pt,
 reprint,
 amsmath,amssymb,
 aps,
floatfix,
]{revtex4-1}

\usepackage{graphicx}
\usepackage{amsmath,amssymb,xurl}
\usepackage{dcolumn}
\usepackage{bm}
\usepackage[fontsize=10pt]{fontsize}
\usepackage{setspace}
\usepackage{booktabs}
\usepackage{array}
\usepackage{algpseudocode}
\usepackage{needspace}
\usepackage{etoolbox}
\usepackage{xcolor}
\usepackage[colorlinks=true,linkcolor=blue,citecolor=blue,urlcolor=blue,
  filecolor=blue,hypertexnames=false,bookmarksnumbered=true]{hyperref}

\renewcommand{\thesection}{\arabic{section}}
\renewcommand{\thesubsection}{\thesection.\arabic{subsection}}
\makeatletter
\renewcommand{\p@subsection}{}
\makeatother
\makeatletter
\AtBeginDocument{%
  \def\section{\@startsection{section}{1}{\z@}{0.45cm \@plus.5ex \@minus.2ex}{0.2cm}%
    {\normalfont\small\bfseries\preprintsty@sw{}{\centering}}}%
  \def\subsection{\@startsection{subsection}{2}{\z@}{0.4cm \@plus.5ex \@minus.2ex}{0.15cm}%
    {\normalfont\small\bfseries\preprintsty@sw{}{\centering}}}%
  \def\subsubsection{\@startsection{subsubsection}{3}{\z@}{0.3cm \@plus.5ex \@minus.2ex}{0.1cm}%
    {\normalfont\small\itshape\preprintsty@sw{}{\centering}}}%
}
\makeatother

\def\MainDocument{}

\makeatletter
\AtBeginDocument{%
  \let\SM@title\title \let\SM@author\author \let\SM@affiliation\affiliation
  \let\SM@email\email \let\SM@thanks\thanks \let\SM@and\and
  \let\SM@maketitle\maketitle \let\SM@AFjoin\@AF@join
  \def\RestoreFrontMatter{%
    \let\title\SM@title \let\author\SM@author \let\affiliation\SM@affiliation
    \let\email\SM@email \let\thanks\SM@thanks \let\and\SM@and
    \let\maketitle\SM@maketitle \let\@AF@join\SM@AFjoin}%
}
\makeatother

\begin{document}

\title{All Paths All at Once: A Quantum Monte Carlo and \\ Neural Committor Approach to Reaction Pathway Search}
\author{Chien-Chai Chang} %
\affiliation{
Department of Engineering and System Science, National Tsing Hua University (NTHU), Hsinchu City, 300044, Taiwan
}

\begin{abstract}
Locating competing reaction pathways is difficult when potential-energy gradients are unavailable. Here we present tReax\_expedition, an energy-only approach combining coupled Monte Carlo exploration, stochastic quench labelling and restrained continuation. Exploration identifies connected states, while continuation resolves candidate channels and records sampled chain maxima. Single-search benchmarks on three two-dimensional potentials recovered an intermediate and competing routes without transition-state guesses. For eight of nine gas-phase reactions on GFN2-xTB, sampled chain maxima differed from independently refined saddle energies by less than 0.76~kJ\,mol$^{-1}$. Alanine dipeptide instead revealed incomplete conformational relaxation, showing that route connectivity can persist despite structural and energetic errors. Neural committor learning and local straddling-pair sampling connect the resulting paths to transition-region sampling, although molecular shooting diagnostics indicate limited committor calibration. These results show how energy-only exploration and continuation recover candidate pathways, while distinguishing route connectivity from geometric localization, barrier resolution and finite-temperature transition statistics.
\end{abstract}

\maketitle
\setstretch{1}

\section{Introduction}\label{sec1}
Transition states determine the rates and mechanisms of chemical reactions and conformational changes\cite{laidler1983development,truhlar1996current,e2010transition}. Locating them requires searching rarely visited regions of a potential energy surface, often along several competing channels\cite{bolhuis2002tps,henin2022enhanced}. This task becomes expensive when accurate electronic-structure methods provide energies without analytic gradients\cite{tikhonov2026two}.

Most transition-state searches use forces. Single-ended methods locate a saddle from one structure\cite{henkelman1999dimer,zimmerman2015single}, whereas chain-of-states methods relax a path between prescribed endpoints\cite{henkelman2000climbing,asgeirsson2021nudged,e2002string,peters2004growing}. Their outcomes depend on the starting structure or path. Generative models instead propose transition states from endpoint structures\cite{duan2023accurate,duan2025optimal}, molecular graphs\cite{kim2024diffusion} or metastable configurations under guided diffusion\cite{lim2026astra}. These proposals depend on training data and require validation on the target energy surface.

Rare-event methods describe the transition through the committor, the probability of reaching a product basin before a reactant basin\cite{du1998transition,berezhkovskii2005one,e2006transition}. Its isosurface $q=1/2$ defines a transition-state ensemble even when several channels compete\cite{bolhuis2002tps}. Neural committors can be learned from path-sampling outcomes\cite{jung2023machine,lazzeri2023molecular} or variational formulations\cite{khoo2019solving,li2019computing,megias2025iterative,contreras2026learning}. Kang et al. used a committor-derived Kolmogorov bias to sample the transition region while learning the committor\cite{kang2024committor}. Trizio et al. combined this bias with on-the-fly probability enhanced sampling to sample basins, intermediates and competing transition channels\cite{trizio2025everything}. Their molecular-dynamics implementation requires forces and specified endpoint basins. Learned slow modes\cite{bonati2021deep}, descriptor-free graph collective variables\cite{zhang2024descriptor} and graph committors\cite{contreras2026learning} reduce reliance on manually chosen inputs. These methods motivate a complementary question: can energy-only exploration identify intermediates and competing routes while retaining explicit statistical targets for subsequent sampling?

An energy-only alternative is the Monte Carlo transition-state search (MCTSS) of Tikhonov and Santra\cite{tikhonov2026two}. Inspired by real-space diffusion Monte Carlo\cite{austin2012quantum}, it drives walkers from opposite endpoints until they meet. The authors demonstrated reaction-path searches with DLPNO-CCSD(T) energies\cite{tikhonov2026two}. MCTSS was designed for search rather than equilibrium sampling: its acceptance includes a kinetic factor, and its geometric pre-selection redraws trials until one passes. Its time parameter also uses harmonic frequencies, while its barriers are estimated from binned trajectory averages. These choices motivate a statistically explicit energy-only search.

Here we present tReax\_expedition for energy-only candidate-route search. Coupled walkers explore between endpoints, stochastic quenches assign states, and restrained continuation resolves candidate channels. We test connectivity and saddle localization on model potentials, gas-phase reactions and a conformational transition, then assess the distinct requirements for committor learning and local transition-region sampling. Supplementary Table~\ref{tab:s-classes} separates fixed-target identities, modelling assumptions and numerical heuristics.

\section{Results}\label{sec2}

\subsection{Benchmark systems and reference quantities}\label{sec:r-systems}
We evaluated channel recovery, chain energies and transition-region structures separately from finite-temperature committor prediction. The benchmark comprised three two-dimensional potentials, nine gas-phase reactions and alanine dipeptide (Supplementary Table~\ref{tab:s-systems}). The two-dimensional models tested a single channel, an intermediate and competing routes\cite{tikhonov2026two,muller1979location,trizio2025everything}. The gas-phase set combined eight Baker--Chan reactions in the SACTS input geometries\cite{baker1996location,suresh2026sacts} with $\mathrm{Cl^-+CH_3Br\to CH_3Cl+Br^-}$. All nine gas-phase searches and their reference comparisons used GFN2-xTB\cite{bannwarth2019gfn2}. Alanine dipeptide tested the $\mathrm{C7_{eq}}\to\mathrm{C7_{ax}}$ conformational transition on the energy-only ff99SB surface\cite{hornak2006ff99sb}. The searches used endpoint structures, without a transition-state guess or potential-energy gradients. The complete benchmark was rerun with the $SE(3)$-corrected implementation (Supplementary Section~\ref{sm:repro}).

Each retained channel supplied two candidates: its highest-energy chain node and the crossing $\mathbf R^\ddagger_0$ obtained by bisecting stochastic quench labels. The crossing estimates a boundary between quench basins, without establishing the physical-temperature isocommittor $q=1/2$. We compared both candidates with same-surface reference saddles. Molecular references were independently refined and checked for one imaginary mode, which verifies local saddle character but not connection to both intended endpoints. These derivative-based calculations served only for evaluation. Table~\ref{tab:r-benchmark} reports crossing RMSDs and activation estimates relative to the same recorded state-$A$ energy; Supplementary Table~\ref{tab:s-2d-comparison} gives the model-potential comparisons.

The sampled chain maximum $E_{\rm up}$ includes link, quench-trajectory and end-quench evaluations (Methods Section~\ref{sec:m-chains}). Endpoint certification checks the states reached by a chain; a continuous-path upper bound additionally requires the uncomputed discretization correction derived in Supplementary Section~\ref{prop:bound}. Each system was searched once, so the reported counts describe these outcomes rather than success probabilities or seed variability.

\subsection{Energy-only chains recovered intermediates and competing routes}\label{sec:r-2d}
\begin{figure*}[!t]
\centering
\includegraphics[width=\textwidth]{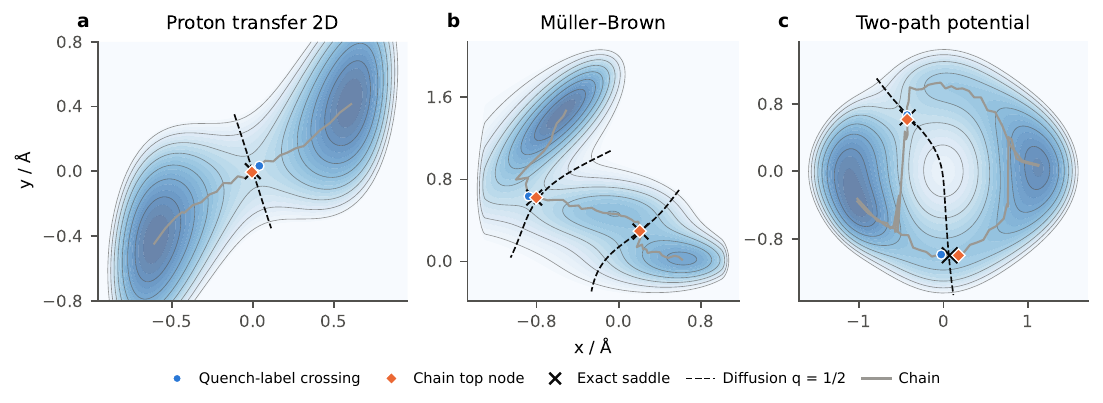}
\caption{\textbf{Recovery of an intermediate and competing reaction channels.} \textbf{a}, Proton transfer coupled to a harmonic bath. \textbf{b}, M\"uller--Brown potential. \textbf{c}, Two-path potential. Grey curves trace retained continuation nodes over potential-energy contours; blue circles denote quench-label crossings, orange diamonds the highest-energy chain nodes (labelled Chain top node) and black crosses the reference saddles. Dashed lines are the $q=1/2$ contours of numerical diffusion references, not learned committors. Coordinates are in \AA. Supplementary Table~\ref{tab:s-2d-comparison} quantifies the saddle-referenced deviations. One search was performed per model.}\label{fig:r-2d}
\end{figure*}

The two-dimensional tests recovered the expected connectivity without supplying an intermediate or a transition-state structure. On the proton-transfer surface, one chain connected the two wells through the saddle region (Fig.~\ref{fig:r-2d}a). On the M\"uller--Brown surface, quench labelling identified the intermediate and separated the route into $A\to I_1$ and $I_1\to B$ chains (Fig.~\ref{fig:r-2d}b). The two-path surface yielded both its lower and upper channels (Fig.~\ref{fig:r-2d}c). The three panels thus test increasing topological complexity: one connection, an intermediate and competing channels. They establish route recovery in these searches, without establishing exhaustive discovery.

Chain-energy agreement and crossing geometry gave different measures of resolution. Across the five two-dimensional saddle comparisons, $E_{\rm up}-E_{\rm saddle}$ ranged from $-0.002$ to $+0.025$~kJ\,mol$^{-1}$ (Supplementary Table~\ref{tab:s-2d-comparison}). Crossing displacements ranged from $0.004$ to $0.101$~\AA. The largest displacement occurred on the nearly flat lower channel of the two-path surface, despite a crossing-energy difference of only $0.006$~kJ\,mol$^{-1}$. Energy agreement therefore did not imply equally precise geometric localization. The small negative chain-energy differences are consistent with finite sampling resolution, as distinct from a continuous-path bound.

\subsection{Most gas-phase chains approached the reference saddle energies}\label{sec:r-gas}
\begin{figure*}[!t]
\centering
\includegraphics[width=\textwidth]{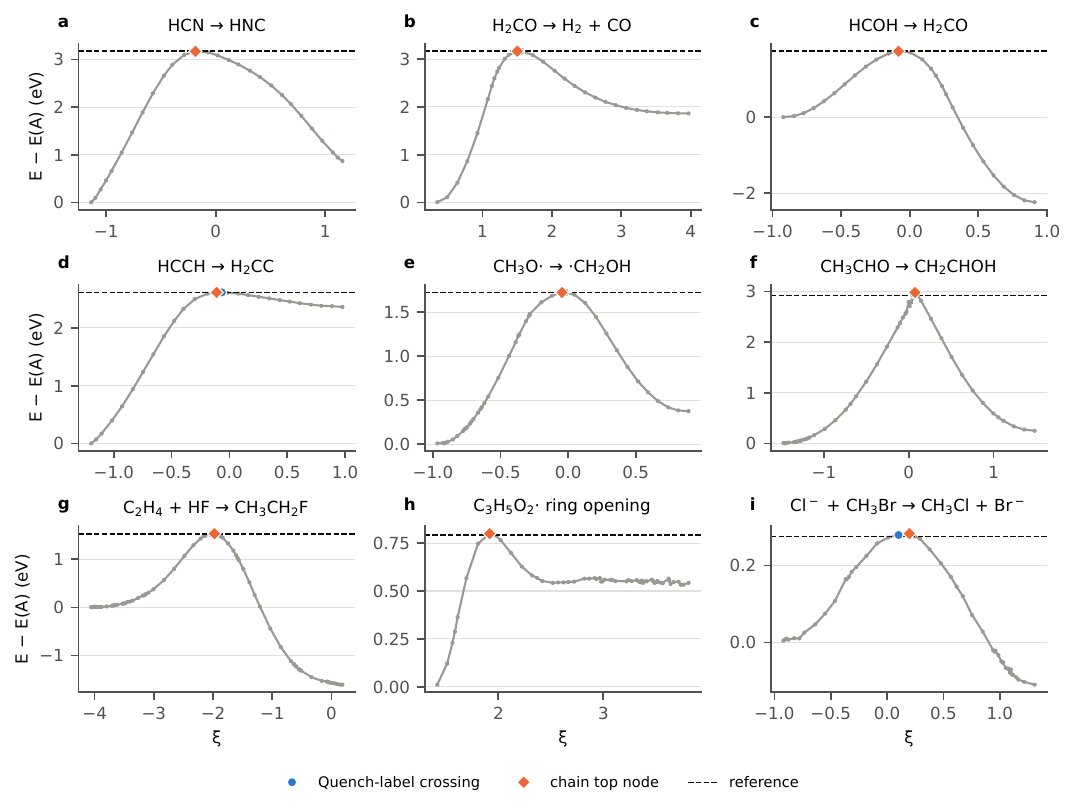}
\caption{\textbf{Gas-phase reaction profiles reconstructed from energy-only continuation.} \textbf{a--i}, Primary continuation chains for HCN isomerization, formaldehyde dissociation, HCOH isomerization, acetylene rearrangement, methoxy isomerization, acetaldehyde tautomerization, HF addition to ethylene, radical ring opening and $\mathrm{S_N2}$ substitution, respectively. Grey points show node energies relative to the lowest recorded state-$A$ minimum, plotted against $\xi$ in \AA, the summed breaking-bond lengths minus the summed forming-bond lengths (Supplementary Section~\ref{sm:coord}). Blue circles mark quench-label crossings, orange diamonds the highest-energy chain nodes and dashed lines the independently refined saddle energies. All energies were evaluated with GFN2-xTB. Lines connect evaluated nodes and do not denote force-verified minimum-energy paths. $E_{\rm up}$ also includes link and quench-trajectory energies and can differ from the orange node maximum (Table~\ref{tab:r-benchmark}). One search was performed per reaction.}\label{fig:r-gas}
\end{figure*}

Eight of the nine GFN2-xTB reactions had $|E_{\rm up}-E_{\rm saddle}|<0.76$~kJ\,mol$^{-1}$ (Table~\ref{tab:r-benchmark}). Their crossing RMSDs were below $0.019$~\AA. In Fig.~\ref{fig:r-gas}a--i, the crossings lie near the principal node maxima across hydrogen transfer, dissociation, addition and ring opening. The local distance comparisons in Supplementary Fig.~\ref{fig:s-r-schematic} complement these energy profiles and the all-atom RMSDs in Table~\ref{tab:r-benchmark}.

Acetaldehyde tautomerization was the exception. Its sampled chain maximum exceeded the reference saddle energy by $6.08$~kJ\,mol$^{-1}$, whereas the crossing lay $3.48$~kJ\,mol$^{-1}$ above it (Fig.~\ref{fig:r-gas}f). Bisection reduced the RMSD from $0.164$~\AA\ at the highest-energy chain node to $0.043$~\AA\ at the crossing. Bisection also improved localization for $\mathrm{S_N2}$ substitution ($0.040$ to $0.018$~\AA) and HF addition to ethylene ($0.023$ to $0.013$~\AA), but did not eliminate the acetaldehyde energy error. Its additional retained chain was symmetry-equivalent and did not establish a different mechanism.

Endpoint relaxation gave a separate energetic check: the nine gas-phase reaction energies differed from independently refined minima by less than $0.08$~kJ\,mol$^{-1}$ (Table~\ref{tab:r-benchmark}). This checks endpoint relaxation on the specified surface, independently of saddle localization. Published HF-3c barriers were excluded from the GFN2-xTB $\mathrm{S_N2}$ comparison (Supplementary Section~\ref{sm:sys-sn2}). The slightly negative HCN chain deviation ($-0.013$~kJ\,mol$^{-1}$) is consistent with discrete undersampling of the saddle, rather than evidence for a lower-energy mechanism. Supplementary Fig.~\ref{fig:s-r-scatter} compares crossing displacement and energy deviation jointly, showing why either measure alone is insufficient.

\subsection{Alanine dipeptide exposed incomplete conformational relaxation}\label{sec:r-alanine}
\begin{figure*}[!t]
\centering
\includegraphics[width=\textwidth]{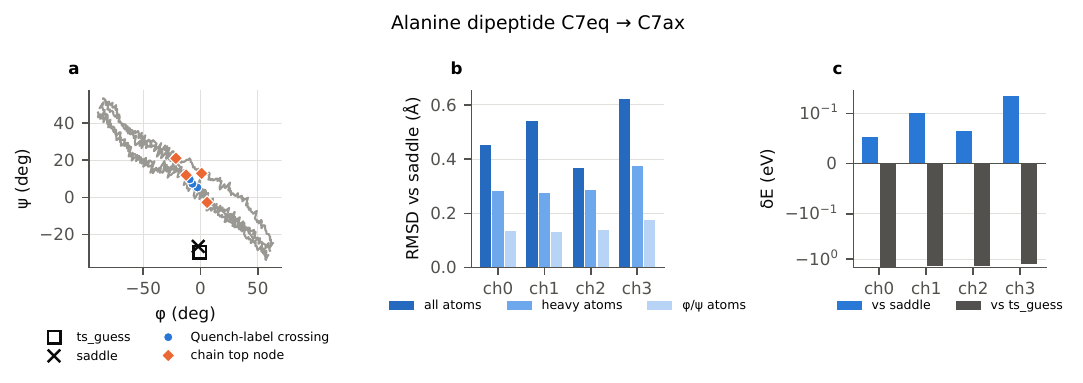}
\caption{\textbf{Residual conformational error in alanine dipeptide.} \textbf{a}, Four chains projected onto the backbone dihedrals $(\phi,\psi)$. Blue circles mark quench-label crossings, orange diamonds the highest-energy chain nodes, the black cross the refined ff99SB saddle and the open square the unrelaxed transition-state guess. \textbf{b}, Crossing RMSDs to the refined saddle, calculated for all atoms, heavy atoms and the atoms defining $\phi/\psi$. \textbf{c}, Signed deviations $\delta E=V(\mathbf R^\ddagger_0)-E_{\rm reference}$ relative to the refined saddle (blue) and the unrelaxed guess (grey), on a symmetric-logarithmic scale. The chain labels ch0--ch3 are shared by b,c; ch0 has the lowest $E_{\rm up}$. All four chains came from one search.}\label{fig:r-alanine}
\end{figure*}

The structural and energetic agreement of the small-molecule tests did not extend to alanine dipeptide. Four chains connected the prescribed conformational states, but their $E_{\rm up}$ values exceeded the refined ff99SB saddle by $11.57$--$25.00$~kJ\,mol$^{-1}$. On the lowest chain, the crossing had all-atom and heavy-atom RMSDs of $0.453$ and $0.284$~\AA, respectively (Fig.~\ref{fig:r-alanine}b). Its energy remained $5.13$~kJ\,mol$^{-1}$ above the reference (Fig.~\ref{fig:r-alanine}c).

The dihedral projection localized the unresolved conformational direction. The lowest-chain crossing had $(\phi,\psi)=(-2.8^\circ,5.4^\circ)$, compared with $(-2.1^\circ,-26.4^\circ)$ at the reference saddle. Across all four crossings, $\psi$ remained between $5.4^\circ$ and $12.6^\circ$ (Fig.~\ref{fig:r-alanine}a). This systematic displacement is consistent with incomplete relaxation along the soft conformational direction, although these data do not isolate its cause. The grey bars in Fig.~\ref{fig:r-alanine}c show how comparison with the unrelaxed guess would conceal this residual error. Endpoint energies exposed a related limitation: $\Delta E$ was $0.122$~eV, compared with $0.062$~eV for the refined minima (Table~\ref{tab:r-benchmark}), an excess of $5.78$~kJ\,mol$^{-1}$ associated with greater residual energy in state $B$.

\subsection{Committor validation remained a separate requirement}\label{sec:r-committor}
Transition-region sampling required validation beyond saddle localization. The insets in Supplementary Fig.~\ref{fig:s-r-sampling}a compare the learned committor with numerical diffusion references. These are diagnostic comparisons: the references use overdamped dynamics and absorbing discs, whereas shooting follows the finite-step Monte Carlo kernel and its prescribed cores.

Molecular shooting tests exposed a narrower validation range. In a development test using the shooting likelihood, the HCOH crossing prediction was $0.43$, against a measured product fraction of $0.60\pm0.11$. HCN predictions remained poorly calibrated. For alanine dipeptide, $14$--$20$ of $20$ probe trajectories failed to reach either core within the prescribed budget. Excluding these trajectories prevents an unconditional committor estimate without assumptions about their eventual outcomes. The molecular evidence therefore supports quench-label localization more strongly than committor calibration. Supplementary Fig.~\ref{fig:s-r-sampling}b,c illustrates a separate sampling distinction: biased walkers cover basins and transition regions, whereas straddling-pair midpoints remain near a chosen crossing surface. The projected free-energy curves in panel d do not extend the molecular calibration evidence.

This capability also required substantial energy evaluation. The nine gas-phase searches used $0.249$--$1.517$ million oracle calls per system; alanine dipeptide used $3.254$ million, for the specified search budgets.

\begin{table*}[!t]
\caption{\textbf{Same-surface activation energies, reaction energies and candidate-structure comparisons.} Both $E_a$ columns use the recorded state-$A$ energy: $E_{\rm saddle}-E_A$ for the independently refined first-order saddle and $E_{\rm up}-E_A$ for tReax\_expedition. Reference $\Delta E$ is the product-minus-reactant energy of independently refined minima; tReax\_expedition $\Delta E=E_B-E_A$ uses the final recorded state minima. All energies are in eV, without zero-point or thermal corrections; $\Delta E$ is a potential reaction energy, not a finite-temperature enthalpy. Crossing RMSD is the all-atom deviation from the reference saddle after alignment and symmetry handling. Gas-phase reactions use GFN2-xTB; alanine dipeptide uses ff99SB. Values are rounded to three decimals, with one search per system.}\label{tab:r-benchmark}
\begin{ruledtabular}
\begin{tabular}{@{}lcrrrrr@{}}
 & & \multicolumn{2}{c}{Reference} & \multicolumn{2}{c}{tReax\_expedition} & \\
\cline{3-4}\cline{5-6}
Reaction & No. of atoms & $E_a$ (eV) & $\Delta E$ (eV) & $E_a$ (eV) & $\Delta E$ (eV) & Crossing RMSD (\AA) \\
\hline
$\mathrm{HCN\to HNC}$ & 3 & 3.175 & 0.868 & 3.175 & 0.868 & 0.013 \\
$\mathrm{H_2CO\to H_2+CO}$ & 4 & 3.167 & 1.862 & 3.168 & 1.863 & 0.003 \\
$\mathrm{HCOH\to H_2CO}$ & 4 & 1.744 & -2.244 & 1.745 & -2.244 & 0.008 \\
$\mathrm{HCCH\to H_2CC}$ & 4 & 2.614 & 2.360 & 2.614 & 2.360 & 0.008 \\
$\mathrm{CH_3O^\bullet\to{}^\bullet CH_2OH}$ & 5 & 1.723 & 0.370 & 1.726 & 0.370 & 0.012 \\
$\mathrm{Cl^-+CH_3Br\to CH_3Cl+Br^-}$ & 6 & 0.275 & -0.117 & 0.283 & -0.116 & 0.018 \\
$\mathrm{CH_3CHO\to CH_2CHOH}$ & 7 & 2.918 & 0.248 & 2.981 & 0.249 & 0.043 \\
$\mathrm{C_2H_4+HF\to CH_3CH_2F}$ & 8 & 1.523 & -1.606 & 1.528 & -1.606 & 0.013 \\
$\mathrm{C_3H_5O_2^\bullet}$ ring opening & 10 & 0.792 & 0.528 & 0.800 & 0.528 & 0.016 \\
\hline
Alanine dipeptide, $\mathrm{C7_{eq}\to C7_{ax}}$ & 22 & 0.347 & 0.062 & 0.467 & 0.122 & 0.453 \\
\end{tabular}
\end{ruledtabular}
\end{table*}

\section{Discussion}\label{sec3}

Energy-only pathway search and statistical characterization of a transition require different evidence. tReax\_expedition connects the exploratory role of MCTSS\cite{tikhonov2026two} to quench-based state assignment, continuation and optional committor learning\cite{kang2024committor,trizio2025everything,jung2023machine}. The benchmarks support recovery of candidate routes without potential-energy gradients, while showing that connectivity alone does not establish a resolved barrier or a calibrated transition probability.

The joint structural and energetic comparisons explain this distinction. A flat saddle region can accommodate geometric displacement with little energy change, whereas an accurate reaction energy checks endpoint relaxation rather than the intervening barrier (Supplementary Fig.~\ref{fig:s-r-scatter}; Table~\ref{tab:r-benchmark}). Alanine extends this limitation to a flexible molecule: its route remains identifiable despite residual errors at both the crossing and the endpoints (Fig.~\ref{fig:r-alanine}). Collective proposals may improve relaxation in soft conformational directions, but the present searches do not isolate the contributions of proposals, coordinates and quench budgets. A comparison under matched oracle budgets would be needed to resolve them.

The sampling displays distinguish two further objectives. Kolmogorov and latent biases enrich transition regions while maintaining broader basin-to-basin coverage\cite{kang2024committor,trizio2025everything}, whereas the local straddling-pair ensemble confines midpoints near a crossing surface (Supplementary Fig.~\ref{fig:s-r-sampling}b,c). Its midpoint weights correct the pair-energy factor but retain the finite-width window and stiff-limit surface measure derived in Supplementary Section~\ref{sm:tse}. The local straddling-pair ensemble therefore cannot supply basin populations or a full latent free-energy profile; the blue segments in panel d mark its restricted support. In the illustrated runs, the side constraint uses the chain coordinate, so proximity to a saddle does not establish either the finite-temperature isocommittor or a reactive-flux distribution.

The green projections in Supplementary Fig.~\ref{fig:s-r-sampling}d instead use MBAR reweighting of biased samples\cite{shirts2008mbar}. Frozen biases make this calculation possible, but do not establish equilibrium, overlap or adequate effective sample size. Training additionally caps weights at the 99th percentile, changing its estimator; whether that cap was used for the supplied free-energy curves is undocumented. The projections consequently illustrate sampling support rather than an uncertainty-controlled free-energy calculation.

Committor interpretation is likewise tied to the chosen dynamics and cores. The network targets a finite-step Monte Carlo first-hitting probability, whereas the variational functional and numerical grid references use a diffusion approximation. Shooting under the actual kernel supplies a separate diagnostic, subject to the initial-core and censoring limitations described in Supplementary Section~\ref{sm:nc-shoot}. The free-particle proposal is quantum Monte Carlo-inspired, but the sampled targets are classical and establish neither nuclear tunnelling nor physical-time reaction rates.

One search per system cannot determine channel-discovery probabilities. The molecular comparisons assess localization on GFN2-xTB or ff99SB, rather than chemical accuracy against experiment, and energies alone do not guarantee a computational advantage over force-based methods. Within these limits, the study provides a way to organize candidate routes while evaluating connectivity, barrier resolution and transition statistics separately.

\section{Methods}\label{sec4}
tReax\_expedition takes two endpoint structures and an energy oracle. Supplementary Algorithm~\ref{alg:search} connects exploration, state assignment, chain construction and the optional committor and ensemble calculations. Sections~\ref{sec:m-kernel}--\ref{sec:m-committor} define the statistical targets and reported quantities. The execution details are concentrated in Supplementary Algorithms~\ref{alg:chain} and \ref{alg:committor} and Supplementary Tables~\ref{tab:s-search}--\ref{tab:s-committor}. The validity domains of the resulting quantities are distinguished in Supplementary Table~\ref{tab:s-classes}.

\subsection{Energy-only sampling in imaginary time}\label{sec:m-kernel}
A configuration of $N$ atoms with masses $m_i$ is written $\mathbf R=(\mathbf r_1,\ldots,\mathbf r_N)\in\mathbb R^{3N}$, and $V(\mathbf R)$ denotes its potential energy. The oracle returns an energy or a failed evaluation. Reported search cost counts oracle calls requested by the algorithm, including failed evaluations. Additional evaluations of pre-screened trials in the two-dimensional models are used only for diagnostics and excluded from this count (Supplementary Section~\ref{sm:sim}).

Structures are compared through the mass-weighted squared distance
\begin{equation}
d^2(\mathbf R,\mathbf R')=\min_{\mathsf Q\in SO(3),\,\mathbf t}\sum_{i=1}^{N} m_i\,\bigl\lVert \mathbf r_i-\mathsf Q\mathbf r'_i-\mathbf t\bigr\rVert^2,
\label{eq:d2}
\end{equation}
where $\mathsf Q$ is a proper rotation and $\mathbf t$ a translation. The minimum is found by centring both structures and applying the mass-weighted Kabsch rotation with the sign correction that excludes reflections\cite{kabsch1976solution,kabsch1978discussion}. $d^2$ is symmetric and invariant under a separate rigid motion of either structure (Supplementary Section~\ref{prop:invariance}). No frame is removed for periodic cells, systems with fixed atoms or the two-dimensional models. For isolated molecules with invariant energies, biases and surrogates, the unaligned kernel is $SE(3)$-equivariant in distribution. The scalar committor is invariant and its Cartesian gradient is rotation-equivariant; frame removal preserves the process of invariant observables (Supplementary Section~\ref{sm:proof-se3}). Coupled walkers retain a common frame, as specified below.

The move is adapted from the Monte Carlo transition-state search (MCTSS) of Tikhonov and Santra\cite{tikhonov2026two}, which obtains it from real-space diffusion Monte Carlo\cite{austin2012quantum} in imaginary time $\tau$ (Supplementary Section~\ref{sm:bg-origin}). A walker at parameter $\tau$ targets
\begin{equation}
\pi_\tau(\mathbf R)=\frac{\exp\bigl[-\beta(\tau)\,V(\mathbf R)\bigr]}{\displaystyle\int\exp\bigl[-\beta(\tau)\,V(\mathbf R')\bigr]\,\mathrm d\mathbf R'},
\label{eq:target}
\end{equation}
where $\beta(\tau)=\tau/\hbar$ is an inverse effective temperature. The integral is over the chosen configuration domain and assumes a finite partition function. For free molecules, translations are removed; dissociating fragments require confinement for an equilibrium interpretation (Supplementary Section~\ref{prop:frame}). The integral is never evaluated, because the acceptance test uses only ratios of $\pi_\tau$. Here $\tau$ sets the effective thermal energy $k_{\rm B}T_{\rm eff}=\hbar/\tau$, $0.66$~eV at $\tau=1$~fs. It is a sampling parameter rather than a physical time, and equation~\eqref{eq:target} is a classical Boltzmann factor. The physical temperature corresponds to $\tau^\ast=\hbar/(k_{\rm B}T_{\rm phys})$, or $25.6$~fs at $298.15$~K.

A trial displaces one uniformly chosen atom $i$ by a Gaussian vector of covariance $\sigma_i^2\mathsf I_3$, with $\sigma_i=\sqrt{\hbar\tau/(2m_i)}$ as in the free-particle kernel of MCTSS\cite{tikhonov2026two}. For a single walker the trial is then re-centred and rotated onto the current structure by the proper Kabsch rotation. Every factor of the acceptance probability is invariant under proper rigid motions, so this frame removal leaves the sampled distribution of every invariant quantity exactly unchanged (Supplementary Section~\ref{prop:frame}). MCTSS also places the kinetic factor in the acceptance probability\cite{tikhonov2026two}. We use it only as the proposal, because the kinetic factor in the acceptance breaks detailed balance with respect to equation~\eqref{eq:target} (Supplementary Section~\ref{prop:kinetic}).

Acceptance follows a delayed-acceptance Metropolis--Hastings rule\cite{metropolis1953equation,hastings1970monte,christen2005mcmc}. For a move $\mathbf R\to\mathbf R'$ the acceptance probability is
\begin{equation}
\begin{split}
\alpha={}&\min\bigl(1,e^{\Delta c}\bigr)\,\min\bigl(1,e^{-\beta\Delta(\tilde V+b)}\bigr)\\
&\times\min\bigl(1,e^{-\beta\Delta(V-\tilde V)}\bigr),
\end{split}
\label{eq:da}
\end{equation}
where $\Delta f=f(\mathbf R')-f(\mathbf R)$ for any function $f$. Here $c$ is the logarithm of a geometric coupling factor that belongs to the target, $b$ an optional bias and $\tilde V$ an optional surrogate energy. The three factors are tested in turn, each with its own uniform random number, and the first rejection ends the move. The oracle is therefore called only after the cheap tests pass. The surrogate, GFN2-xTB\cite{bannwarth2019gfn2} screening ORCA energies, is used only where stated. Without a surrogate the last two factors merge into $\min(1,e^{-\beta\Delta(V+b)})$. The product of the three ratios equals the Metropolis--Hastings ratio of the target
\begin{equation}
\pi(\mathbf R)=\frac{\exp\bigl[c(\mathbf R)-\beta V(\mathbf R)-\beta b(\mathbf R)\bigr]}{\displaystyle\int\exp\bigl[c(\mathbf R')-\beta V(\mathbf R')-\beta b(\mathbf R')\bigr]\,\mathrm d\mathbf R'},
\label{eq:targetcb}
\end{equation}
where the denominator normalizes $\pi$ over all configurations. For a symmetric unaligned proposal, the move satisfies detailed balance for any fixed $c$, $b$ and $\tilde V$. Frame removal additionally requires each acceptance factor to be rigid-motion invariant (Supplementary Sections~\ref{prop:da} and \ref{prop:frame}). A failed oracle call rejects the trial, as if $V=+\infty$.

\subsection{Bidirectional exploration with coupled walkers}\label{sec:m-explore}
Exploration starts two walkers $\mathbf R_1$ and $\mathbf R_2$ at the endpoints, relaxed by the quench of Section~\ref{sec:m-chains}. It updates them alternately with the move of Section~\ref{sec:m-kernel} on the pair target
\begin{equation}
\begin{split}
\Pi(\mathbf R_1,\mathbf R_2)={}&\frac{1}{Z_\Pi}\exp\bigl\{-\beta_{\rm b}\bigl[V(\mathbf R_1)+V(\mathbf R_2)\bigr]\bigr\}\\
&\times\exp\Bigl[-\frac{d_{\rm c}^2(\mathbf R_1,\mathbf R_2)}{\hbar\tau_{12}}\Bigr],
\end{split}
\label{eq:pair}
\end{equation}
where $Z_\Pi$ is the integral of the numerator over $\mathbf R_1$ and $\mathbf R_2$, $\beta_{\rm b}=\tau_{\rm b}/\hbar$ for the bead parameter $\tau_{\rm b}$, and $\tau_{12}=n_{12}\tau_{\rm b}$ is the coupling parameter. The coupling distance $d_{\rm c}^2(\mathbf R_1,\mathbf R_2)=\sum_im_i\lVert\bar{\mathbf r}_{1,i}-\bar{\mathbf r}_{2,i}\rVert^2$ is taken between the centred walkers $\bar{\mathbf R}_1$ and $\bar{\mathbf R}_2$ in their common frame, without rotating one onto the other. $\Pi$ is therefore invariant under a common rigid motion of both walkers and under a separate translation of either, but not under a separate rotation. Trials are centred and never rotated. For free molecules the relative orientation is instead sampled once per sweep by a rigid rotation of $\mathbf R_2$ about a uniformly drawn axis, by a Gaussian angle of standard deviation $\min\{0.5,(\hbar\tau_{12}/\bar I)^{1/2}\}$~rad. Here $\bar I$ is the mean principal moment of inertia of $\mathbf R_2$. The rotation leaves $V$ unchanged, so it is accepted on the coupling alone and costs no oracle call. A coupling on the rotation-minimized $d^2$ would make $\Pi$ invariant under a separate rotation of either walker and would spoil the midpoint identity below (Supplementary Section~\ref{prop:pairframe}).

MCTSS uses the same free-particle factor to pre-select trials that move towards the other walker and redraws a trial until it passes\cite{tikhonov2026two}. In tReax\_expedition each trial is tested once, as the first factor of equation~\eqref{eq:da} with $\Delta c=-[d_{\rm c}^2(\mathbf R_1',\mathbf R_2)-d_{\rm c}^2(\mathbf R_1,\mathbf R_2)]/(\hbar\tau_{12})$, and a failed test leaves the walker in place. The coupling thus becomes part of the target. Redrawing until a trial passes would instead tilt the stationary distribution by the pass probability (Supplementary Section~\ref{prop:prescreen}).

Exploration stops at a join and therefore does not provide an equilibrium sample of $\Pi$. For a stationary pair, define the midpoint $\mathbf M=(\bar{\mathbf R}_1+\bar{\mathbf R}_2)/2$ and separation $\mathbf D=\bar{\mathbf R}_1-\bar{\mathbf R}_2$. The weight
\begin{equation}
w=\exp\bigl\{\beta_{\rm b}\bigl[V(\mathbf R_1)+V(\mathbf R_2)-2V(\mathbf M)\bigr]\bigr\}
\label{eq:midweight}
\end{equation}
gives a midpoint marginal proportional to $\exp[-2\beta_{\rm b}V(\mathbf M)]$, provided the normalization exists and the relative-coordinate integration domain is independent of the midpoint. The latter holds on the full centred linear configuration space; hard domain restrictions can introduce an additional support factor. The identity follows from a linear change of variables, not a harmonic approximation (Supplementary Section~\ref{prop:pairframe}). Thus $\tau_{\rm mid}=2\tau_{\rm b}$. It does not make stopped exploration trajectories equilibrium samples; Supplementary Table~\ref{tab:s-classes} distinguishes these uses.

A run ends at the MCTSS meeting criterion\cite{tikhonov2026two}, $d^2(\mathbf R_1,\mathbf R_2)/(\hbar\tau_{\rm b})\le\tfrac32Ns_{\rm meet}$ with $s_{\rm meet}=1$ for molecules. This test uses the rotation-minimized $d^2$ of equation~\eqref{eq:d2}, because the walkers meet when their shapes do. A run also ends when the two walkers overlap along the mass-weighted axis between the endpoints, or after 5000 steps. The first two outcomes count as a join. The path consists of the walker-1 configurations followed by the walker-2 configurations in reverse order, written in one frame by a common rotation of both walkers.

The bead parameter is chosen from join outcomes on a ladder that first searches for a successful rung and then doubles $\tau_{\rm b}$ until failure. Supplementary Algorithm~\ref{alg:search} gives the stopping logic; Supplementary Table~\ref{tab:s-search} specifies its run budgets and thresholds. The coldest successful rung defines $\tau_{\rm x}$, the lowest effective temperature at which the walkers join within the search budget. This heuristic depends only on the observed join outcomes. $\tau_{\rm x}$ also fixes subsequent proposal widths and $\ell=\sqrt{3\hbar\tau_{\rm x}/2}$, the root-mean-square mass-weighted length of one single-atom move. The resulting joined paths enter the state and channel analysis of Section~\ref{sec:m-chains}.

\subsection{States, candidate channels and sampled chain maxima}\label{sec:m-chains}
States are assigned by quenching with the single-walker move of Section~\ref{sec:m-kernel} over decreasing effective temperatures. Proposal widths adapt to acceptance, so a quench is a stochastic optimizer rather than an equilibrium sampler. Supplementary Table~\ref{tab:s-quench} separates the schedules for relaxation, labelling and chain construction. Their budgets and temperature scaling are defined in Supplementary Section~\ref{sm:quench}.

The quenched configuration is assigned a state by its fingerprint. For chemical reactions this is the bond graph, with atoms $i$ and $j$ bonded when $r_{ij}<r_i^{\rm cov}+r_j^{\rm cov}+0.4$~\AA. For alanine dipeptide it is the backbone dihedral pair $(\phi,\psi)$ matched within $60^\circ$, and for the two-dimensional models the position of the minimum. The molecular fingerprints are invariant under proper rigid motions (Supplementary Section~\ref{prop:invariance}). The endpoints define states $A$ and $B$, a new fingerprint registers an intermediate $I_k$, and every state keeps the lowest minimum found for it (Supplementary Section~\ref{sm:states}).

Up to eight joined exploration paths (twelve for the two-dimensional models), taken from rungs with $\tau_{\rm b}\le\tau_{\rm x}$, are labelled by bisection. The middle frame of every interval whose end labels differ is quenched until each label change lies between adjacent frames. Each such pair of frames is a transition between two states. Excursions and intermediates between frames with the same assigned label can be missed. Candidate state pairs are subsequently checked by continuation and endpoint quenches.

Each connected pair of states $(S_a,S_b)$ receives a progress coordinate $\xi(\mathbf R)$ built from the two fingerprints. For bond graphs, $\xi$ is the summed length of the bonds that break minus that of the bonds that form. For dihedral fingerprints, $\xi$ projects the wrapped dihedral change onto the change between the two minima. Otherwise $\xi$ is the latent output of a classifier trained on the path labels (Supplementary Section~\ref{sm:coord}). Chains are grown from up to four transition frames (eight in two dimensions), chosen by farthest-point selection in $d$.

The interval of $\xi$ between the two minima is divided into slices of width $\Delta\xi$. Node $k$ is obtained by quenching under a harmonic restraint, as in umbrella sampling\cite{torrie1977nonphysical},
\begin{equation}
b_k(\mathbf R)=\tfrac12\kappa\bigl[\xi(\mathbf R)-\xi_k\bigr]^2,
\label{eq:restraint}
\end{equation}
where $\xi_k$ is the slice centre and $\kappa=\kappa_0/\Delta\xi^2$. Each node starts from its predecessor. A gradual restraint ramp and, when needed, restraint stiffening keep it near its target slice. Large gaps receive intermediate nodes. Unrestrained end quenches identify the states reached by the chain. Supplementary Algorithm~\ref{alg:chain} specifies this sequence; Supplementary Table~\ref{tab:s-search} fixes its resolution and tolerances.

Consecutive nodes are joined by two connected discrete paths. On the straight link of mass-weighted length $L$, $V$ is evaluated at $\max(4,\lceil4L/\ell\rceil)$ interior points. The quench trajectory, the sequence of accepted configurations that grew a node from its parent, supplies the second path at no extra cost. Keeping the lower maximum of each segment gives the sampled chain maximum
\begin{equation}
\begin{split}
E_{\rm up}=\max\Bigl\{&V(\mathbf R_{\rm seed}),\,\max_{k}\min\bigl(E_k^{\rm link},E_k^{\rm traj}\bigr),\\
&E_a^{\rm desc},\,E_b^{\rm desc}\Bigr\},
\end{split}
\label{eq:eup}
\end{equation}
where $\mathbf R_{\rm seed}$ is the first node, $E_k^{\rm link}$ and $E_k^{\rm traj}$ are the largest energies on the link and quench trajectory of segment $k$, and $E_a^{\rm desc}$ and $E_b^{\rm desc}$ are the largest energies encountered during the two end quenches. A chain is certified when these quenches reach $S_a$ and $S_b$. $E_{\rm up}$ is then the largest energy on a connected sequence of evaluated configurations between the two basins. If $V$ is $L_V$-Lipschitz on the interpolating segments, the continuous minimax energy is at most $E_{\rm up}+\tfrac12L_Vh$. Here $h$ includes accepted displacements before frame removal and link-evaluation spacings (Supplementary Section~\ref{prop:bound}). Neither $L_V$ nor this correction is computed. Certification therefore concerns endpoint connectivity and evaluated energies, rather than an unconditional continuous-path bound (Supplementary Table~\ref{tab:s-classes}). Barriers are reported as $\Delta E_{\rm up}=E_{\rm up}-E(S_a)$ relative to the lowest minimum found for $S_a$.

Certified chains whose highest-energy chain nodes lie within $3\ell$ of each other's nodes form one channel, represented by its lowest chain. On the state graph weighted by $E_{\rm up}$, the route from $A$ to $B$ minimizes the largest sampled edge value\cite{pollack1960maximum}. A continuous-path upper bound requires the discretization correction for every edge and connections between the end configurations assigned to each intermediate state (Supplementary Section~\ref{prop:route}). If committor learning is requested and no chain of a pair is certified, the committor is learned from the transition frames first and the chains are regrown along its latent $z_\theta$.

\subsection{Quench-label crossing and local ensemble}\label{sec:m-crossing}
The majority-quench label of Section~\ref{sec:m-chains} is bisected along the lowest chain of each channel, first over its nodes and then four times along the straight link that holds the change. Each label is the majority of three quenches with the label schedule of Supplementary Table~\ref{tab:s-quench}, to reduce sensitivity to an individual stochastic quench near the ridge. The midpoint of the final interval is the crossing $\mathbf R^\ddagger_0$, with energy $E^\ddagger_0$. It is the midpoint of an interval with opposing majority-quench labels, a finite-budget stochastic construction that need not coincide with the isocommittor surface $q=1/2$ at $T_{\rm phys}$. The energy curvature along the chain tangent at the highest-energy chain node gives a harmonic frequency $\omega$ and the crossover temperature $T_{\rm c}=\hbar\lvert\omega\rvert/(2\pi k_{\rm B})$ as diagnostics (Supplementary Section~\ref{sm:crossing}).

On request, a local straddling-pair ensemble at $T_{\rm phys}$ is sampled by a pair that straddles the surface $\xi=\xi(\mathbf R^\ddagger_0)$. The pair follows equation~\eqref{eq:pair} of Section~\ref{sec:m-explore}, including the rotation of its relative orientation, with $\tau_{\rm b}=\tau^\ast/2$ and $\tau_{12}=0.05\,\tau_{\rm b}$. Each walker is confined to its own side of the surface. With the weights of equation~\eqref{eq:midweight}, the midpoint targets $\exp(-\beta^\ast V)$, where $\beta^\ast=\tau^\ast/\hbar$, within a window around the surface whose width is set by the coupling. For stiff coupling and locally linear $\xi$, its limiting density is proportional to $e^{-\beta^\ast V}|\nabla\xi|_{\mathsf W^{-1}}\delta(\xi-\xi^\ast)$, where $\xi^\ast=\xi(\mathbf R^\ddagger_0)$ and $\mathsf W=\mathrm{diag}(m_1,\ldots,m_N)\otimes\mathsf I_3$ (Supplementary Section~\ref{sm:tse}). Every ensemble reports the Kish effective sample size of its weights\cite{kish1965survey}. This diagnoses weight concentration and does not account for autocorrelation.

\subsection{Committor learning at the physical temperature}\label{sec:m-committor}
Committor learning uses the states, progress coordinates and chains of Section~\ref{sec:m-chains}. For a pair $(S_a,S_b)$, the core $C_a$ contains configurations with the fingerprint of $S_a$ and normalized progress $u=(\xi-\xi_a)/(\xi_b-\xi_a)\le0.2$, and $C_b$ those of $S_b$ with $u\ge0.8$. The committor is the first-hitting probability
\begin{equation}
q(\mathbf R)=\Pr\nolimits_{\mathbf R}\bigl[t_{C_b}<t_{C_a}\bigr],
\label{eq:committor}
\end{equation}
where $t_C$ is the first step at which an unbiased chain of the move of Section~\ref{sec:m-kernel} at $\tau^\ast$, started at $\mathbf R$, enters $C$\cite{du1998transition,bolhuis2002tps,e2010transition}. It is therefore the committor of the Monte Carlo chain, not of Newtonian dynamics. For small steps, the move approximates overdamped Langevin dynamics with diffusion constants $D_i=\hbar\tau^\ast/(4Nm_i)$ per step\cite{kikuchi1991metropolis} (Supplementary Section~\ref{prop:generator}). The committor of this diffusion minimizes
\begin{equation}
K[q]=\bigl\langle\lvert\nabla_{\!u}q\rvert^2\bigr\rangle_{\beta^\ast}
\label{eq:ritz}
\end{equation}
under the conditions $q=0$ on $C_a$ and $q=1$ on $C_b$\cite{e2006transition,kang2024committor}, where the average is over $\exp(-\beta^\ast V)$ and $\lvert\nabla_{\!u}q\rvert^2=\sum_{j=1}^{3N}m_j^{-1}(\partial q/\partial x_j)^2$ is the squared gradient in the inverse mass metric. Here $m_j$ is the mass of Cartesian coordinate $x_j$, and the inverse masses follow from the diffusion constants (Supplementary Section~\ref{prop:ritz}). We treat the small-step correspondence as a modelling assumption. Shooting provides a separate diagnostic of equation~\eqref{eq:committor}, with initial-core and finite-budget limitations described in Supplementary Section~\ref{sm:nc-shoot}.

The model is $q_\theta(\mathbf R)=\sigma\bigl(s\,z_\theta(\mathbf R)\bigr)$, where $\sigma$ is the logistic function, $s=3$ and $z_\theta$ is the latent output of a fully connected network, following fully connected neural committor models\cite{khoo2019solving,li2019computing,kang2024committor}. The network reads all interatomic distances for molecules and the coordinates in two dimensions (Supplementary Section~\ref{sm:nc-net}), so the molecular $q_\theta$ is invariant under proper rigid motions and its Cartesian gradient is rotation-equivariant (Supplementary Section~\ref{prop:equivariance}). The intended full-data training objective is
\begin{equation}
\begin{split}
\mathcal L(\theta)={}&\sum_n w_n\lvert\nabla_{\!u}q_\theta(\mathbf R_n)\rvert^2+\gamma\,\ell_{\rm shot}(\theta)\\
&+\alpha\bigl[\langle q_\theta^2\rangle_{C_a}+\langle(1-q_\theta)^2\rangle_{C_b}\bigr],
\end{split}
\label{eq:loss}
\end{equation}
where $w_n$ are the normalized, capped reweighting estimates used in training, $\alpha=1$ weights the boundary penalty and $\gamma$ weights the shooting term $\ell_{\rm shot}$. The gradient term and the boundary penalty follow Kang et al.\cite{kang2024committor}. The term $\ell_{\rm shot}$ is the binomial negative log-likelihood of shooting outcomes per trajectory. It goes back to the likelihood maximization of Peters and Trout\cite{peters2006obtaining} and is used in the same form in artificial intelligence for molecular mechanism discovery (AIMMD)\cite{jung2023machine} and in the path sampling built on it\cite{lazzeri2023molecular}. Its weight $\gamma$ equals $\Gamma$ times the ratio of the gradient term to $\ell_{\rm shot}$ at the start of each fit, with $\Gamma=1$ for molecules and $\Gamma=0$ in two dimensions. With the boundary conditions imposed as a penalty, the minimizer depends on $\alpha$ and $\mathcal L$ is not a strict variational bound\cite{petersen2026trial}. Normalizing weights within each minibatch introduces a further estimator approximation (Supplementary Section~\ref{sm:nc-loss}). Shooting outcomes provide diagnostics separate from the training loss.

Training data are generated iteratively, following Kang et al.\cite{kang2024committor} and Trizio et al.\cite{trizio2025everything}. Unbiased walks from both minima at $\tau^\ast$ provide the core configurations and a first data block. In each of six iterations, walkers started from the minima and from the highest-energy chain node of each channel (Section~\ref{sec:m-chains}) are driven by $V_{\rm bias}=V_{\rm K}+V_z$, which enters the move of Section~\ref{sec:m-kernel} as the bias $b$. The Kolmogorov bias\cite{kang2024committor,trizio2025everything} is
\begin{equation}
\begin{split}
V_{\rm K}={}&-\frac{\lambda}{\beta^\ast}\Bigl[\ln\bigl(\lvert\nabla_{\!u}z_\theta\rvert^2+\epsilon\bigr)-\ln\epsilon\\
&+2\ln(q_\theta+\delta)+2\ln(1-q_\theta+\delta)\Bigr],
\end{split}
\label{eq:vk}
\end{equation}
where $\lambda$ sets the bias strength and $\epsilon,\delta$ regularize the logarithms (Supplementary Table~\ref{tab:s-committor}). For vanishing regularizers $V_{\rm K}$ equals $-(\lambda/\beta^\ast)\ln\lvert\nabla_{\!u}q_\theta\rvert^2$ up to a constant, so it concentrates sampling on the transition region (Supplementary Section~\ref{sm:pinc-vk}). $V_z$ is a well-tempered metadynamics bias\cite{laio2002escaping,barducci2008welltempered} along $z_\theta$ that drives crossings between the basins. Biasing along a learned coordinate is now common in enhanced sampling\cite{mehdi2024enhanced}, and this role is played by on-the-fly probability enhanced sampling\cite{invernizzi2020rethinking} in Trizio et al.\cite{trizio2025everything}. The walkers first build $V_z$. The bias is then frozen and a production block is recorded under it. The code assigns approximate weights using the multistate Bennett acceptance ratio (MBAR)\cite{shirts2008mbar}, followed by percentile capping (Supplementary Section~\ref{sm:pinc-mbar}). Capping changes the estimator; its bias is not quantified. The move needs only bias values. The search and learning stages use no derivatives of the potential-energy oracle; network derivatives enter the bias and loss. Independent reference refinement and Hessian calculations use potential-energy derivatives for evaluation only.

After each production block, shooting points are drawn without replacement from the chain nodes and the new block. Candidate $n$ is drawn with probability
\begin{equation}
P_n=\frac{\bigl[1+(z_n/0.5)^2\bigr]^{-1}/\rho(z_n)}{\sum_{m}\bigl[1+(z_m/0.5)^2\bigr]^{-1}/\rho(z_m)},
\label{eq:select}
\end{equation}
where $z_n=z_\theta(\mathbf R_n)$ and $\rho$ is the histogram density of the candidate latents. This is the Lorentzian selection around $q=1/2$ of AIMMD\cite{jung2023machine}, which concentrates shooting on the transition region as in shooting from the top\cite{jung2017transition}, here flattened against the sampling density. Two unbiased chains at $\tau^\ast$ run from each point until they enter a core, following the shooting logic of transition path sampling\cite{dellago1998efficient,bolhuis2002tps,lazzeri2023molecular}. Their outcomes score the current model before refitting (Supplementary Section~\ref{sm:nc-shoot}). Finally, 20 chains from each crossing $\mathbf R^\ddagger_0$ give the complete-case fraction $p_B=n_B/(n_A+n_B)$ and its binomial standard error for comparison with $q_\theta(\mathbf R^\ddagger_0)$. Supplementary Algorithm~\ref{alg:committor} gives the iterative loop, Supplementary Table~\ref{tab:s-committor} lists its defaults, and Supplementary Table~\ref{tab:s-sysnc} records system-dependent descriptor sizes and shot budgets.

\subsection{Computational details}\label{sec:m-systems}

The workstation had an Intel Xeon 6520P CPU, 1.2~TB of memory and two NVIDIA RTX PRO 6000 Blackwell GPUs, each with 96~GB of VRAM. tReax\_expedition is written in Python 3.12 with NumPy\cite{harris2020numpy} and SciPy\cite{virtanen2020scipy}. Structures were handled with ASE\cite{larsen2017ase}, GFN2-xTB energies were evaluated with tblite, and HF-3c energies\cite{sure2013hf3c} with ORCA\cite{neese2020orca}. Networks were built in PyTorch\cite{paszke2019pytorch} in double precision and trained with Adam\cite{kingma2015adam}. Default settings, random-seed conventions and the software environment are given in Supplementary Sections~\ref{sm:hyper} and \ref{sm:repro} and Supplementary Tables~\ref{tab:s-search}--\ref{tab:s-committor}.

\section*{Data availability}
\phantomsection\label{sec:data-availability}
Benchmark models and starting structures were obtained from three sources. The two-dimensional proton-transfer model and the $\mathrm{Cl^-+CH_3Br\to CH_3Cl+Br^-}$ substitution inputs were adapted from the Supplementary Material of Tikhonov and Santra\cite{tikhonov2026two}. The M\"uller--Brown and two-path potentials and alanine dipeptide inputs were based on the models and data of Trizio, Kang and Parrinello\cite{trizio2025everything}, whose supporting data are available in their published repository\cite{trizio2025data}. Input geometries for the eight remaining gas-phase reactions were obtained from the SACTS benchmark of Suresh and Vogiatzis\cite{suresh2026sacts}.

\section*{Code availability}
The tReax\_expedition implementation is at \url{https://github.com/FrancisPhillipsBuger/tReax_expedition}. 

\providecommand{\bibcommenthead}{}
\bibliographystyle{nature}
\bibliography{literature}

\newpage
\ifdefined\MainDocument\else
\documentclass[%
 10pt,
 preprint,
 amsmath,amssymb,
 aps,
floatfix,
]{revtex4-1}

\usepackage{graphicx}
\usepackage{amsmath,amssymb,xurl}
\usepackage{dcolumn}
\usepackage{bm}
\usepackage[fontsize=8.5pt]{fontsize}
\usepackage{setspace}
\usepackage{booktabs}
\usepackage{array}
\usepackage{algpseudocode}
\usepackage{needspace}
\usepackage{etoolbox}
\usepackage{xcolor}
\usepackage{xr-hyper}
\usepackage[colorlinks=true,linkcolor=blue,citecolor=blue,urlcolor=blue,
  filecolor=blue,bookmarksnumbered=true]{hyperref}
\ifdefined\MainDocument\else\externaldocument[main-]{main}\fi
\ifdefined\MainDocument\else
\makeatletter
\AtBeginDocument{%
  \def\section{\@startsection{section}{1}{\z@}{0.35cm \@plus.2ex \@minus.1ex}{0.1cm}%
    {\normalfont\small\bfseries\preprintsty@sw{}{\centering}}}%
  \def\subsection{\@startsection{subsection}{2}{\z@}{0.3cm \@plus.2ex \@minus.1ex}{0.05cm}%
    {\normalfont\small\bfseries\preprintsty@sw{}{\centering}}}%
  \def\subsubsection{\@startsection{subsubsection}{3}{\z@}{0.2cm \@plus.2ex \@minus.1ex}{0.05cm}%
    {\normalfont\small\itshape\preprintsty@sw{}{\centering}}}%
}
\makeatother
\fi


\begin{document}
\fi

\ifdefined\MainDocument
\onecolumngrid
\clearpage
\RestoreFrontMatter
\setcounter{section}{0}
\setcounter{figure}{0}
\setcounter{table}{0}
\setcounter{equation}{0}
\fi

\title{Supplementary Information for All Paths All at Once: A Quantum Monte Carlo and Neural Committor Approach to Reaction Pathway Search}
\author{Chien-Chai Chang} %
\affiliation{
Department of Engineering and System Science, National Tsing Hua University (NTHU), Hsinchu City, 300044, Taiwan
}

\ifdefined\MainDocument
\makeatletter
\patchcmd{\titleblock@produce}{\label{FirstPage}}{\label{FirstPageSupplement}}{}%
  {\PackageWarning{supplement}{supplement title label patch failed}}
\makeatother
\fi
\maketitle
\setstretch{1.0}
\raggedbottom
\makeatletter
\@booleantrue\raggedcolumn@sw
\makeatother
\emergencystretch=1em

\renewcommand{\thefigure}{S.\arabic{figure}}
\renewcommand{\thetable}{S.\arabic{table}}
\renewcommand{\theequation}{S.\arabic{equation}}

\ifdefined\MainDocument
\newcommand{\MainEq}[1]{\eqref{#1}}
\newcommand{\MainSec}[1]{\ref{#1}}
\newcommand{\MainTable}[1]{\ref{#1}}
\newcommand{\MainFig}[1]{\ref{#1}}
\else
\newcommand{\MainEq}[1]{\eqref{main-#1}}
\newcommand{\MainSec}[1]{\ref{main-#1}}
\newcommand{\MainTable}[1]{\ref{main-#1}}
\newcommand{\MainFig}[1]{\ref{main-#1}}
\fi

\setcounter{subsection}{0}
\renewcommand{\thesubsection}{\thesection.\arabic{subsection}}
\newcommand{\TabCell}[2]{\parbox[t]{#1}{\raggedright #2\strut}}
\newcounter{suppalg}
\renewcommand{\thesuppalg}{\arabic{suppalg}}
\newcommand{\SuppAlgorithm}[2]{\refstepcounter{suppalg}\label{#1}%
  \noindent\textbf{Supplementary Algorithm~\thesuppalg.} #2\par\vspace{2pt}\hrule\vspace{3pt}}
\makeatletter
\def\ALG@noputindents{\hskip\ALG@tlm\hangindent\dimexpr\ALG@tlm+1em\relax\hangafter1\relax}
\patchcmd{\ALG@doentity}{\noindent\hskip\ALG@tlm}%
  {\noindent\hskip\ALG@tlm\hangindent\dimexpr\ALG@tlm+1em\relax\hangafter1\relax}{}{\PackageWarning{supplement}{algorithm hang patch failed}}
\makeatother
\newcommand{\SuppTableCaption}[2]{\refstepcounter{table}\label{#2}%
  {\small\noindent\MakeUppercase{\tablename}~\thetable. #1\par}\vspace{4pt}}
\newcommand{\SuppFigureCaption}[2]{\par\vspace{6pt}\refstepcounter{figure}\label{#2}%
  {\small\raggedright\noindent\MakeUppercase{\figurename}~\thefigure. #1\par}}
\newenvironment{SuppTable}%
  {\ifdefined\MainDocument\onecolumngrid\fi
   \par\addvspace{8pt}\noindent\begin{minipage}{\textwidth}\small}%
  {\end{minipage}\par\addvspace{8pt}%
   \ifdefined\MainDocument\twocolumngrid\fi}
\newenvironment{SuppColTable}[2]%
  {\par\addvspace{8pt}\noindent\begin{minipage}{\columnwidth}%
   \SuppTableCaption{#2}{#1}\footnotesize}%
  {\end{minipage}\par\addvspace{8pt}}
\newenvironment{SuppAlgBlock}%
  {\par\addvspace{8pt}\noindent\begin{minipage}{\linewidth}%
   \small\raggedright\hrule\vspace{3pt}}%
  {\par\vspace{2pt}\hrule\end{minipage}\par\addvspace{8pt}}
\newenvironment{SuppFigure}%
  {\par\addvspace{8pt}\noindent\begin{minipage}{\linewidth}\centering}%
  {\end{minipage}\par\addvspace{8pt}}
\makeatletter
\renewcommand{\p@subsection}{}
\makeatother

\section{Background}\label{sm:bg}
\subsection{Notation, units and effective temperatures}\label{sm:units}
Energies are in eV, lengths in \AA, masses in atomic mass units (amu) and the time parameters $\tau$ in fs. The inverse effective temperature of one walker is $\beta(\tau)=\tau/\hbar$, so $k_{\rm B}T_{\rm eff}=\hbar/\tau$. With $\hbar=0.6582$~eV\,fs, $\tau=1$~fs corresponds to $0.658$~eV ($63.5$~kJ\,mol$^{-1}$), and each doubling of $\tau$ halves this value. The physical temperature enters only through $\tau^\ast=\hbar/(k_{\rm B}T_{\rm phys})$, which is $25.62$~fs at $298.15$~K and $25.46$~fs at $300$~K. For the two-dimensional models in reduced units, one energy unit is $k_{\rm B}\times300$~K and $\tau^\ast=25.46$~fs. For the proton-transfer model, whose energies are in cm$^{-1}$, $\tau^\ast=10$~fs places the barrier near $8\,k_{\rm B}T^\ast$.

The Gaussian proposal of the kernel has standard deviation $\sigma_i=\lambda_{\rm s}\sqrt{\hbar\tau/(2m_i)}$ per Cartesian component. One single-atom move therefore has mean mass-weighted squared length $3m_i\sigma_i^2=\tfrac32\hbar\tau\lambda_{\rm s}^2$, independent of the atom, which defines $\ell=\sqrt{3\hbar\tau_{\rm x}/2}$ for $\lambda_{\rm s}=1$. At $\tau=1$~fs, $\ell=0.098$~amu$^{1/2}$\,\AA.

\subsection{Origin of the Monte Carlo move}\label{sm:bg-origin}
MCTSS derives its move from real-space diffusion Monte Carlo\cite{tikhonov2026two,austin2012quantum}. An imaginary time step $\Delta t=-{\rm i}\tau$ and the squared second-order Trotter kernel of $\exp(-\tau\hat H/\hbar)$ give the unnormalized transition weight used by MCTSS after rigid-motion minimization
\begin{equation}
\begin{split}
\tilde p(\mathbf R\to\mathbf R')={}&\exp\Bigl[-\frac{\tau V(\mathbf R)}{\hbar}-\frac{\tau V(\mathbf R')}{\hbar}\Bigr]\\
&\times\exp\Bigl[-\frac{d^2(\mathbf R,\mathbf R')}{\hbar\tau}\Bigr],
\end{split}
\label{eq:s-mctss}
\end{equation}
where $d^2$ is the distance of equation~\MainEq{eq:d2}. The last factor is the squared free-particle propagator\cite{feynman1965quantum}. MCTSS accepts $\mathbf R'$ with $\min[1,\tilde p(\mathbf R\to\mathbf R')/\tilde p(\mathbf R\to\mathbf R)]$ and draws $\mathbf R'$ by displacing one atom with a Gaussian of variance $\hbar\tau/(2m_i)$ per component\cite{tikhonov2026two}. For a single-atom move of atom $i$ by $\boldsymbol\delta$, the density of that proposal is
\begin{equation}
g(\boldsymbol\delta)=\Bigl(\frac{m_i}{\pi\hbar\tau}\Bigr)^{3/2}\exp\Bigl[-\frac{m_i\lVert\boldsymbol\delta\rVert^2}{\hbar\tau}\Bigr],
\end{equation}
The proposal has the Gaussian form of a free-particle kinetic kernel. Its raw displacement exponent is not identical to the rotation-minimized distance factor in equation~\eqref{eq:s-mctss}: centring and alignment can reduce $d^2$. MCTSS therefore uses a kinetic-inspired proposal together with an additional distance factor in acceptance. tReax\_expedition keeps the Gaussian proposal and removes the kinetic factor from the acceptance, which leaves the Boltzmann factor $\exp[-\beta(\tau)V]$ of equation~\MainEq{eq:target} as the stationary distribution. Supplementary Section~\ref{prop:kinetic} shows that the kinetic factor in the acceptance breaks detailed balance with respect to this factor. The directional pre-selection of MCTSS, $\exp[-d^2/(\hbar\tau_{12})]$ with $\tau_{12}=N_{\rm inter}\tau$ in their notation, becomes the coupling factor of the pair target, equation~\MainEq{eq:pair}. In tReax\_expedition this factor uses the common-frame distance $d_{\rm c}^2$ instead of $d^2$ (Supplementary Section~\ref{prop:pairframe}). The meeting criterion is taken over unchanged (Supplementary Section~\ref{sm:pair}).

\subsection{Classes of modelling elements}\label{sm:bg-classes}
The method combines exact statistical-mechanical relations with algorithmic devices whose parameters have no physical meaning. Supplementary Table~\ref{tab:s-classes} summarizes the main definitions, guarantees and approximations; the linked sections give their full assumptions.

\begin{SuppTable}
\SuppTableCaption{\textbf{Key definitions, guarantees and approximations.} The final column gives the condition needed for each interpretation. Detailed arguments appear in the cited sections; numerical settings are collected in Supplementary Tables~\ref{tab:s-search}--\ref{tab:s-committor}.}{tab:s-classes}
\begin{ruledtabular}
\begin{tabular}{@{}lll@{}}
\TabCell{0.26\textwidth}{Element} & \TabCell{0.19\textwidth}{Status} & \TabCell{0.40\textwidth}{Key condition or interpretation} \\
\hline
\TabCell{0.26\textwidth}{$V$, masses, $T_{\rm phys}$; equilibrium target and $q$} & \TabCell{0.19\textwidth}{Definitions} & \TabCell{0.40\textwidth}{Fixed energy model; $q$ refers to the specified Monte Carlo chain and cores.} \\
\TabCell{0.26\textwidth}{Fixed-target kernel and frame removal} & \TabCell{0.19\textwidth}{Exact properties} & \TabCell{0.40\textwidth}{Fixed acceptance factors; invariant observables and acceptance for frame removal (Sections~\ref{prop:da}, \ref{prop:frame}, \ref{prop:pairframe}).} \\
\TabCell{0.26\textwidth}{Weighted pair midpoints} & \TabCell{0.19\textwidth}{Exact identity} & \TabCell{0.40\textwidth}{Stationary, normalizable pair target in a common frame (Section~\ref{prop:pairframe}).} \\
\TabCell{0.26\textwidth}{Functional $K[q]$} & \TabCell{0.19\textwidth}{Diffusion limit} & \TabCell{0.40\textwidth}{Committor minimization applies in the small-step limit (Sections~\ref{prop:generator}, \ref{prop:ritz}).} \\
\TabCell{0.26\textwidth}{Quench labels and crossing $\mathbf R^\ddagger_0$} & \TabCell{0.19\textwidth}{Numerical estimates} & \TabCell{0.40\textwidth}{Stochastic, finite-budget state assignment; crossing need not lie on $q=1/2$.} \\
\TabCell{0.26\textwidth}{Chains, $E_{\rm up}$ and minimax route} & \TabCell{0.19\textwidth}{Sampled search outputs} & \TabCell{0.40\textwidth}{Continuous bounds require the Lipschitz correction $L_Vh/2$; routes also require intermediate-state connections (Sections~\ref{prop:bound}, \ref{prop:route}).} \\
\TabCell{0.26\textwidth}{Local straddling-pair surface measure} & \TabCell{0.19\textwidth}{Limiting approximation} & \TabCell{0.40\textwidth}{Stiff coupling and locally linear $\xi$ (Section~\ref{sm:tse}).} \\
\TabCell{0.26\textwidth}{$q_\theta$ and $\xi_{\rm label}$} & \TabCell{0.19\textwidth}{Learned approximations} & \TabCell{0.40\textwidth}{Finite-capacity models; committor boundary penalties and shooting checks (Sections~\ref{sm:coord}, \ref{sm:nc}).} \\
\TabCell{0.26\textwidth}{MBAR weights} & \TabCell{0.19\textwidth}{Statistical estimator} & \TabCell{0.40\textwidth}{Equilibrated samples, fixed biases and overlap; capping adds bias (Section~\ref{sm:pinc-mbar}).} \\
\end{tabular}
\end{ruledtabular}
\end{SuppTable}

\section{Proof of SE(3) Invariance and Equivariance}\label{sm:proof-se3}
This section establishes the rigid-motion properties used by the sampling kernel and committor model. For an isolated molecule, the energy and interatomic-distance descriptors are invariant under proper rigid motions, whereas the network gradients entering the Kolmogorov bias and loss are equivariant. These properties imply that removing the frame of a single walker, or retaining a coupled pair in a common frame, preserves the sampled distribution of invariant observables. The derivation below makes the required assumptions explicit; the released unit tests provide numerical checks of the same identities.

\subsection{Notation}\label{sm:se3-notation}
Let $\mathsf Q\in SO(3)$ be a proper rotation matrix and $\mathbf t\in\mathbb R^3$ a translation vector. The rigid motion $g=(\mathsf Q,\mathbf t)$ acts on a configuration as $g\mathbf R=(\mathsf Q\mathbf r_1+\mathbf t,\ldots,\mathsf Q\mathbf r_N+\mathbf t)$, or in $3N$-dimensional notation
\begin{equation}
g\mathbf R=\hat{\mathsf Q}\mathbf R+\mathbf 1_N\otimes\mathbf t,
\label{eq:s-action}
\end{equation}
where $\hat{\mathsf Q}=\mathsf I_N\otimes\mathsf Q$ is block diagonal and $\mathbf 1_N$ is the vector of $N$ ones. The mass matrix $\mathsf W=\mathrm{diag}(m_1,\ldots,m_N)\otimes\mathsf I_3$ multiplies each atomic block by a scalar, so that
\begin{equation}
\hat{\mathsf Q}^{\mathsf T}\mathsf W\hat{\mathsf Q}=\mathsf W,\qquad \hat{\mathsf Q}^{\mathsf T}\mathsf W^{-1}\hat{\mathsf Q}=\mathsf W^{-1}.
\label{eq:s-wcommute}
\end{equation}
We write $\lVert\mathbf x\rVert_{\mathsf W}^2=\mathbf x^{\mathsf T}\mathsf W\mathbf x$. The centring map is $C(\mathbf R)=\mathbf R-\mathbf 1_N\otimes\bar{\mathbf r}(\mathbf R)$, where $\bar{\mathbf r}=\sum_im_i\mathbf r_i/\sum_im_i$ is the centre of mass. Since $\bar{\mathbf r}(g\mathbf R)=\mathsf Q\bar{\mathbf r}(\mathbf R)+\mathbf t$,
\begin{equation}
C(g\mathbf R)=\hat{\mathsf Q}\,C(\mathbf R).
\label{eq:s-centre}
\end{equation}
For an isolated molecule we require of the oracle and of any surrogate that
\begin{equation}
V(g\mathbf R)=V(\mathbf R),\qquad \tilde V(g\mathbf R)=\tilde V(\mathbf R),
\label{eq:s-oracle}
\end{equation}
for all $g$. GFN2-xTB, HF-3c and the force field of alanine dipeptide satisfy equation~\eqref{eq:s-oracle}. Frame removal is switched off for periodic cells, for systems with fixed atoms and for the two-dimensional models, where equation~\eqref{eq:s-oracle} is not needed.

\subsection{Invariance of distances, fingerprints and the committor model}\label{prop:invariance}
We prove that every scalar quantity used by the method is invariant to rigid motions.

\textit{Rotation-minimized distance.} The optimal translation in equation~\MainEq{eq:d2} superposes the two centres of mass, so that
\begin{equation}
d^2(\mathbf R,\mathbf R')=\min_{\mathsf Q''\in SO(3)}\bigl\lVert C(\mathbf R)-\hat{\mathsf Q}''C(\mathbf R')\bigr\rVert_{\mathsf W}^2 .
\label{eq:s-d2centred}
\end{equation}
For two rigid motions $g=(\mathsf Q,\mathbf t)$ and $g'=(\mathsf Q',\mathbf t')$, equations~\eqref{eq:s-wcommute} and \eqref{eq:s-centre} give
\begin{equation}
\begin{split}
\bigl\lVert C(g\mathbf R)-\hat{\mathsf Q}''C(g'\mathbf R')\bigr\rVert_{\mathsf W}&=\bigl\lVert\hat{\mathsf Q}C(\mathbf R)-\hat{\mathsf Q}''\hat{\mathsf Q}'C(\mathbf R')\bigr\rVert_{\mathsf W}\\
&=\bigl\lVert C(\mathbf R)-\hat{\mathsf Q}^{\mathsf T}\hat{\mathsf Q}''\hat{\mathsf Q}'C(\mathbf R')\bigr\rVert_{\mathsf W}.
\end{split}
\end{equation}
As $\mathsf Q''$ runs over $SO(3)$, so does $\mathsf Q^{\mathsf T}\mathsf Q''\mathsf Q'$. The two minima therefore coincide, and $d^2(g\mathbf R,g'\mathbf R')=d^2(\mathbf R,\mathbf R')$ for separate rigid motions of the two structures. Since $\lVert C(\mathbf R)-\hat{\mathsf Q}''C(\mathbf R')\rVert_{\mathsf W}=\lVert\hat{\mathsf Q}''^{\mathsf T}C(\mathbf R)-C(\mathbf R')\rVert_{\mathsf W}$ and $\mathsf Q''^{\mathsf T}\in SO(3)$, $d^2$ is also symmetric.

\textit{Common-frame distance.} The coupling of equation~\MainEq{eq:pair} uses $d_{\rm c}^2(\mathbf R,\mathbf R')=\lVert C(\mathbf R)-C(\mathbf R')\rVert_{\mathsf W}^2$. For a common rigid motion,
\begin{equation}
d_{\rm c}^2(g\mathbf R,g\mathbf R')=\bigl\lVert\hat{\mathsf Q}\bigl[C(\mathbf R)-C(\mathbf R')\bigr]\bigr\rVert_{\mathsf W}^2=d_{\rm c}^2(\mathbf R,\mathbf R').
\end{equation}
A translation of one structure does not change its centred coordinates, so $d_{\rm c}^2$ is also invariant to separate translations. It is not invariant to a rotation of one structure alone, which is why the relative orientation of a pair is sampled by the explicit move of Supplementary Section~\ref{prop:pairframe}.

\textit{Interatomic distances and descriptors.} For every pair of atoms,
\begin{equation}
\bigl\lVert(\mathsf Q\mathbf r_i+\mathbf t)-(\mathsf Q\mathbf r_j+\mathbf t)\bigr\rVert=\bigl\lVert\mathsf Q(\mathbf r_i-\mathbf r_j)\bigr\rVert=\lVert\mathbf r_i-\mathbf r_j\rVert .
\end{equation}
The bond graph compares $r_{ij}$ with fixed thresholds, $\xi_{\rm bond}$ is a signed sum of $r_{ij}$, and the descriptors are $(r_{ij}^2+10^{-12})^{1/2}$. The network outputs $z_\theta$ and $q_\theta$ depend on $\mathbf R$ only through the descriptors. Therefore all of them are invariant.

\textit{Dihedral angles.} For atoms $a,b,c,d$ let $\mathbf b_1=\mathbf r_b-\mathbf r_a$, $\mathbf b_2=\mathbf r_c-\mathbf r_b$, $\mathbf b_3=\mathbf r_d-\mathbf r_c$, $\mathbf n_1=\mathbf b_1\times\mathbf b_2$ and $\mathbf n_2=\mathbf b_2\times\mathbf b_3$. The dihedral angle is
\begin{equation}
\phi=\operatorname{atan2}\Bigl(\frac{(\mathbf n_1\times\mathbf n_2)\cdot\mathbf b_2}{\lVert\mathbf b_2\rVert},\;\mathbf n_1\cdot\mathbf n_2\Bigr).
\end{equation}
Translations cancel in the $\mathbf b_k$. For $\det\mathsf Q=+1$, $\mathsf Q(\mathbf u\times\mathbf v)=(\mathsf Q\mathbf u)\times(\mathsf Q\mathbf v)$ and scalar products are preserved, so both arguments of $\operatorname{atan2}$ are unchanged. Therefore the dihedral fingerprint and $\xi_{\rm dih}$ are invariant.

\textit{Progress along the endpoint axis.} The progress $p(\mathbf R)$ of the overlap criterion (Supplementary Section~\ref{sm:pair}) is evaluated after aligning $\mathbf R$ onto a fixed endpoint. The aligned structure is the element of the orbit $\{g\mathbf R\}$ closest to that endpoint, which is the same for $\mathbf R$ and for $g\mathbf R$. Therefore $p$ is invariant.

\textit{Reflections and permutations.} Under an improper rotation, distances and bond graphs are unchanged, whereas dihedral angles change sign. The committor model and the bond-graph fingerprint therefore cannot distinguish mirror images, but the dihedral fingerprint can. The chain never applies a reflection. Moves are displacements, and the Kabsch rotation is restricted to $\det\mathsf Q=+1$ by its sign correction\cite{kabsch1978discussion}. Atoms carry fixed labels, and bond graphs are not canonicalized under permutations of identical atoms. A permutation changes the registered state when it changes the labelled graph; graph automorphisms leave that graph unchanged. Symmetry-equivalent structures or channels can therefore be registered separately.

\subsection{Equivariance of gradients and invariance of biases}\label{prop:equivariance}
We now prove that the gradient of an invariant function is equivariant to rotations and that its norm in the inverse mass metric is invariant.

\textit{Equivariance.} Let $f:\mathbb R^{3N}\to\mathbb R$ be differentiable with $f(g\mathbf R)=f(\mathbf R)$ for all $g$. Differentiating $f(\hat{\mathsf Q}\mathbf R+\mathbf 1_N\otimes\mathbf t)=f(\mathbf R)$ with respect to $\mathbf R$ gives $\hat{\mathsf Q}^{\mathsf T}\nabla f(g\mathbf R)=\nabla f(\mathbf R)$, and multiplication by $\hat{\mathsf Q}$ gives
\begin{equation}
\nabla f(g\mathbf R)=\hat{\mathsf Q}\,\nabla f(\mathbf R).
\label{eq:s-equivariant}
\end{equation}

\textit{Invariance of the gradient norm.} With equation~\eqref{eq:s-wcommute}, the squared gradient in the inverse mass metric, $\lvert\nabla_{\!u}f\rvert^2=\nabla f^{\mathsf T}\mathsf W^{-1}\nabla f$, satisfies
\begin{equation}
\begin{split}
\lvert\nabla_{\!u}f\rvert^2(g\mathbf R)&=\nabla f(\mathbf R)^{\mathsf T}\hat{\mathsf Q}^{\mathsf T}\mathsf W^{-1}\hat{\mathsf Q}\,\nabla f(\mathbf R)\\
&=\lvert\nabla_{\!u}f\rvert^2(\mathbf R).
\end{split}
\end{equation}

\textit{Biases and targets.} Taking $f=z_\theta$, which is invariant by Supplementary Section~\ref{prop:invariance}, shows that $\lvert\nabla_{\!u}z_\theta\rvert^2$ is invariant. The Kolmogorov bias of equation~\MainEq{eq:vk} depends on $\mathbf R$ only through $\lvert\nabla_{\!u}z_\theta\rvert^2$ and $q_\theta$, the adaptive bias $V_z$ only through $z_\theta$, and the restraints $b_k$ and the side constraint of the local straddling-pair ensemble only through $\xi$. Together with equation~\eqref{eq:s-oracle}, this gives
\begin{equation}
\begin{split}
\pi(g\mathbf R)&=\pi(\mathbf R),\\
\Pi(g\mathbf R_1,g\mathbf R_2)&=\Pi(\mathbf R_1,\mathbf R_2),\\
\Pi(\mathbf R_1+\mathbf 1_N\otimes\mathbf t_1,\mathbf R_2+\mathbf 1_N\otimes\mathbf t_2)&=\Pi(\mathbf R_1,\mathbf R_2),
\end{split}
\label{eq:s-targetinv}
\end{equation}
for the targets of equations~\MainEq{eq:targetcb} and \MainEq{eq:pair}. Therefore, we can conclude that every target of the method is invariant to rigid motions, and $\Pi$ is invariant to a common rigid motion of both walkers and to separate translations. The same argument shows that the loss of equation~\MainEq{eq:loss} does not change when any training configuration is moved rigidly, so augmenting the data with rotated copies would add no information.

\subsection{Exactness of frame removal for a single walker}\label{prop:frame}
We prove that aligning each trial onto the current structure leaves the distribution of every invariant quantity unchanged. Let $P$ be the kernel of the move of Methods Section~\MainSec{sec:m-kernel} without frame removal. It selects atom $i$ with probability $1/N$, proposes $\mathbf y=\mathbf R+\mathbf e_i\otimes\boldsymbol\delta$ with $\boldsymbol\delta\sim\mathcal N(\mathbf 0,\sigma_i^2\mathsf I_3)$, where $\mathbf e_i$ is the $i$-th unit vector of $\mathbb R^N$, and accepts with $\alpha(\mathbf R,\mathbf y)$ of equation~\MainEq{eq:da}. Let $P_{\rm A}$ be the implemented kernel, in which the trial is replaced by $A_{\mathbf R}(\mathbf y)=\hat{\mathsf Q}^\ast C(\mathbf y)$, where $\mathsf Q^\ast\in SO(3)$ is the Kabsch rotation of $C(\mathbf y)$ onto $C(\mathbf R)$. A set $B$ is called invariant when $gB=B$ for all $g$. It is a union of orbits $[\mathbf R]=\{g\mathbf R\}$, and an invariant observable is a function of the orbit alone.

\textit{Acceptance.} The acceptance probability depends on the trial only through $c$, $V$, $b$ and $\tilde V$, which are invariant (Supplementary Section~\ref{prop:equivariance}). Since $A_{\mathbf R}(\mathbf y)$ lies in the orbit of $\mathbf y$,
\begin{equation}
\alpha\bigl(\mathbf R,A_{\mathbf R}(\mathbf y)\bigr)=\alpha(\mathbf R,\mathbf y).
\end{equation}

\textit{Invariant sets.} For invariant $B$, $A_{\mathbf R}(\mathbf y)\in B$ if and only if $\mathbf y\in B$. A rejected trial leaves $\mathbf R$ in place under both kernels, so
\begin{equation}
P_{\rm A}(\mathbf R,B)=\mathbb E\bigl[\alpha\,\mathbb 1_B(\mathbf y)\bigr]+\bigl(1-\mathbb E[\alpha]\bigr)\mathbb 1_B(\mathbf R)=P(\mathbf R,B),
\label{eq:s-lump1}
\end{equation}
where the expectation is over the atom and the displacement.

\textit{Equivariance of the kernel.} From $g\mathbf R$ the trial is
\begin{equation}
g\mathbf R+\mathbf e_i\otimes\boldsymbol\delta=g\bigl(\mathbf R+\mathbf e_i\otimes\mathsf Q^{\mathsf T}\boldsymbol\delta\bigr).
\end{equation}
The displacement is isotropic, so $\mathsf Q^{\mathsf T}\boldsymbol\delta$ has the same distribution as $\boldsymbol\delta$, and the trial from $g\mathbf R$ is distributed as $g$ applied to the trial from $\mathbf R$. With $\alpha(g\mathbf R,g\mathbf y)=\alpha(\mathbf R,\mathbf y)$ this gives
\begin{equation}
P(g\mathbf R,B)=P(\mathbf R,g^{-1}B)=P(\mathbf R,B).
\label{eq:s-lump2}
\end{equation}
Hence $P(\mathbf R,B)$ depends on $\mathbf R$ only through its orbit $[\mathbf R]$.

\textit{Induction over steps.} Let the chains $\mathbf R_n$ of $P$ and $\mathbf R_n^{\rm A}$ of $P_{\rm A}$ start from the same configuration $\mathbf R_0$. We show by induction that their orbits have the same distribution at every step.
\begin{itemize}
\item \textit{Base case ($n=0$):} Both chains start at $\mathbf R_0$, so $\Pr\bigl[\mathbf R_0^{\rm A}\in B\bigr]=\Pr\bigl[\mathbf R_0\in B\bigr]$ for every invariant $B$.
\item \textit{Induction hypothesis:} At step $n=k$, $\Pr\bigl[\mathbf R_k^{\rm A}\in B\bigr]=\Pr\bigl[\mathbf R_k\in B\bigr]$ for every invariant $B$.
\item \textit{Induction step:} By equation~\eqref{eq:s-lump1},
\begin{equation}
\Pr\bigl[\mathbf R_{k+1}^{\rm A}\in B\bigr]=\mathbb E\bigl[P_{\rm A}(\mathbf R_k^{\rm A},B)\bigr]=\mathbb E\bigl[P(\mathbf R_k^{\rm A},B)\bigr].
\end{equation}
By equation~\eqref{eq:s-lump2}, $P(\cdot,B)$ is an invariant function, so its expectation depends only on the distribution of the orbit, which is the same for both chains by the induction hypothesis. Hence
\begin{equation}
\Pr\bigl[\mathbf R_{k+1}^{\rm A}\in B\bigr]=\mathbb E\bigl[P(\mathbf R_k,B)\bigr]=\Pr\bigl[\mathbf R_{k+1}\in B\bigr].
\end{equation}
\end{itemize}
By induction, the orbits of the two chains have the same distribution at every step. Applying the same argument to joint events at several steps shows that invariant observables also have the same time correlations. The orbit process is a lumped Markov chain\cite{kemeny1960finite} with kernel $\bar P([\mathbf R],B)=P(\mathbf R,B)$. $P$ satisfies detailed balance with respect to $\pi$ (Supplementary Section~\ref{prop:da}), and integrating it over invariant sets $B_1$ and $B_2$ gives
\begin{equation}
\int_{B_1}\pi(\mathbf R)P(\mathbf R,B_2)\,\mathrm d\mathbf R=\int_{B_2}\pi(\mathbf R)P(\mathbf R,B_1)\,\mathrm d\mathbf R,
\end{equation}
which is detailed balance of the lumped chain with respect to the image of $\pi$. Frame removal therefore preserves the process of invariant observables.

\textit{Remarks.} (a) The proof does not require the aligned proposal to be symmetric. Frame removal is therefore exact rather than an approximation. (b) Along translations $\pi$ is not normalizable, so the statement is applied to centred configurations. The image of $\pi$ on molecular shapes is then normalizable when $\exp(-\beta V)$ is integrable over shapes. This fails for fragments that separate in free space, as in $\mathrm{H_2CO\to H_2+CO}$, a property of the gas-phase ensemble rather than of frame removal. (c) Linear configurations, such as HCN and HCCH near their minima, have lower-dimensional orbits and a Kabsch rotation that is not unique. They form a set of measure zero, and any choice keeps the trial in its orbit. (d) The result applies to every single-walker chain of the method, namely quenches, restrained nodes, biased walks, shots and validation chains. (e) Without frame removal, the symmetric-proposal kernel satisfies detailed balance for any fixed surrogate (Supplementary Section~\ref{prop:da}). With frame removal, the acceptance-invariance argument additionally requires the surrogate to satisfy equation~\eqref{eq:s-oracle}.

\subsection{Exactness of the common frame for a coupled pair}\label{prop:pairframe}
For the full centred linear configuration space, we prove that the pair sweep leaves $\Pi$ of equation~\MainEq{eq:pair} invariant and that the midpoint weights are exact. One sweep consists of two kinds of move. (a) For $n=1,2$, walker $n$ is displaced as in Supplementary Section~\ref{prop:frame}, centred without rotation, and tested with equation~\MainEq{eq:da} with $c=-d_{\rm c}^2/(\hbar\tau_{12})$. (b) The centred walker $\bar{\mathbf R}_2$ is rotated to $\hat{\mathsf Q}_\omega\bar{\mathbf R}_2$, where $\mathsf Q_\omega$ rotates by the angle $\omega\sim\mathcal N(0,\sigma_{\rm rot}^2)$ about an axis drawn uniformly on the unit sphere and $\sigma_{\rm rot}$ is given by equation~\eqref{eq:s-rotwidth}.

\textit{Displacements.} At fixed $\mathbf R_m$, the displacement of walker $n$ with the coupling test is the move of Supplementary Section~\ref{prop:da} for the conditional of $\Pi$, so it satisfies detailed balance with respect to that conditional. Centring is a translation of walker $n$ alone, under which $\Pi$ is invariant by equation~\eqref{eq:s-targetinv}. The argument of Supplementary Section~\ref{prop:frame}, with the translations of walker $n$ in place of $SE(3)$, shows that centring leaves this conclusion unchanged. An update that preserves the conditional of $\Pi$ preserves $\Pi$.

\textit{Relative rotation.} The inverse of $\mathsf Q_\omega$ is the rotation by $-\omega$ about the same axis, which has the same probability because the distribution of $\omega$ is symmetric. $\sigma_{\rm rot}$ depends on $\bar{\mathbf R}_2$ only through the trace of its inertia tensor, which a rotation does not change. The proposal is therefore symmetric with respect to the rotation-invariant measure on the orbit of $\bar{\mathbf R}_2$. Because $V(\hat{\mathsf Q}_\omega\bar{\mathbf R}_2)=V(\bar{\mathbf R}_2)$, the Metropolis--Hastings ratio of $\Pi$ reduces to
\begin{equation}
\frac{\Pi(\mathbf R_1,\hat{\mathsf Q}_\omega\bar{\mathbf R}_2)}{\Pi(\mathbf R_1,\bar{\mathbf R}_2)}=\exp\Bigl[-\frac{d_{\rm c}^2(\mathbf R_1,\hat{\mathsf Q}_\omega\bar{\mathbf R}_2)-d_{\rm c}^2(\mathbf R_1,\bar{\mathbf R}_2)}{\hbar\tau_{12}}\Bigr],
\end{equation}
which is the ratio used in the test. Detailed balance with respect to $\Pi$ follows. A composition of kernels that each preserve $\Pi$ preserves $\Pi$\cite{tierney1994markov}, so the whole sweep does.

\textit{Midpoint identity.} Let $\mathbf M=(\bar{\mathbf R}_1+\bar{\mathbf R}_2)/2$ and $\mathbf D=\bar{\mathbf R}_1-\bar{\mathbf R}_2$. The map $(\bar{\mathbf R}_1,\bar{\mathbf R}_2)\mapsto(\mathbf M,\mathbf D)$ is linear with a Jacobian of unit modulus, and $d_{\rm c}^2=\mathbf D^{\mathsf T}\mathsf W\mathbf D$ without any rotation. Multiplying $\Pi$ by the weight $w$ of equation~\MainEq{eq:midweight} replaces $\exp\{-\beta_{\rm b}[V(\mathbf R_1)+V(\mathbf R_2)]\}$ by $\exp[-2\beta_{\rm b}V(\mathbf M)]$, which gives
\begin{equation}
\Pi(\mathbf R_1,\mathbf R_2)\,w=\frac{1}{Z_\Pi}\exp\bigl[-2\beta_{\rm b}V(\mathbf M)\bigr]\exp\Bigl[-\frac{\mathbf D^{\mathsf T}\mathsf W\mathbf D}{\hbar\tau_{12}}\Bigr],
\label{eq:s-pairexact}
\end{equation}
where $Z_\Pi$ is the normalization of $\Pi$. When the normalization exists, the weighted midpoint marginal is proportional to $\exp[-2\beta_{\rm b}V(\mathbf M)]$. This identity presumes stationary sampling of $\Pi$ and a midpoint-independent integration domain for $\mathbf D$. Hard confinement or excluded oracle-failure regions can leave a midpoint-dependent support factor; finite normalization alone does not remove it. Smooth confinement included in $V$ on the full space is distinct from a hard domain restriction.

\textit{Remark on the choice of coupling.} With the rotation-minimized $d^2$ in place of $d_{\rm c}^2$, $\Pi$ would be invariant to a separate rotation of either walker, and the relative orientation would carry no weight. The midpoint would then have to be formed after rotating one walker onto the other. This map is not linear, and equation~\eqref{eq:s-pairexact} would hold only approximately. The common frame, together with move (b), restores the exact identity at the cost of one oracle-free rotation per sweep. The output frame of a pair, in which both walkers are turned by one common rotation before they are written, leaves $\Pi$ unchanged by equation~\eqref{eq:s-targetinv}.

\section{Proof of Stationarity and of the Diffusion Limit}\label{sm:proof-stat}
The following proofs separate detailed balance of the fixed-target kernel from its small-step approximation to overdamped dynamics. The latter motivates equation~\MainEq{eq:ritz} but does not establish equality of finite-step and diffusion committors.

\subsection{Detailed balance of delayed acceptance}\label{prop:da}
We prove that the move satisfies detailed balance with respect to $\pi$ of equation~\MainEq{eq:targetcb} for any fixed $c$, $b$ and $\tilde V$. Let $g(\mathbf y\mid\mathbf x)=g(\mathbf x\mid\mathbf y)$ be a symmetric proposal and write $\alpha=\prod_{k=1}^3\min(1,r_k)$ with
\begin{equation}
r_1=e^{\Delta c},\qquad r_2=e^{-\beta\Delta(\tilde V+b)},\qquad r_3=e^{-\beta\Delta(V-\tilde V)},
\end{equation}
where $\Delta f=f(\mathbf y)-f(\mathbf x)$. The reverse move has ratios $1/r_k$, and $\min(1,r)/\min(1,1/r)=r$ for every $r>0$. Hence
\begin{equation}
\frac{\alpha(\mathbf x,\mathbf y)}{\alpha(\mathbf y,\mathbf x)}=r_1r_2r_3=\exp\bigl[\Delta c-\beta\Delta V-\beta\Delta b\bigr]=\frac{\pi(\mathbf y)}{\pi(\mathbf x)},
\end{equation}
because $\tilde V$ cancels between $r_2$ and $r_3$. With $g$ symmetric this gives, for all $\mathbf x\ne\mathbf y$,
\begin{equation}
\pi(\mathbf x)\,g(\mathbf y\mid\mathbf x)\,\alpha(\mathbf x,\mathbf y)=\pi(\mathbf y)\,g(\mathbf x\mid\mathbf y)\,\alpha(\mathbf y,\mathbf x),
\label{eq:s-db}
\end{equation}
which is detailed balance\cite{christen2005mcmc}. Each move of a fixed sequence preserves $\pi$, and so does their composition\cite{tierney1994markov}. A failed oracle call sets $V(\mathbf y)=+\infty$ and $r_3=0$. If the oracle fails on a fixed set $\mathcal F$, equation~\eqref{eq:s-db} then holds for the target $\pi\,\mathbb 1_{\mathcal F^{\rm c}}$, and the chain samples $\pi$ restricted to the complement of $\mathcal F$. The order of the three tests does not enter the proof and is chosen so that the cheapest test comes first. If the oracle fails at random rather than on a fixed set, this restriction is not defined, and the retries of Supplementary Section~\ref{sm:kernel} serve to make such failures rare.

\subsection{Single versus repeated pre-screen}\label{prop:prescreen}
We prove that redrawing trials until one passes the pre-screen changes the stationary distribution. Consider the update of walker $n$ at fixed $\mathbf R_m$, with conditional target $\pi_n(\mathbf x)=Z_n^{-1}\exp[c(\mathbf x)-\beta_{\rm b}V(\mathbf x)]$, where $c(\mathbf x)=-d_{\rm c}^2(\mathbf x,\mathbf R_m)/(\hbar\tau_{12})$ and $Z_n$ normalizes $\pi_n$. A trial passes the pre-screen with probability $a(\mathbf x,\mathbf y)=\min(1,e^{c(\mathbf y)-c(\mathbf x)})$, and one draw passes with probability
\begin{equation}
A(\mathbf x)=\int g(\mathbf y\mid\mathbf x)\,a(\mathbf x,\mathbf y)\,\mathrm d\mathbf y .
\end{equation}
Suppose trials are redrawn until one passes, for at most $K$ draws, and a passing trial is then accepted with $m(\mathbf x,\mathbf y)=\min(1,e^{-\beta_{\rm b}\Delta V})$. A trial $\mathbf y\ne\mathbf x$ is accepted at the $k$-th draw with probability $[1-A(\mathbf x)]^{k-1}g(\mathbf y\mid\mathbf x)a(\mathbf x,\mathbf y)m(\mathbf x,\mathbf y)$. Summing over $k\le K$ gives the transition density
\begin{equation}
T(\mathbf x,\mathbf y)=g(\mathbf y\mid\mathbf x)\,a(\mathbf x,\mathbf y)\,m(\mathbf x,\mathbf y)\,\frac{1-[1-A(\mathbf x)]^K}{A(\mathbf x)} .
\end{equation}
Now define
\begin{equation}
\pi_n^{(K)}(\mathbf x)=\frac{1}{Z_n^{(K)}}\,\pi_n(\mathbf x)\,\frac{A(\mathbf x)}{1-[1-A(\mathbf x)]^K},
\label{eq:s-tilt}
\end{equation}
where $Z_n^{(K)}$ normalizes $\pi_n^{(K)}$. Then
\begin{equation}
\begin{split}
Z_nZ_n^{(K)}\,\pi_n^{(K)}(\mathbf x)\,T(\mathbf x,\mathbf y)={}&g(\mathbf y\mid\mathbf x)\min\bigl(e^{c(\mathbf x)},e^{c(\mathbf y)}\bigr)\\
&\times\min\bigl(e^{-\beta_{\rm b}V(\mathbf x)},e^{-\beta_{\rm b}V(\mathbf y)}\bigr),
\end{split}
\end{equation}
which is symmetric in $\mathbf x$ and $\mathbf y$. The update therefore satisfies detailed balance with respect to $\pi_n^{(K)}$. For $K=1$, the single test of tReax\_expedition, $\pi_n^{(1)}=\pi_n$. For $K\to\infty$, $\pi_n^{(\infty)}$ is $\pi_n$ tilted by the pass probability $A$. $A$ is not constant. Near the other walker almost every displacement increases $d_{\rm c}^2$, whereas far from it about half of the displacements decrease it. MCTSS uses the repeated pre-screen as a search device\cite{tikhonov2026two}, for which such a tilt is harmless. The midpoint weights and the local straddling-pair ensemble require $\Pi$ exactly, which the single test provides at the price of about $1/A$ more steps.

\subsection{Kinetic factor in the acceptance}\label{prop:kinetic}
We prove that placing the kinetic factor in the acceptance breaks detailed balance with respect to $\pi_\tau$ of equation~\MainEq{eq:target}. The rule obtained from equation~\eqref{eq:s-mctss} accepts with
\begin{equation}
\alpha_{\rm M}(\mathbf x,\mathbf y)=\min\bigl\{1,\exp\bigl[-\beta\Delta V-k(\mathbf x,\mathbf y)\bigr]\bigr\},
\end{equation}
where $k=d^2(\mathbf x,\mathbf y)/(\hbar\tau)\ge0$, with equality for rigidly equivalent structures. Let $Z$ be the normalization of $\pi_\tau$. Then
\begin{equation}
\begin{split}
Z\,\pi_\tau(\mathbf x)\,\alpha_{\rm M}(\mathbf x,\mathbf y)&=\min\bigl\{e^{-\beta V(\mathbf x)},\,e^{-\beta V(\mathbf y)-k}\bigr\},\\
Z\,\pi_\tau(\mathbf y)\,\alpha_{\rm M}(\mathbf y,\mathbf x)&=\min\bigl\{e^{-\beta V(\mathbf y)},\,e^{-\beta V(\mathbf x)-k}\bigr\}.
\end{split}
\end{equation}
For a pair with $d(\mathbf x,\mathbf y)>0$ and $\beta V(\mathbf x)\ge\beta V(\mathbf y)+k$, the first line equals $e^{-\beta V(\mathbf x)}$ and the second $e^{-\beta V(\mathbf x)-k}$, which is smaller. Multiplying both by the symmetric proposal density shows that the flux from $\mathbf x$ to $\mathbf y$ exceeds the reverse flux. Removing the kinetic factor from the acceptance gives $\alpha=\min(1,e^{-\beta\Delta V})$, a special case of Supplementary Section~\ref{prop:da}. The kinetic factor then acts once, through the proposal (Supplementary Section~\ref{sm:bg-origin}).

\subsection{Small-step limit of the move}\label{prop:generator}
We derive the small-displacement expansion at fixed $\beta$. A fixed-temperature diffusion limit scales proposal widths by $\lambda_{\rm s}\to0$ and rescales the step count by $\lambda_{\rm s}^{-2}$; varying $\tau$ alone also changes $\beta$ and is not this limit. Let $U=V+b$ be twice continuously differentiable, let $f$ have bounded continuous derivatives up to third order, and let $P$ be the kernel of the single-walker move at parameter $\tau$ without coupling or surrogate. Fix atom $i$ and write $\mathbf a=\nabla_if(\mathbf R)$ and $\mathbf u=\nabla_iU(\mathbf R)$, where $\nabla_i$ acts on the coordinates of atom $i$. Then $\Delta U=\mathbf u\cdot\boldsymbol\delta+O(\lVert\boldsymbol\delta\rVert^2)$, and $\min(1,e^{-x})=1-x_++O(x^2)$ with $x_+=\max(x,0)$ gives $\alpha=1-\beta(\mathbf u\cdot\boldsymbol\delta)_++O(\lVert\boldsymbol\delta\rVert^2)$. Expanding $f$ to second order,
\begin{equation}
\begin{split}
\mathbb E\bigl[\alpha\,\Delta f\bigr]={}&\mathbb E[\mathbf a\cdot\boldsymbol\delta]-\beta\,\mathbb E\bigl[(\mathbf a\cdot\boldsymbol\delta)(\mathbf u\cdot\boldsymbol\delta)_+\bigr]\\
&+\tfrac12\,\mathbb E\bigl[\boldsymbol\delta^{\mathsf T}\nabla_i\nabla_i^{\mathsf T}f\,\boldsymbol\delta\bigr]+O(\sigma_i^3).
\end{split}
\end{equation}
The first term vanishes. For the second, $(\mathbf u\cdot\boldsymbol\delta)_+=\tfrac12(\mathbf u\cdot\boldsymbol\delta)+\tfrac12\lvert\mathbf u\cdot\boldsymbol\delta\rvert$, and $\mathbb E[(\mathbf a\cdot\boldsymbol\delta)\lvert\mathbf u\cdot\boldsymbol\delta\rvert]=0$ because the integrand is odd under $\boldsymbol\delta\to-\boldsymbol\delta$. Hence $\mathbb E[(\mathbf a\cdot\boldsymbol\delta)(\mathbf u\cdot\boldsymbol\delta)_+]=\tfrac12\sigma_i^2\,\mathbf a\cdot\mathbf u$. The third term equals $\tfrac12\sigma_i^2\nabla_i^2f$. Averaging over the atom with probability $1/N$ gives
\begin{equation}
(Pf-f)(\mathbf R)=\sum_{i=1}^{N}D_i\bigl[\nabla_i^2f-\beta\,\nabla_iU\cdot\nabla_if\bigr](\mathbf R)+O(\sigma_{\max}^3),
\label{eq:s-generator}
\end{equation}
where $\sigma_{\max}=\max_i\sigma_i$ and
\begin{equation}
D_i=\frac{\sigma_i^2}{2N}=\frac{\hbar\tau}{4Nm_i}
\label{eq:s-diff}
\end{equation}
is the diffusion constant per step at the nominal scale $\lambda_{\rm s}=1$; scaling the widths multiplies $D_i$ by $\lambda_{\rm s}^2$. The right-hand side of equation~\eqref{eq:s-generator} is the generator of overdamped Langevin dynamics with mass-dependent diffusion constants\cite{kikuchi1991metropolis,lelievre2016partial}. For invariant $f$, $f(A_{\mathbf R}(\mathbf y))=f(\mathbf y)$, so the implemented kernel $P_{\rm A}$ has the same expansion. The error terms are small when $\beta\sigma_i\lVert\nabla_iU\rVert\ll1$ and $\beta\sigma_i^2\lVert\nabla_i\nabla_i^{\mathsf T}U\rVert\ll1$. At $\tau^\ast=25.6$~fs a hydrogen atom has $\sigma=0.28$~\AA, so the correspondence is approximate for molecules.

\subsection{Variational form of the committor}\label{prop:ritz}
We prove that the committor of the limiting diffusion minimizes $K$ of equation~\MainEq{eq:ritz}. Let
\begin{equation}
\begin{split}
\mathcal Lf&=\sum_iD_i\bigl(\nabla_i^2f-\beta^\ast\nabla_iV\cdot\nabla_if\bigr)\\
&=e^{\beta^\ast V}\sum_iD_i\nabla_i\cdot\bigl(e^{-\beta^\ast V}\nabla_if\bigr),
\end{split}
\end{equation}
with $D_i=\hbar\tau^\ast/(4Nm_i)$, and let $\Omega$ be the complement of the cores $C_a$ and $C_b$, with no flux through its outer boundary. The committor $q$ of this diffusion solves $\mathcal Lq=0$ in $\Omega$ with $q=0$ on $\partial C_a$ and $q=1$ on $\partial C_b$\cite{e2010transition}. For admissible $f$, which take these boundary values and are constant on each core,
\begin{equation}
I[f]=\int_\Omega e^{-\beta^\ast V}\sum_iD_i\lVert\nabla_if\rVert^2\,\mathrm d\mathbf R=\frac{\hbar\tau^\ast}{4N}\,Z^\ast K[f],
\end{equation}
where $Z^\ast=\int e^{-\beta^\ast V}\mathrm d\mathbf R$, since $\sum_iD_i\lVert\nabla_if\rVert^2=\tfrac{\hbar\tau^\ast}{4N}\lvert\nabla_{\!u}f\rvert^2$. Let $h$ vanish on $\partial C_a$ and $\partial C_b$. Integration by parts, with no boundary terms, gives
\begin{equation}
\int_\Omega e^{-\beta^\ast V}\sum_iD_i\nabla_iq\cdot\nabla_ih\,\mathrm d\mathbf R=-\int_\Omega h\,e^{-\beta^\ast V}\mathcal Lq\,\mathrm d\mathbf R=0 .
\end{equation}
Hence $I[q+h]=I[q]+I[h]\ge I[q]$. Therefore, $q$ minimizes $I$ and hence $K$, and the prefactor $\hbar\tau^\ast/(4N)$ does not affect the minimizer\cite{e2006transition,bovier2004metastability}. The relation between the Monte Carlo committor, equation~\MainEq{eq:committor}, and $K[q]$ therefore relies on the small-step limit of Supplementary Section~\ref{prop:generator}. The diffusion functional approximates the finite-step Monte Carlo target. Shooting provides a separate diagnostic under the specified kernel and cores, subject to the implementation and finite-budget limitations in Supplementary Section~\ref{sm:nc-shoot}.

\section{Conditional continuous-path energy bound}\label{sm:proof-bound}
Equation~\MainEq{eq:eup} records evaluated energies. A continuous-path bound additionally requires control of the unevaluated segments; the derivation below identifies this correction.

\subsection{Segments}\label{sm:bound-lemmas}
The minimax energy between two configurations is
\begin{equation}
E^\ast(\mathbf x,\mathbf y)=\inf_{\gamma}\max_{s\in[0,1]}V\bigl(\gamma(s)\bigr),
\label{eq:s-minimax}
\end{equation}
where the infimum runs over continuous paths $\gamma$ with $\gamma(0)=\mathbf x$ and $\gamma(1)=\mathbf y$. $V$ is called $L_V$-Lipschitz on a set $\mathcal S$ when $\lvert V(\mathbf x)-V(\mathbf y)\rvert\le L_V\lVert\mathbf x-\mathbf y\rVert_{\mathsf W}$ for all $\mathbf x$ and $\mathbf y$ in $\mathcal S$.

\textit{Straight segment.} Let $V$ be $L_V$-Lipschitz on the segment from $\mathbf x$ to $\mathbf y$, of length $\ell_{xy}=\lVert\mathbf y-\mathbf x\rVert_{\mathsf W}$. At parameter $s$, $V\le V(\mathbf x)+L_Vs\ell_{xy}$ and $V\le V(\mathbf y)+L_V(1-s)\ell_{xy}$. The smaller of the two bounds is at most their mean, so
\begin{equation}
\begin{split}
\max_{s\in[0,1]}V\bigl(\mathbf x+s(\mathbf y-\mathbf x)\bigr)&\le\tfrac12\bigl[V(\mathbf x)+V(\mathbf y)+L_V\ell_{xy}\bigr]\\
&\le\max\{V(\mathbf x),V(\mathbf y)\}+\tfrac12L_V\ell_{xy}.
\end{split}
\label{eq:s-segment}
\end{equation}

\textit{Rigid motion.} Let $\mathbf y'=g\mathbf y$ with $g=(\mathsf Q,\mathbf t)$. $SO(3)$ is path-connected, since $\mathsf Q=\exp(\boldsymbol\Omega)$ for a skew-symmetric $\boldsymbol\Omega$. The path $s\mapsto g(s)\mathbf y$ with $g(s)=(\exp(s\boldsymbol\Omega),s\mathbf t)$ joins $\mathbf y$ to $\mathbf y'$, and by equation~\eqref{eq:s-oracle}
\begin{equation}
V\bigl(g(s)\mathbf y\bigr)=V(\mathbf y)\qquad\text{for all } s\in[0,1].
\label{eq:s-rigidpath}
\end{equation}

\subsection{Chains}\label{prop:bound}
We prove that, for a certified chain with sampled maximum $E_{\rm up}$ and end configurations $\mathbf x_a$ and $\mathbf x_b$ of its two end quenches,
\begin{equation}
E^\ast(\mathbf x_a,\mathbf x_b)\le E_{\rm up}+\tfrac12L_Vh .
\label{eq:s-certified}
\end{equation}
Here $h$ is the largest mass-weighted length of a single accepted move, taken before frame removal, and of the spacing between consecutive evaluated points of a link, over all quenches and links that enter equation~\MainEq{eq:eup}. $V$ is assumed to be $L_V$-Lipschitz on a set that contains all these segments.

\textit{Construction of the path.} The path $\Gamma$ traverses the end quench of $S_a$ backwards from $\mathbf x_a$ to the end node. It then follows the parent tree of the main nodes to the seed, continues to the end node on the other side and follows the end quench of $S_b$ to $\mathbf x_b$. Each segment between a node and its parent is traversed along its link or along its quench trajectory, whichever has the lower maximum, as in equation~\MainEq{eq:eup}. Let $\mathbf y_0=\mathbf x_a,\mathbf y_1,\ldots,\mathbf y_M=\mathbf x_b$ be the evaluated configurations in the order in which $\Gamma$ visits them. Two consecutive configurations are joined in one of two ways. For an accepted single-atom move, $\mathbf y_{k+1}=A_{\mathbf y_k}(\mathbf y_k+\mathbf e_i\otimes\boldsymbol\delta)$, and $\Gamma$ follows the straight segment to $\mathbf y_k+\mathbf e_i\otimes\boldsymbol\delta$, of length at most $h$, and then the rigid path of equation~\eqref{eq:s-rigidpath}. Along a link, $\Gamma$ follows the straight segment between consecutive evaluated points, again of length at most $h$. A link written in the frame of its first node is joined to the next node by a rigid path.

\textit{Induction along the path.} Let $\Gamma_k$ be the part of $\Gamma$ from $\mathbf y_0$ to $\mathbf y_k$. We show by induction that
\begin{equation}
\max_{\Gamma_k}V\le\max_{j\le k}V(\mathbf y_j)+\tfrac12L_Vh .
\label{eq:s-induction}
\end{equation}
\begin{itemize}
\item \textit{Base case ($k=0$):} $\Gamma_0$ is the single point $\mathbf y_0$, and $\max_{\Gamma_0}V=V(\mathbf y_0)$.
\item \textit{Induction hypothesis:} Equation~\eqref{eq:s-induction} holds for $k$.
\item \textit{Induction step:} $\Gamma_{k+1}$ is $\Gamma_k$ followed by the piece from $\mathbf y_k$ to $\mathbf y_{k+1}$. On the straight part of this piece, equation~\eqref{eq:s-segment} bounds $V$ by $\max\{V(\mathbf y_k),V(\mathbf y_{k+1})\}+\tfrac12L_Vh$, since the displaced trial and $\mathbf y_{k+1}$ have the same energy. On the rigid part, $V$ is constant by equation~\eqref{eq:s-rigidpath}. Together with the induction hypothesis,
\begin{equation}
\max_{\Gamma_{k+1}}V\le\max_{j\le k+1}V(\mathbf y_j)+\tfrac12L_Vh .
\end{equation}
\end{itemize}
By induction, equation~\eqref{eq:s-induction} holds for $k=M$. The largest evaluated energy along $\Gamma$ is the seed energy, the maximum over segments of the chosen link or trajectory maximum, or an end-quench maximum, which is $E_{\rm up}$ by equation~\MainEq{eq:eup}. Restraints change which moves are accepted but not the energies recorded along $\Gamma$. Therefore, $\max_\Gamma V\le E_{\rm up}+\tfrac12L_Vh$, and equation~\eqref{eq:s-certified} follows from equation~\eqref{eq:s-minimax}.

The code does not compute $L_V$ or $h$. Under the stated assumptions, $\tfrac12L_Vh$ is the discretization correction. Smaller proposal widths and link spacings can reduce $h$, but their nominal values do not provide a deterministic bound on the largest accepted Gaussian move. The evaluated straight-link spacing is at most $\ell/4$. The bound refers to $\mathbf x_a$ and $\mathbf x_b$. The end quench may reach a minimum of $S_a$ other than the lowest one recorded for that state, so barriers are quoted against the lowest minimum found at the end of the search.

\subsection{Routes}\label{prop:route}
We finally extend the bound to routes through intermediate states. Concatenating a path from $\mathbf x$ to $\mathbf y$ with one from $\mathbf y$ to $\mathbf z$ gives a path from $\mathbf x$ to $\mathbf z$ whose maximum is the larger of the two maxima, so that
\begin{equation}
E^\ast(\mathbf x,\mathbf z)\le\max\bigl\{E^\ast(\mathbf x,\mathbf y),E^\ast(\mathbf y,\mathbf z)\bigr\}.
\label{eq:s-ultra}
\end{equation}
Let a route $A=S_0,S_1,\ldots,S_K=B$ use certified chains with sampled maxima $E_{{\rm up},k}$ and spacings $h_k$, and let $U=\max_k(E_{{\rm up},k}+\tfrac12L_Vh_k)$. Suppose that, within every intermediate state $S_k$, the end configurations of the two adjacent chains satisfy $E^\ast\le U$. Applying equation~\eqref{eq:s-ultra} along the route, with equation~\eqref{eq:s-certified} for every chain, shows that the end configurations of the route in $A$ and $B$ satisfy $E^\ast\le U$. A route that minimizes the largest edge value on the graph of states is found by a variant of Dijkstra's algorithm for maximum-capacity paths\cite{pollack1960maximum}. The connection assumption within intermediate states is not checked by the code. It can fail when one bond graph contains several conformers separated by a barrier higher than $U$.

\section{Quantum-Inspired Monte Carlo Simulation}\label{sm:qmc}
This section supplies the implementation details for Methods Sections~\MainSec{sec:m-kernel}--\MainSec{sec:m-crossing}. Supplementary Sections~\ref{sm:kernel} and \ref{sm:pair} specify the single-walker kernel and coupled-pair ensemble used in Methods Sections~\MainSec{sec:m-kernel} and \MainSec{sec:m-explore}; their statistical properties are established in Supplementary Sections~\ref{sm:proof-se3} and \ref{sm:proof-stat}. Supplementary Sections~\ref{sm:quench}--\ref{sm:chains} give the quench, state-assignment and chain-construction details for Methods Section~\MainSec{sec:m-chains}. Supplementary Tables~\ref{tab:s-search} and \ref{tab:s-quench} collect the corresponding settings.

\subsection{Proposal and delayed acceptance}\label{sm:kernel}
This section describes how the move of Methods Section~\MainSec{sec:m-kernel} is implemented. Its exactness is proved in Supplementary Section~\ref{sm:proof-stat} and its invariance in Supplementary Section~\ref{sm:proof-se3}.

\paragraph*{Target and tests.} The target of one walker is equation~\MainEq{eq:targetcb}. A symmetric proposal $g(\mathbf y\mid\mathbf x)=g(\mathbf x\mid\mathbf y)$ generates the trial $\mathbf y$, which is accepted with
\begin{equation}
\alpha(\mathbf x,\mathbf y)=\prod_{k=1}^{3}\min(1,r_k),
\end{equation}
where $r_1=e^{\Delta c}$, $r_2=e^{-\beta\Delta(\tilde V+b)}$, $r_3=e^{-\beta\Delta(V-\tilde V)}$ and $\Delta f=f(\mathbf y)-f(\mathbf x)$. Each test draws its own uniform number, and the first rejection ends the move. Absent terms are zero, and without a surrogate the second and third tests merge into $\min(1,e^{-\beta\Delta(V+b)})$. Supplementary Section~\ref{prop:da} shows that the unaligned symmetric-proposal kernel preserves $\pi$ for any fixed surrogate when $c$, $b$ and $\tilde V$ are fixed. Frame removal additionally requires invariant acceptance factors (Supplementary Section~\ref{prop:frame}).

\paragraph*{Failures.} A failed oracle call is treated as $V(\mathbf y)=+\infty$, so $r_3=0$ and the trial is rejected. A failed surrogate evaluation returns $+\infty$ and is rejected by the second test without an oracle call. A chain is not started from a configuration where the surrogate is infinite, since it could never leave it. During coupled-walker exploration the oracle is called up to four times before a failure is declared, which absorbs transient failures of file-driven programs.

\paragraph*{Single walker.} One atom is chosen with probability $1/N$ and displaced by $\boldsymbol\delta\sim\mathcal N(\mathbf 0,\sigma_i^2\mathsf I_3)$. For free molecules the trial is then centred at its centre of mass and rotated onto the current structure by the proper mass-weighted Kabsch rotation\cite{kabsch1976solution,kabsch1978discussion}. By Supplementary Section~\ref{prop:frame} this frame removal leaves the distribution of every invariant quantity unchanged. Frame removal is skipped for periodic cells and for systems with fixed atoms.

\paragraph*{Coupled walkers.} For a pair, walker $n$ is updated with the conditional target
\begin{equation}
\pi(\mathbf R_n\mid\mathbf R_m)=\frac{1}{Z_n}\exp\Bigl[-\beta_{\rm b}V(\mathbf R_n)-\frac{d_{\rm c}^2(\mathbf R_n,\mathbf R_m)}{\hbar\tau_{12}}\Bigr],
\end{equation}
where $Z_n$ is the integral of the exponential over $\mathbf R_n$ at fixed $\mathbf R_m$ and $d_{\rm c}^2$ is the common-frame distance of equation~\MainEq{eq:pair}. The coupling enters as $c$. Trials of either walker are centred but never rotated, so both walkers stay in one frame. After both walkers have been updated, the second walker receives the rigid rotation of Supplementary Section~\ref{prop:pairframe}, whose angle has standard deviation
\begin{equation}
\sigma_{\rm rot}=\min\Bigl\{0.5,\Bigl(\frac{\hbar\tau_{12}}{\bar I}\Bigr)^{1/2}\Bigr\}~\mathrm{rad},
\label{eq:s-rotwidth}
\end{equation}
where $\bar I=\tfrac23\sum_im_i\lVert\bar{\mathbf r}_{2,i}\rVert^2$ is the mean principal moment of inertia of the centred second walker. A rotation by this angle changes $d_{\rm c}^2/(\hbar\tau_{12})$ by about one. The rotation keeps the energy, so it is tested on the coupling alone and needs no oracle call. The pre-screen is tested once, and a failed test leaves the walker in place. MCTSS instead repeats the pre-selection until a trial passes, for up to 100 attempts\cite{tikhonov2026two}, which tilts the stationary distribution (Supplementary Section~\ref{prop:prescreen}). The squared distance is carried between updates, because it changes only when a move is accepted.

\paragraph*{Adaptive widths.} Quenches and the tuning phase of local straddling-pair ensemble sampling adapt the proposal width. Such chains do not satisfy detailed balance and are used as optimizers or discarded. For ensemble sampling, the width is frozen before recording. The resulting fixed kernel preserves the target; equilibrium estimates additionally require equilibration and adequate mixing.

\subsection{Pair ensemble and midpoint weights}\label{sm:pair}
Let $\bar{\mathbf R}_1$ and $\bar{\mathbf R}_2$ be the centred walkers in their common frame and write $\mathbf M=(\bar{\mathbf R}_1+\bar{\mathbf R}_2)/2$ and $\mathbf D=\bar{\mathbf R}_1-\bar{\mathbf R}_2$, a linear change of variables with a Jacobian of unit modulus. Then $d_{\rm c}^2=\mathbf D^{\mathsf T}\mathsf W\mathbf D$, where $\mathsf W$ is the diagonal mass matrix. Multiplying $\Pi$ of equation~\MainEq{eq:pair} by the weight $w$ of equation~\MainEq{eq:midweight} gives equation~\eqref{eq:s-pairexact}, in which the two factors separate. After normalization, the weighted marginal of $\mathbf M$ is exactly
\begin{equation}
p_w(\mathbf M)=\frac{\exp\bigl[-2\beta_{\rm b}V(\mathbf M)\bigr]}{\displaystyle\int\exp\bigl[-2\beta_{\rm b}V(\mathbf M')\bigr]\,\mathrm d\mathbf M'},
\end{equation}
the ensemble of $\tau_{\rm mid}=2\tau_{\rm b}$. The identity holds for free molecules as well, because the midpoint is formed in the common frame without rotating one walker onto the other (Supplementary Section~\ref{prop:pairframe}). Recorded midpoints are rotated onto the first one before they are written, which changes only the frame of the output.

\paragraph*{Unweighted marginal.} Without weights, expanding to second order in $\mathbf D$ gives $V(\mathbf R_1)+V(\mathbf R_2)=2V(\mathbf M)+\tfrac14\mathbf D^{\mathsf T}\mathsf H\mathbf D+O(\lVert\mathbf D\rVert^4)$, where $\mathsf H$ is the Hessian at $\mathbf M$. The Gaussian integral over $\mathbf D$ and its determinant are restricted to the centred subspace, with $d=3N-3$ for molecules and $d=2$ for the model potentials. When the restricted quadratic form is positive definite, it yields
\begin{equation}
\begin{split}
p(\mathbf M)\approx{}&\frac{\pi^{d/2}}{Z_\Pi}\,e^{-2\beta_{\rm b}V(\mathbf M)}\\
&\times\det\Bigl[\frac{\mathsf W}{\hbar\tau_{12}}+\frac{\beta_{\rm b}}{4}\mathsf H(\mathbf M)\Bigr]^{-1/2}.
\end{split}
\end{equation}
The determinant is the factor that the weights remove.

\paragraph*{Distribution of the log-weights.} To leading order in $\tau_{12}$, $\mathbf D$ is Gaussian with covariance $\boldsymbol\Sigma_D=\tfrac12\hbar\tau_{12}\mathsf W^{-1}$ on the centred subspace (Supplementary Section~\ref{sm:tse}), and $\ln w\approx\tfrac14\beta_{\rm b}\mathbf D^{\mathsf T}\mathsf H\mathbf D$. The moments of a Gaussian quadratic form then give
\begin{equation}
\begin{split}
\langle\ln w\rangle&\approx\tfrac14\beta_{\rm b}\,\mathrm{tr}(\mathsf H\boldsymbol\Sigma_D)=\frac{n_{12}\tau_{\rm b}^2}{8}\sum_k\omega_k^2,\\
\mathrm{Var}(\ln w)&\approx\tfrac18\beta_{\rm b}^2\,\mathrm{tr}\bigl[(\mathsf H\boldsymbol\Sigma_D)^2\bigr]=\frac{(n_{12}\tau_{\rm b}^2)^2}{32}\sum_k\omega_k^4,
\end{split}
\label{eq:s-logw}
\end{equation}
where $\omega_k^2$ are the eigenvalues of $\mathsf W^{-1/2}\mathsf H\mathsf W^{-1/2}$ and $\beta_{\rm b}\hbar\tau_{12}=n_{12}\tau_{\rm b}^2$ was used. For a Gaussian log-weight distribution, the large-sample Kish fraction tends to $\exp[-\mathrm{Var}(\ln w)]$. The finite-sample quantity is defined in Supplementary Section~\ref{sm:tse} and does not account for autocorrelation. Useful weights require control of the total log-weight variance summed over modes; small individual contributions alone are insufficient in a large system. For a C--H stretch near 3000~cm$^{-1}$ at $\tau_{\rm b}=\tau^\ast/2=12.8$~fs, $n_{12}(\omega\tau_{\rm b})^2\approx2.6$ for $n_{12}=0.05$, compared with about 160 for the exploration value $n_{12}=3$. This motivates the stiff coupling used for the local straddling-pair ensemble. It also shows that the weights of exploration pairs, with $n_{12}=3$, are dominated by few samples, so the weighted estimates of Supplementary Section~\ref{sm:barrier} are reported only as heuristics.

\paragraph*{Meeting criterion.} If all atoms were displaced at once, the mean mass-weighted squared displacement per step would be $\tfrac32N\hbar\tau$\cite{tikhonov2026two}. Following MCTSS, a molecular run stops when $d^2(\mathbf R_1,\mathbf R_2)/(\hbar\tau_{\rm b})\le\tfrac32Ns_{\rm meet}$ with $s_{\rm meet}=1$. For the two-dimensional models the test is $d^2/(\hbar\tau_{\rm b})\le s_{\rm meet}$ with $s_{\rm meet}=0.5$.

\paragraph*{Overlap criterion.} The progress of a structure is $p(\mathbf R)=\langle\mathbf R-\mathbf R_A^0,\mathbf R_B^0-\mathbf R_A^0\rangle_{\mathsf W}/\lVert\mathbf R_B^0-\mathbf R_A^0\rVert_{\mathsf W}^2$, evaluated after alignment onto an endpoint, with $\langle\mathbf a,\mathbf b\rangle_{\mathsf W}=\mathbf a^{\mathsf T}\mathsf W\mathbf b$. It is 0 at the reactant and 1 at the product. A run also stops once the largest progress reached by walker 1 is at least the smallest progress reached by walker 2. For the two-dimensional models the axis is the unweighted line between the two starting points.

\subsection{Coupled-walker exploration ladder}\label{sm:explore}
Supplementary Algorithm~\ref{alg:search} gives the ladder. Each rung comprises four independent coupled-walker runs of at most 5000 steps. Molecular runs use $n_{12}=3$ and the two-dimensional runs $n_{12}=1$. Every step of a molecular run updates both walkers and then rotates the second one as in Supplementary Section~\ref{sm:kernel}. A run joins when it meets or overlaps. Both walkers are written in one frame, obtained by the common rotation that takes their midpoint onto the starting midpoint, so consecutive frames of the joined path share one orientation. The exploration parameter $\tau_{\rm x}$ is the largest $\tau_{\rm b}$ of a successful rung before the first failure. Its only role downstream is to set proposal widths and the length $\ell$. It is not used as a temperature in any later step. The paths labelled by quenching come from joined runs of all rungs with $\tau_{\rm b}\le\tau_{\rm x}$, coldest rung first and alternating between rungs, so that more than one effective temperature contributes.

\subsection{Quenches}\label{sm:quench}
A quench runs the move of Methods Section~\MainSec{sec:m-kernel} for one walker at each thermal energy of its schedule, starting from the hottest. Its initial proposal parameter is given in Supplementary Table~\ref{tab:s-quench}. After every block of 25 steps, the proposal parameter is halved if the block acceptance was below 0.2 and multiplied by 1.5 if it exceeded 0.5. When patience $P>0$ is set, a level ends after $P$ consecutive blocks in which the lowest value of $V+b$ fell by less than $0.1$~meV. A quench is flagged as converged when its coldest level ended in this way. Its result is the configuration held at the end, together with the highest energy held along the way.

Single-atom moves relax one atom at a time, so budgets grow with the number of degrees of freedom $d$ ($3N$ for molecules, 2 in two dimensions). With $r=\max(1,d/18)$, every schedule has $r$ times the steps per level listed in Supplementary Table~\ref{tab:s-quench}. The schedules marked as cooled also divide their thermal energies by $r$, so that the excess energy $\tfrac d2k_{\rm B}T$ of the hottest level stays that of an 18-dimensional system. The optional per-system multipliers of budget and patience were not changed from their defaults.

\subsection{States and majority-quench labels}\label{sm:states}
\paragraph*{Fingerprints.} For reactions, atoms $i$ and $j$ are bonded when $r_{ij}<r_i^{\rm cov}+r_j^{\rm cov}+0.4$~\AA, with the covalent radii of ASE\cite{larsen2017ase}. An additive tolerance separates H--H distances of 0.74 to 0.80~\AA\ and stretched from broken ring bonds, which a multiplicative factor cannot do for both. Two bond graphs belong to the same state only if they are identical. For alanine dipeptide the fingerprint is $(\phi,\psi)$, and two configurations share a state when both periodic differences are at most $60^\circ$. For the two-dimensional models two minima share a state when they lie within the basin radius of Supplementary Table~\ref{tab:s-systems}.

\paragraph*{Registration.} A quenched configuration that matches a known state is assigned to it, and the state keeps the lower of the two minima. An unmatched configuration registers a new intermediate $I_k$. Two exceptions apply. If the quench did not converge, or if new states are disallowed as for alanine dipeptide, the configuration is assigned to the nearest known state wherever the fingerprint defines a finite distance. Bond graphs have no such distance, so a non-converged quench with a new bond graph still registers a new state. A label with several quenches takes the majority, and ties go to the label met first.

\paragraph*{Bisection.} The two ends of an exploration path carry the labels $A$ and $B$. For every interval whose end labels differ, the middle frame is quenched and both halves are examined again. The recursion stops at adjacent frames. The number of quenches grows logarithmically with the path length for each label change. Frames that lie between two quenched frames of the same state inherit that state and provide the training data of the label classifier.

\subsection{Progress coordinates}\label{sm:coord}
Let $\mathcal B_a$ and $\mathcal B_b$ be the bond sets of the minima of $S_a$ and $S_b$. The bond-change coordinate is
\begin{equation}
\xi_{\rm bond}(\mathbf R)=\sum_{(i,j)\in\mathcal B_a\setminus\mathcal B_b}r_{ij}-\sum_{(k,l)\in\mathcal B_b\setminus\mathcal B_a}r_{kl},
\end{equation}
which generalizes the $\mathrm{S_N2}$ coordinate $r(\mathrm{C{-}Br})-r(\mathrm{C{-}Cl})$. For a dihedral fingerprint $\boldsymbol\theta$, let $\boldsymbol\Delta=\mathrm{wrap}(\boldsymbol\theta_b-\boldsymbol\theta_a)$ with $\mathrm{wrap}$ mapping angles into $[-180^\circ,180^\circ)$. Then
\begin{equation}
\xi_{\rm dih}(\mathbf R)=\frac{\mathrm{wrap}\bigl(\boldsymbol\theta(\mathbf R)-\boldsymbol\theta_a\bigr)\cdot\boldsymbol\Delta}{\boldsymbol\Delta\cdot\boldsymbol\Delta},
\end{equation}
which is 0 at $S_a$ and 1 at $S_b$. When neither coordinate exists, for instance for the two-dimensional models, $\xi_{\rm label}$ is the latent $z$ of a classifier with the architecture of the committor network (Supplementary Section~\ref{sm:nc-net}). It is trained on the path frames labelled $S_a$ or $S_b$ and on the two minima, with targets 0 and 1. The loss is the binary cross-entropy of $\sigma(3z)$ with the two classes weighted equally. Optimization uses 400 full-batch epochs of Adam\cite{kingma2015adam} with learning rate $3\times10^{-3}$ and weight decay $10^{-3}$, which keeps $z$ finite on separable data so that its level sets remain usable as slices. When chains are regrown after committor learning, the coordinate is the committor latent $z_\theta$.

\subsection{Continuation chains and sampled energy maxima}\label{sm:chains}
This section completes the chain construction and sampled maximum $E_{\rm up}$ of Methods Section~\MainSec{sec:m-chains}, and Supplementary Algorithm~\ref{alg:chain} summarizes it. The seed is restrained at its own coordinate, clipped to the interval between the two minima. Slices farther than $\Delta\xi/4$ from the seed on either side are visited in order, each node starting from the previous one. For each node, the restraint centre moves from the current $\xi$ to the target in five equal steps. Each step is a short quench at the second-coldest level of the node schedule with half its step budget. The full node schedule then follows, and the restraint is stiffened fourfold if the node misses its slice by more than $\Delta\xi/4$, up to three times. The proposal parameter carried to the next node is the larger of the value the quench ended with and a quarter of the value it started with. An inserted node targets the mid-point of its two neighbours with half the slice width and the current restraint. Densification stops after at most three recursive levels.

\paragraph*{Candidates for the sampled maximum.} The main nodes are the seed node and the nodes on the slice grid. Every main node $m$ other than the seed has a parent $p$, the main node from which it was grown. The segment between $m$ and $p$ spans the node indices $a=\min(m,p)$ to $b=\max(m,p)$, including inserted nodes. Its straight-path maximum $E^{\rm link}$ is the largest node energy in $[a,b]$ and interior link energy between them. Links are evaluated in the frame of their first node. Its trajectory maximum $E^{\rm traj}$ is the largest energy held along all quenches that grew $m$ from $p$, including the starting energy. The sampled chain maximum is equation~\MainEq{eq:eup}, and the location of its maximum is reported as a node, a straight segment, a trajectory segment or an end quench.

\paragraph*{Continuous-path correction.} Supplementary Section~\ref{prop:bound} shows that the minimax energy between the end configurations of a certified chain is at most $E_{\rm up}+\tfrac12L_Vh$, where $L_V$ is a Lipschitz constant of $V$ and $h$ the largest spacing between consecutive evaluated configurations. Certification checks endpoint connectivity and the largest evaluated energy; the code does not compute $L_V$ or the discretization correction. The reported barrier convention is defined in Methods Section~\MainSec{sec:m-chains}; it uses the lowest recorded state minimum even when a chain ends at another conformer.

\paragraph*{Channels and routes.} Two certified chains form one channel when the highest-energy chain node of each lies within $3\ell$ of some node of the other. The lowest chain of each channel is written out and used to locate the quench-label crossing and to sample the local straddling-pair ensemble. On the undirected graph whose edges carry the lowest $E_{\rm up}$ of each pair, a Dijkstra-type search that minimizes the largest edge value along the path\cite{pollack1960maximum} returns the $A$-to-$B$ route and its sampled score. The reported graph value is a sampled minimax score. The continuous-path bound additionally requires the edge discretization corrections and intermediate-state connections of Supplementary Section~\ref{prop:route}.

\subsection{Quench-label crossing and curvature}\label{sm:crossing}
Crossing labels are majority votes of three quenches with the label schedule. Bisection first narrows the pair of adjacent nodes whose labels differ, starting from the known end labels. The straight link between them is then bisected four times, leaving a bracket with one sixteenth of the original link length. The candidate is its midpoint, and $1/32$ denotes half this geometric width rather than uncertainty in a basin boundary. Because the crossing lies on a straight link, $E^\ddagger_0$ is not bounded by $E_{\rm up}$, which may be set by a trajectory segment. The energy curvature along the chain is
\begin{equation}
k=\frac{V(\mathbf R_k+h\hat{\mathbf e})+V(\mathbf R_k-h\hat{\mathbf e})-2V(\mathbf R_k)}{h^2},
\end{equation}
where $\mathbf R_k$ is the highest-energy chain node, $\hat{\mathbf e}$ the unit mass-weighted tangent through its neighbours and $h=0.05$~amu$^{1/2}$\,\AA. The implied frequency is $\omega=\sqrt{\lvert k\rvert}$, reported with the sign of $k$ as a wavenumber, and $T_{\rm c}=\hbar\lvert\omega\rvert/(2\pi k_{\rm B})$. The tangent is not a normal mode, so both values are diagnostics.

The geometric bisection interval does not quantify stochastic label uncertainty. With three random quench votes per query, inconsistent labels can alter the bracket; its midpoint is a protocol-defined candidate rather than verified membership of a unique separatrix.

\subsection{Local straddling-pair ensemble}\label{sm:tse}
The output label TSE denotes the local straddling-pair ensemble defined here, not a verified isocommittor ensemble. Let $s(\mathbf R)=\pm[\xi(\mathbf R)-\xi^\ast]$, where $\xi^\ast=\xi(\mathbf R^\ddagger_0)$ and the sign is chosen so that $s<0$ at the minimum of $S_a$. The ensemble is sampled from $\Pi$ of equation~\MainEq{eq:pair} with $\tau_{\rm b}=\tau^\ast/2$ and $\tau_{12}=0.05\tau_{\rm b}$, multiplied by the indicator $\mathbb 1[s(\mathbf R_1)<0<s(\mathbf R_2)]$. The indicator enters the geometric test of the move, so a move that breaks it costs no oracle call. The walkers share one frame and their relative orientation is sampled by the rotation of Supplementary Section~\ref{sm:kernel}. Because $s$ is invariant (Supplementary Section~\ref{prop:equivariance}), the rotation never violates the indicator. During 2000 tuning sweeps the proposal parameter starts at $\tau_{12}$ and is multiplied by $\exp[2(a-0.3)]$ after every 100 moves with acceptance $a$. The width is then frozen and 20000 sweeps are recorded. The midpoint energy is evaluated every 10 recorded sweeps.

For molecular pairs, $\mathbf D$ lies in the centre-of-mass-zero subspace. Gaussian covariances and inverse masses in the following calculation are restricted to that subspace. In ambient coordinates the covariance is
\begin{equation}
\boldsymbol\Sigma_D=\frac{\hbar\tau_{12}}{2}\left[\left(\mathrm{diag}(m_i^{-1})-\frac{\mathbf 1_N\mathbf 1_N^{\mathsf T}}{M_{\rm tot}}\right)\otimes\mathsf I_3\right],
\end{equation}
where $M_{\rm tot}=\sum_i m_i$. Translation-invariant gradients have zero summed Cartesian components, so their covariance contractions equal those obtained with $(\hbar\tau_{12}/2)\mathsf W^{-1}$. The same convention applies to the harmonic log-weight contractions in Supplementary Section~\ref{sm:pair}.

\paragraph*{Weighted joint density.} Since $2\beta_{\rm b}=\beta^\ast$, equation~\eqref{eq:s-pairexact} gives the weighted joint density
\begin{equation}
\begin{split}
\Pi\,w\,\mathbb 1={}&\frac{1}{Z_\Pi}\,e^{-\beta^\ast V(\mathbf M)}\exp\Bigl[-\frac{\mathbf D^{\mathsf T}\mathsf W\mathbf D}{\hbar\tau_{12}}\Bigr]\\
&\times\mathbb 1\bigl[s(\mathbf M+\tfrac12\mathbf D)<0<s(\mathbf M-\tfrac12\mathbf D)\bigr],
\end{split}
\label{eq:s-tsejoint}
\end{equation}
where $\mathbf M$ and $\mathbf D$ are defined in Supplementary Section~\ref{sm:pair}. The marginal derivation below additionally assumes the full-space relative-coordinate integration used in Supplementary Section~\ref{prop:pairframe}; within that setting, it linearizes $s$ over the scale of $\mathbf D$.

\paragraph*{Marginal of the midpoint.} At fixed $\mathbf M$, the second factor of equation~\eqref{eq:s-tsejoint} makes $\mathbf D$ Gaussian with covariance $\boldsymbol\Sigma_D$ as specified above (or $\tfrac12\hbar\tau_{12}\mathsf W^{-1}$ for the uncentred two-dimensional models). With $\mathbf g=\nabla s(\mathbf M)$ and $s(\mathbf M\pm\tfrac12\mathbf D)\approx s(\mathbf M)\pm\tfrac12\mathbf g\cdot\mathbf D$, the indicator requires $-\mathbf g\cdot\mathbf D>2\lvert s(\mathbf M)\rvert$. The scalar $\mathbf g\cdot\mathbf D$ is Gaussian with mean zero and variance $\sigma_g^2=\mathbf g^{\mathsf T}\boldsymbol\Sigma_D\mathbf g=\tfrac12\hbar\tau_{12}\,\mathbf g^{\mathsf T}\mathsf W^{-1}\mathbf g$, so the indicator holds with probability $\tfrac12\operatorname{erfc}(\sqrt2\lvert s(\mathbf M)\rvert/\sigma_g)$. The marginal of $\mathbf M$ is therefore
\begin{equation}
p(\mathbf M)=\frac{1}{Z_s}\,e^{-\beta^\ast V(\mathbf M)}\,\tfrac12\operatorname{erfc}\Bigl(\frac{\sqrt2\,\lvert s(\mathbf M)\rvert}{\sigma_g(\mathbf M)}\Bigr),
\end{equation}
where $Z_s$ is the integral of the numerator over $\mathbf M$. The erfc factor is a window of width of order $\sigma_g$ around the dividing surface.

\paragraph*{Stiff limit.} With $\int_0^\infty\operatorname{erfc}(u)\,\mathrm du=1/\sqrt\pi$,
\begin{equation}
\int_{-\infty}^{\infty}\tfrac12\operatorname{erfc}\Bigl(\frac{\sqrt2\,\lvert s\rvert}{\sigma_g}\Bigr)\mathrm ds=\frac{\sigma_g}{\sqrt2}\int_0^\infty\operatorname{erfc}(u)\,\mathrm du=\frac{\sigma_g}{\sqrt{2\pi}} .
\end{equation}
As $\tau_{12}\to0$ the window therefore tends to $(\sigma_g/\sqrt{2\pi})\,\delta(s)$, and
\begin{equation}
p(\mathbf M)\to\frac{\sigma_g(\mathbf M)\,e^{-\beta^\ast V(\mathbf M)}\,\delta\bigl(\xi(\mathbf M)-\xi^\ast\bigr)}{\displaystyle\int\sigma_g(\mathbf M')\,e^{-\beta^\ast V(\mathbf M')}\,\delta\bigl(\xi(\mathbf M')-\xi^\ast\bigr)\,\mathrm d\mathbf M'} ,
\end{equation}
where $\sigma_g=(\tfrac12\hbar\tau_{12})^{1/2}\lvert\nabla\xi\rvert_{\mathsf W^{-1}}$ with $\lvert\nabla\xi\rvert_{\mathsf W^{-1}}=(\mathbf g^{\mathsf T}\mathsf W^{-1}\mathbf g)^{1/2}$. By the coarea formula, $\lvert\nabla\xi\rvert_{\mathsf W^{-1}}\delta(\xi-\xi^\ast)\,\mathrm d\mathbf M$ is, up to a constant Jacobian, the intrinsic surface element in mass-weighted coordinates. It is unchanged by a smooth monotone relabelling of the same dividing surface, since the gradient factor cancels the delta-function Jacobian. Averages over $\delta(\xi-\xi^\ast)\,e^{-\beta^\ast V}$ without this factor would require an additional reweighting by $\lvert\nabla\xi\rvert_{\mathsf W^{-1}}^{-1}$, which is not applied.

\paragraph*{Reported quantities.} The run reports normalized weights. Their concentration is summarized by the Kish fraction $\mathrm{ESS}/n=(n\sum_iw_i^2)^{-1}$\cite{kish1965survey}. It also reports the weighted mean midpoint energy and the weighted root-mean-square deviation of $\xi(\mathbf M)$ from $\xi^\ast$. The last quantity measures the width of the window and so indicates how far the run is from the stiff limit.

\section{Committor-learning formulation}\label{sm:pinc}
This section gives the physical and statistical basis of the committor procedure in Methods Section~\MainSec{sec:m-committor}. The biased walks, shooting trajectories and validation chains all use the kernel of Methods Section~\MainSec{sec:m-kernel} at $\tau^\ast$. Their core sets are constructed from the fingerprints and progress coordinates defined in Methods Section~\MainSec{sec:m-chains}.

\subsection{Committor of the Monte Carlo chain}\label{sm:pinc-dyn}
The core sets and finite-step hitting probability are defined in Methods Section~\MainSec{sec:m-committor}. Supplementary Sections~\ref{prop:generator} and \ref{prop:ritz} derive the diffusion approximation and explain the inverse masses in the variational functional. These results do not remove finite-step error. Shooting follows the implemented chain and supplies its validation reference (Supplementary Section~\ref{sm:nc-shoot}).

\subsection{Kolmogorov bias}\label{sm:pinc-vk}
With $q=\sigma(sz)$, $\nabla q=s\,q(1-q)\nabla z$, so $\ln\lvert\nabla_{\!u}q\rvert^2=2\ln s+\ln\lvert\nabla_{\!u}z\rvert^2+2\ln q+2\ln(1-q)$. Equation~\MainEq{eq:vk} regularizes each logarithm separately and shifts the first by $-\ln\epsilon$ so that it is non-negative\cite{trizio2025everything}. The gradient is that of $z$, since the factor $q(1-q)$ is already present in the other two terms. Because the first bracketed term is non-negative and $\ln(q+\delta)+\ln(1-q+\delta)\ge\ln\delta$ to first order in $\delta$, $V_{\rm K}\le2\lambda\lvert\ln\delta\rvert/\beta^\ast$.

\paragraph*{Biased ensemble.} Exponentiating equation~\MainEq{eq:vk} gives
\begin{equation}
e^{-\beta^\ast V_{\rm K}}=\Bigl[\frac{\lvert\nabla_{\!u}z_\theta\rvert^2+\epsilon}{\epsilon}\Bigr]^{\lambda}(q_\theta+\delta)^{2\lambda}(1-q_\theta+\delta)^{2\lambda}.
\label{eq:s-vkexp}
\end{equation}
For $\epsilon,\delta\to0$ and with $\lvert\nabla_{\!u}q_\theta\rvert^2=s^2q_\theta^2(1-q_\theta)^2\lvert\nabla_{\!u}z_\theta\rvert^2$, the right-hand side tends to $(s^2\epsilon)^{-\lambda}\lvert\nabla_{\!u}q_\theta\rvert^{2\lambda}$. For $\lambda=1$ the walks without $V_z$ thus sample
\begin{equation}
p_{\rm K}(\mathbf R)=\frac{\lvert\nabla_{\!u}q_\theta(\mathbf R)\rvert^2e^{-\beta^\ast V(\mathbf R)}}{\displaystyle\int\lvert\nabla_{\!u}q_\theta(\mathbf R')\rvert^2e^{-\beta^\ast V(\mathbf R')}\,\mathrm d\mathbf R'},
\label{eq:s-pk}
\end{equation}
which is the normalized integrand of $K[q_\theta]$ in equation~\MainEq{eq:ritz}. This observation underlies the Kolmogorov bias of Kang et al.\cite{kang2024committor}. Sampling concentrates where the current model places the transition region, which is where the functional is least constrained by unbiased data. By Supplementary Section~\ref{prop:equivariance}, $V_{\rm K}$ and $p_{\rm K}$ are invariant under rigid motions, so the biased walks remain covered by Supplementary Section~\ref{prop:frame}.

\subsection{Adaptive bias along the latent}\label{sm:pinc-vz}
The grid of $V_z$ spans the range of $z_\theta$ over all stored configurations, widened by 2 on each side, with 201 points. Its range is recomputed after each refit because the fitted latent values may change; the sigmoid slope remains fixed at $s=3$. A deposition at $z_0$ adds $h\exp[-(z-z_0)^2/(2\times0.25^2)]$, with $h=\tfrac12k_{\rm B}T^\ast\exp[-V_z(z_0)/(k_{\rm B}T^\ast(\gamma_{\rm wt}-1))]$ and bias factor $\gamma_{\rm wt}=10$\cite{barducci2008welltempered}.

\subsection{Reweighting biased blocks}\label{sm:pinc-mbar}
Block $k$ holds $N_k$ configurations sampled with frozen bias $b_k$, and the basin block has $b_0=0$. With reduced biases $u_k(\mathbf R)=\beta^\ast b_k(\mathbf R)$, the MBAR free energies\cite{shirts2008mbar} solve
\begin{equation}
e^{-f_k}=\sum_{n}\frac{e^{-u_k(\mathbf R_n)}}{\sum_lN_le^{f_l-u_l(\mathbf R_n)}},
\end{equation}
where $f_k$ is the reduced free energy of block $k$, the sum over $n$ runs over all stored configurations and $f_0=0$ fixes the additive constant. The equations are iterated until the largest change is below $10^{-10}$, for at most 500 iterations. Under equilibrium and overlap assumptions, the reweighting estimate for sample $n$ is
\begin{equation}
w_n=\frac{\bigl[\sum_lN_le^{f_l-u_l(\mathbf R_n)}\bigr]^{-1}}{\sum_m\bigl[\sum_lN_le^{f_l-u_l(\mathbf R_m)}\bigr]^{-1}} .
\end{equation} Weights above the percentile specified in Supplementary Table~\ref{tab:s-committor} are capped before normalization. This regularization trades bias for variance reduction.

\section{Hyperparameter configuration}\label{sm:hyper}
The proposal widths, free-particle pair coupling and molecular meeting criterion follow MCTSS\cite{tikhonov2026two}. The neural architecture, sigmoid slope and Kolmogorov-bias regularizers follow Trizio et al.\cite{trizio2025everything}; shooting-point selection and likelihood follow AIMMD\cite{jung2023machine}. The remaining budgets, tolerances and resolution settings are numerical heuristics. Supplementary Tables~\ref{tab:s-search}--\ref{tab:s-committor} collect the core parameters. Detailed move, restraint and ensemble protocols are given in Supplementary Section~\ref{sm:qmc}; neural and bias protocols are specified in Supplementary Sections~\ref{sm:pinc} and \ref{sm:nc}.

Defaults are shared across systems, with molecular/two-dimensional differences shown by slashes in the tables. Degree-of-freedom scaling sets quench and shooting budgets, $\tau_{\rm x}$ sets proposal widths and $\ell$, and $T_{\rm phys}$ sets $\tau^\ast$. The inexpensive two-dimensional potentials use longer sampling walks, more labelled paths and chains, and the distinct meeting criterion in Supplementary Section~\ref{sm:pair}.

The shooting-loss multiplier is $\Gamma=1$ for molecules and 0 for the two-dimensional models. Thus, two-dimensional shots contribute to diagnostics but not to refitting. No setting was tuned separately for an individual system; optional quench-budget and patience multipliers remained at their defaults.

\begin{SuppColTable}{tab:s-search}{\textbf{Exploration and continuation parameters.} Slash-separated values denote molecules / two-dimensional models; single values are shared. $\ell=\sqrt{3\hbar\tau_{\rm x}/2}$ sets geometric resolution and $L$ is the link length. The rung join fraction selects an exploration scale, not a benchmark success probability. Local-ensemble settings are given in Supplementary Section~\ref{sm:tse}, and seed conventions in Supplementary Section~\ref{sm:repro}.}
\begin{ruledtabular}
\begin{tabular}{@{}ll@{}}
\TabCell{0.5\columnwidth}{Parameter} & \TabCell{0.44\columnwidth}{Value} \\
\hline
\TabCell{0.5\columnwidth}{Initial $\tau_0$; rung ratio} & \TabCell{0.44\columnwidth}{1~fs; 2} \\
\TabCell{0.5\columnwidth}{Runs per rung; maximum rungs; halvings} & \TabCell{0.44\columnwidth}{4; 8; 3} \\
\TabCell{0.5\columnwidth}{Rung join fraction; steps per run} & \TabCell{0.44\columnwidth}{0.5; 5000} \\
\TabCell{0.5\columnwidth}{Pair coupling $\tau_{12}/\tau_{\rm b}$} & \TabCell{0.44\columnwidth}{3 / 1} \\
\TabCell{0.5\columnwidth}{Meeting threshold for $d^2/(\hbar\tau_{\rm b})$} & \TabCell{0.44\columnwidth}{$3N/2$ / 0.5} \\
\TabCell{0.5\columnwidth}{Labelled paths; chains per state pair} & \TabCell{0.44\columnwidth}{8 / 12; 4 / 8} \\
\TabCell{0.5\columnwidth}{Slices; restraint $\kappa_0$} & \TabCell{0.44\columnwidth}{24; 1~eV} \\
\TabCell{0.5\columnwidth}{Slice tolerance; restraint stiffening} & \TabCell{0.44\columnwidth}{$\Delta\xi/4$; $\times4$, up to 3 times} \\
\TabCell{0.5\columnwidth}{Maximum node gap; channel radius} & \TabCell{0.44\columnwidth}{$\ell$; $3\ell$} \\
\TabCell{0.5\columnwidth}{Interior link points} & \TabCell{0.44\columnwidth}{$\max(4,\lceil4L/\ell\rceil)$} \\
\TabCell{0.5\columnwidth}{Crossing votes; link bisections} & \TabCell{0.44\columnwidth}{3; 4} \\
\end{tabular}
\end{ruledtabular}
\end{SuppColTable}

\begin{SuppColTable}{tab:s-quench}{\textbf{Quench annealing schedules.} Steps are attempted moves per temperature level before multiplication by $r=\max(1,d/18)$. Endpoint, node and ramp thermal energies are also divided by $r$; label and chain-end levels are unscaled. Patience is four blocks of 25 steps, except zero for the ramp, with a 0.1~meV improvement threshold. Proposal-width adaptation is specified in Supplementary Section~\ref{sm:quench}.}
\setlength{\tabcolsep}{3pt}%
\begin{ruledtabular}
\begin{tabular}{@{}llll@{}}
\TabCell{0.21\columnwidth}{Use} & \TabCell{0.28\columnwidth}{$k_{\rm B}T$ levels (meV)} & \TabCell{0.1\columnwidth}{Steps} & \TabCell{0.28\columnwidth}{Initial parameter} \\
\hline
\TabCell{0.21\columnwidth}{Endpoint relaxation} & \TabCell{0.28\columnwidth}{20, 8, 3, 1, 0.5} & \TabCell{0.1\columnwidth}{400} & \TabCell{0.28\columnwidth}{$\tau_0$} \\
\TabCell{0.21\columnwidth}{Path and crossing labels} & \TabCell{0.28\columnwidth}{5, 2, 1, 0.5} & \TabCell{0.1\columnwidth}{150} & \TabCell{0.28\columnwidth}{$\tau_{\rm x}/16$} \\
\TabCell{0.21\columnwidth}{Chain node} & \TabCell{0.28\columnwidth}{20, 8, 3, 1} & \TabCell{0.1\columnwidth}{250} & \TabCell{0.28\columnwidth}{$\tau_{\rm x}$ initially, then carried over} \\
\TabCell{0.21\columnwidth}{Restraint ramp step} & \TabCell{0.28\columnwidth}{3} & \TabCell{0.1\columnwidth}{125} & \TabCell{0.28\columnwidth}{carried over} \\
\TabCell{0.21\columnwidth}{Chain ends} & \TabCell{0.28\columnwidth}{5, 2, 1, 0.5} & \TabCell{0.1\columnwidth}{300} & \TabCell{0.28\columnwidth}{$\tau_{\rm x}/16$} \\
\end{tabular}
\end{ruledtabular}
\end{SuppColTable}

\begin{SuppColTable}{tab:s-committor}{\textbf{Committor-training and sampling parameters.} Slash-separated values denote molecules / two-dimensional models. Biases are frozen before production; reweighting and weight capping precede refitting. Shooting diagnostics are evaluated separately from refitting. Network architectures and shot limits are given in Supplementary Table~\ref{tab:s-sysnc}, and deposition and storage intervals in Supplementary Section~\ref{sm:nc-iter}.}
\begin{ruledtabular}
\begin{tabular}{@{}ll@{}}
\TabCell{0.5\columnwidth}{Parameter} & \TabCell{0.44\columnwidth}{Value} \\
\hline
\TabCell{0.5\columnwidth}{Iterations after initial fit} & \TabCell{0.44\columnwidth}{6} \\
\TabCell{0.5\columnwidth}{Initial basin steps per minimum} & \TabCell{0.44\columnwidth}{4000 / 20000} \\
\TabCell{0.5\columnwidth}{Biased exploration; production steps per walker} & \TabCell{0.44\columnwidth}{1500 / 5000; 12000 / 20000} \\
\TabCell{0.5\columnwidth}{Kolmogorov bias $\lambda$; $\epsilon,\delta$} & \TabCell{0.44\columnwidth}{1; $10^{-7}$} \\
\TabCell{0.5\columnwidth}{Latent-bias width; initial height; bias factor} & \TabCell{0.44\columnwidth}{0.25; $k_{\rm B}T^\ast/2$; 10} \\
\TabCell{0.5\columnwidth}{Epochs; minibatch; learning rate} & \TabCell{0.44\columnwidth}{300; 256 / 512; $10^{-3}$} \\
\TabCell{0.5\columnwidth}{Loss weights $\alpha$; $\Gamma$} & \TabCell{0.44\columnwidth}{1; 1 / 0} \\
\TabCell{0.5\columnwidth}{Reweighting cap} & \TabCell{0.44\columnwidth}{99th percentile} \\
\TabCell{0.5\columnwidth}{Shooting points per iteration; shots per point} & \TabCell{0.44\columnwidth}{16; 2} \\
\TabCell{0.5\columnwidth}{Probe shots per crossing} & \TabCell{0.44\columnwidth}{20} \\
\end{tabular}
\end{ruledtabular}
\end{SuppColTable}

\section{Neural committor}\label{sm:nc}
This section specifies the model, loss and data-generation loop of Methods Section~\MainSec{sec:m-committor}. Supplementary Algorithm~\ref{alg:committor} shows their execution order, Supplementary Table~\ref{tab:s-committor} lists the shared settings, and Supplementary Table~\ref{tab:s-sysnc} gives the system-dependent descriptors and shot budgets.

\subsection{Network and descriptors}\label{sm:nc-net}
The descriptors are all $N(N-1)/2$ interatomic distances, each computed as $\sqrt{r_{ij}^2+10^{-12}}$, or the two coordinates of the two-dimensional models. The network maps them through two hidden layers of 32 units with hyperbolic-tangent activations to the scalar latent $z_\theta$, with default PyTorch initialization, and $q_\theta=\sigma(3z_\theta)$. The same network instance is refitted in every iteration.

\subsection{Loss and optimization}\label{sm:nc-loss}
The mass-weighted gradient is obtained by automatic differentiation with respect to Cartesian coordinates, $\lvert\nabla_{\!u}q\rvert^2=\sum_jm_j^{-1}(\partial q/\partial x_j)^2$. The first term of equation~\MainEq{eq:loss} normalizes weights within each uniformly shuffled minibatch. These updates need not be unbiased gradients of the full-data objective. The boundary and shooting terms use all core configurations and all committed shots, respectively. The shot term is
\begin{equation}
\ell_{\rm shot}=-\frac{\sum_k\bigl[n_B^{(k)}\ln q_\theta(\mathbf R_k)+n_A^{(k)}\ln\bigl(1-q_\theta(\mathbf R_k)\bigr)\bigr]}{\sum_k\bigl(n_A^{(k)}+n_B^{(k)}\bigr)},
\end{equation}
evaluated through $\ln\sigma(\pm3z)$ for numerical stability\cite{jung2023machine}. At the start of each fit, $\gamma=\Gamma\,\mathcal L_{\rm v}^{(0)}/\ell_{\rm shot}^{(0)}$, where $\mathcal L_{\rm v}^{(0)}$ is the first term over the full data set. Without this scaling a few dozen shots would dominate the Boltzmann average of $\lvert\nabla_{\!u}q\rvert^2$, which is orders of magnitude smaller. Optimization uses Adam\cite{kingma2015adam}, with epochs, learning rate and minibatch sizes specified in Supplementary Table~\ref{tab:s-committor}. Data are shuffled from a fixed seed.

\subsection{Iterative data generation}\label{sm:nc-iter}
Supplementary Algorithm~\ref{alg:committor} gives the loop. The walkers are the two minima and the highest-energy chain node of the lowest chain of every channel. In the exploration phase they take turns growing one shared $V_z$, depositing every 20 steps (50 in two dimensions). The bias is then frozen and every walker runs a production walk in parallel, storing every fifth configuration (every tenth in two dimensions). Walkers continue from their last configuration in the next iteration.

\subsection{Shooting diagnostics}\label{sm:nc-shoot}
Candidates are the chain nodes and the configurations of the newest block. Their latents are histogrammed with $\min[30,\max(2,\lfloor n/10\rfloor)]$ bins, and candidate $n$ is drawn with the probability of equation~\MainEq{eq:select}, with the density $\rho(z_n)$ given by the count $h(z_n)$ of its bin. Sixteen points are drawn without replacement. From each, two independent unbiased chains at $\tau^\ast$ stop at the first accepted configuration inside a core. Their budgets are $2000\max(1,3N/18)$ steps for molecules and 2000 steps in two dimensions. The implementation checks core membership only after an accepted move, so core-start shots need not satisfy the target boundary values. Before refitting, the current model is scored on these outcomes by $\ell_{\rm shot}$ and by the ratio between the observed fraction of points whose two chains reached different states and its expectation $\langle2q_\theta(1-q_\theta)\rangle$\cite{jung2023machine}. Both scores use only chains that committed.

\paragraph*{Why these scores.} For an initial point outside both cores and correctly implemented uncensored first-hitting outcomes, the two chains are independent and each reaches $C_b$ first with probability $q(\mathbf R)$ of equation~\MainEq{eq:committor}. They reach different cores with probability
\begin{equation}
\Pr[\text{split}\mid\mathbf R]=q(1-q)+(1-q)q=2q(\mathbf R)\bigl[1-q(\mathbf R)\bigr],
\end{equation}
so the ratio is one in expectation when $q_\theta=q$ on the shooting points and outcomes are uncensored. Conditioning on commitment within a finite budget can change this relation. For one point, the expected contribution of its chains to $\ell_{\rm shot}$ is the cross-entropy $-q\ln q_\theta-(1-q)\ln(1-q_\theta)$. For uncensored outcomes, Gibbs' inequality makes this expectation minimal only at $q_\theta=q$, so the binomial likelihood is a strictly proper scoring rule\cite{gneiting2007strictly}. A model that predicts $q_\theta=1/2$ everywhere scores $\ell_{\rm shot}=\ln2$, which serves as the chance level.

Crossing probes report the complete-case fraction and binomial standard error defined in Methods Section~\MainSec{sec:m-committor}, using the count in Supplementary Table~\ref{tab:s-committor}. Unresolved shots are counted separately and excluded from $p_B$. This fraction need not estimate the unconditional hitting probability if censoring depends on the eventual outcome.

\section{Simulation details}\label{sm:sim}
This section records the energy function, endpoints, committor settings and independent evaluation references for each benchmark in Results Section~\MainSec{sec:r-systems}. Supplementary Table~\ref{tab:s-systems} summarizes the systems, and Supplementary Table~\ref{tab:s-sysnc} lists the settings that vary with system size. All remaining settings use the defaults in Supplementary Section~\ref{sm:hyper}.

The three two-dimensional models share the following details. A single particle moves in the $(x,y)$ plane, with coordinates in \AA. Both coordinates are displaced at every step, and no frame is removed. The endpoints are fixed guesses of the two minima, relaxed by the endpoint quench of Supplementary Table~\ref{tab:s-quench}. Exploration runs also evaluate trials rejected by the pre-screen, for diagnostics only, and these evaluations are not counted as cost. The committor network reads $(x,y)$ and has the architecture 2--32--32--1. Reference minima and first-order saddles are polished from fixed guesses with analytic gradients and checked with the Hessian. The reference committor is the finite-volume solution described in Supplementary Section~\ref{sm:reference}, on the grid given for each model.

The molecular systems share the following details. One calculator instance is kept per walker, so that each self-consistent field calculation restarts from the walker's own previous solution. GFN2-xTB\cite{bannwarth2019gfn2} is evaluated in process with tblite at accuracy 1, electronic temperature 300~K and at most 250 self-consistent iterations. The committor network reads all $N(N-1)/2$ interatomic distances. The endpoints are relaxed by the endpoint quench before exploration, so that non-optimized input structures are accepted.

\subsection{Proton transfer coupled to a harmonic bath}\label{sm:sys-pt}
\paragraph*{Energy function.} The model of MCTSS\cite{tikhonov2026two} is a symmetric double well along $x$, coupled bilinearly to a harmonic bath coordinate $y$,
\begin{equation}
\begin{split}
V(x,y)={}&\frac{V_{\rm BH}}{x_0^4}(x-x_0)^2(x+x_0)^2\\
&+\tfrac12fy^2-\alpha fxy,
\end{split}
\end{equation}
where $V_{\rm BH}=2000$~cm$^{-1}$ is the uncoupled double-well barrier parameter and $x_0=0.5$~\AA\ its uncoupled minimum coordinate, $f=4\pi^2mc^2\tilde\nu^2$ the force constant of the bath for the wavenumber $\tilde\nu=1000$~cm$^{-1}$, $\alpha=0.7$ the coupling strength and $m=1$~amu the particle mass. Energies are in cm$^{-1}$ and no walls are added.

\paragraph*{Simulation details.} The endpoints are $(\mp x_0,0)$. The physical parameter is $\tau^\ast=10$~fs, which places the barrier near $8\,k_{\rm B}T^\ast$. Two minima share a state when they lie within 0.22~\AA\ of each other.

\paragraph*{Reference quantities.} The bilinear coupling gives minima near $(x,y)=\pm(0.603,0.422)$~\AA. The origin is the only saddle, with an actual barrier of about $4229$~cm$^{-1}$ relative to those coupled minima. The reference committor is computed on a $161\times121$ grid over $-1.2\le x\le1.2$~\AA\ and $-0.8\le y\le0.8$~\AA.

\subsection{M\"uller--Brown potential}\label{sm:sys-mb}
\paragraph*{Energy function.} The M\"uller--Brown potential\cite{muller1979location} is used in the scaled form of Trizio et al.\cite{trizio2025everything},
\begin{equation}
\begin{split}
V(x,y)={}&0.15\Bigl[146.7+\sum_{k=1}^{4}A_ke^{g_k(x,y)}\Bigr],\\
g_k(x,y)={}&a_k(x-x_k)^2+b_k(x-x_k)(y-y_k)\\
&+c_k(y-y_k)^2,
\end{split}
\end{equation}
where, for $k=1$ to 4, $A_k$ is $-200$, $-100$, $-170$ and $15$; $a_k$ is $-1$, $-1$, $-6.5$ and $0.7$; $b_k$ is $0$, $0$, $11$ and $0.6$; $c_k$ is $-10$, $-10$, $-6.5$ and $0.7$; $x_k$ is $1$, $0$, $-0.5$ and $-1$; and $y_k$ is $0$, $0.5$, $1.5$ and $1$. Energies are in units of $k_{\rm B}T$ at 300~K. The exponent $g_k$ is capped at 50 far outside the walls. Harmonic walls $1000[\min(x+1.3,0)^2+\max(x-1.0,0)^2]$ confine $x$, as in the reference simulations\cite{trizio2025everything}.

\paragraph*{Simulation details.} The particle mass is 4~amu and $T_{\rm phys}=300$~K, so $\tau^\ast=25.46$~fs. The endpoints are the guesses $(-0.558,1.442)$ for $A$ and $(0.623,0.028)$ for $B$. The surface has a third minimum near $(-0.050,0.467)$, which the search must find as an intermediate. Two minima share a state when they lie within 0.15~\AA\ of each other.

\paragraph*{Reference quantities.} The two saddles are polished from guesses near $(-0.822,0.624)$ and $(0.212,0.293)$. The reference committor is computed on a $181\times181$ grid over $-1.5\le x\le1.2$ and $-0.5\le y\le2.2$.

\subsection{Two-path potential}\label{sm:sys-tc}
\paragraph*{Energy function.} The two-path potential of Trizio et al.\cite{trizio2025everything} is a quartic double well with two channels between the basins, the upper one raised and the lower one deepened by Gaussians,
\begin{equation}
\begin{split}
V(x,y)={}&10\bigl[2+\tfrac43x^4-2y^2+y^4\bigr]+\tfrac{100}{3}x^2(y^2-1)\\
&+\sum_{k=1}^{3}A_k\exp\Bigl[-\frac{(x-x_k)^2+(y-y_k)^2}{0.4^2}\Bigr],
\end{split}
\end{equation}
where the Gaussian amplitudes are $(A_1,A_2,A_3)=(7,1,-6)$ and their centres $(x_k,y_k)$ are $(-0.7,0.8)$, $(1.0,-0.3)$ and $(-1.0,-0.6)$. Energies are in units of $k_{\rm B}T$ at 300~K, and harmonic walls $1000[\min(x+2,0)^2+\max(x-2,0)^2]$ confine $x$.

\paragraph*{Simulation details.} The particle mass, $T_{\rm phys}$ and $\tau^\ast$ are those of the M\"uller--Brown potential. The endpoints are the guesses $(-1.033,-0.350)$ for $A$ and $(1.122,0.043)$ for $B$. Two minima share a state when they lie within 0.20~\AA\ of each other. The test of this system is whether both channels are found and certified as separate channels (Methods Section~\MainSec{sec:m-chains}).

\paragraph*{Reference quantities.} The saddle of the upper channel is polished from a guess near $(-0.424,0.637)$ and that of the lower channel, which is almost flat, from a guess near $(0.075,-0.995)$. The reference committor is computed on a $181\times181$ grid over $-2\le x,y\le2$.

\subsection{$\mathrm{S_N2}$ reaction $\mathrm{Cl^-+CH_3Br\to CH_3Cl+Br^-}$}\label{sm:sys-sn2}
\paragraph*{Energy function.} The benchmark in Results uses GFN2-xTB energies and the corresponding independently refined reference saddle. The system has six atoms, charge $-1$ and multiplicity 1.

\paragraph*{Simulation details.} The supplied endpoints were optimized at the HF-3c level, then relaxed by the energy-only endpoint quench on GFN2-xTB. States are bond graphs, and $\xi_{\rm bond}$ reduces to $r(\mathrm{C{-}Br})-r(\mathrm{C{-}Cl})$. $T_{\rm phys}=298.15$~K. The input-file calculator default must be overridden when reproducing the GFN2-xTB comparison.

\paragraph*{Reference quantities.} The molecular reference saddle was refined on GFN2-xTB for comparison only, as described in Results. Energy differences in Table~\MainTable{tab:r-benchmark} use that same-surface saddle. The published HF-3c barriers of $5.5$ and $143.5$~kJ\,mol$^{-1}$\cite{tikhonov2026two} refer to a different energy surface and are not used to assess these GFN2-xTB results.

\subsection{Gas-phase reactions of the Baker test set}\label{sm:sys-sacts}
\paragraph*{Energy function.} Eight reactions from the standard transition-state test set of Baker and Chan\cite{baker1996location}, in the geometries used by Suresh and Vogiatzis\cite{suresh2026sacts}, are computed with GFN2-xTB. Their charges and multiplicities are listed in Supplementary Table~\ref{tab:s-systems}.

\paragraph*{Reactions.} Supplementary Table~\ref{tab:s-systems} lists all eight transformations with their sizes, charges and multiplicities. They cover closed-shell and radical hydrogen shifts, keto--enol tautomerization, dissociation, addition and radical ring opening. These classes test distinct changes in the bond-graph fingerprint rather than eight repetitions of one reaction type.

\paragraph*{Simulation details.} The endpoints are those of Suresh and Vogiatzis\cite{suresh2026sacts} and were not optimized. States are bond graphs, and $T_{\rm phys}=298.15$~K for all eight reactions. Each reaction therefore tests the search from unrelaxed input structures on a fast semiempirical oracle.

\paragraph*{Reference quantities.} The reference structures are the transition states reported by Suresh and Vogiatzis, located with xtb and the Berny optimizer\cite{suresh2026sacts}. For this benchmark, these geometries initialized independent Sella reference refinement on GFN2-xTB, followed by a finite-difference Hessian check. Only these refined same-surface references are used for the saddle comparisons in Results; the input guesses are evaluated separately.

\subsection{Alanine dipeptide}\label{sm:sys-ala}
\paragraph*{Energy function.} Alanine dipeptide (Ace-Ala-Nme, 22 atoms) in vacuum is described by an energy-only reimplementation of AMBER ff99SB\cite{hornak2006ff99sb}. Its parameters were exported from the GROMACS run input of Trizio et al.\cite{trizio2025everything}. The energy contains harmonic bonds, including the X--H bonds, harmonic angles, periodic proper and improper dihedrals, scaled 1--4 Lennard-Jones and Coulomb terms, and all remaining non-bonded pairs without cutoff. The calculator returns energies only.

\paragraph*{Simulation details.} The endpoints are the starting frames of the $\mathrm{C7_{eq}}$ and $\mathrm{C7_{ax}}$ simulations of the same reference and were not optimized. States are defined by $(\phi,\psi)$, with $\phi$ through atoms 5--7--9--15 and $\psi$ through atoms 7--9--15--17 in the GROMACS order. Two configurations share a state when both periodic differences are at most $60^\circ$. New states are disallowed, so every quench is assigned to $\mathrm{C7_{eq}}$ or $\mathrm{C7_{ax}}$. $T_{\rm phys}=300$~K, as in the reference simulations.

\paragraph*{Committor model details.} The network reads all 231 interatomic distances. This differs from the 45 heavy-atom distances used by Trizio et al.\cite{trizio2025everything}, because the descriptor layer of tReax\_expedition is the same for every molecule.

\paragraph*{Reference quantities.} The reference is a first-order saddle of the same force field near $(\phi,\psi)=(-2.1^\circ,-26.4^\circ)$, located independently of the search, with a barrier that depends on the reactant energy convention. Relative to the lowest state-$A$ minimum recorded by the retained search, the barrier is $33.46$~kJ\,mol$^{-1}$; the earlier $36.4$~kJ\,mol$^{-1}$ value used a different minimum. Results primarily compares absolute saddle-referenced differences to avoid this offset. The reference serves only for evaluation.

\subsection{Evaluation against reference quantities}\label{sm:reference}
None of the following enters the search. For the two-dimensional models, the reference committor is the finite-volume solution of $\nabla\cdot(e^{-\beta^\ast V}\nabla q)=0$ on the grids of Supplementary Sections~\ref{sm:sys-pt} to \ref{sm:sys-tc}, with harmonic means of $e^{-\beta^\ast V}$ on cell faces. The outer boundaries reflect, and discs around the exact minima nearest to the states found by the search absorb. The discs have the basin radii of Supplementary Table~\ref{tab:s-systems}, and $\beta^\ast(V-V_{\min})$ is capped at 300 to avoid underflow. The learned committor is compared on grid points with $0.05<q<0.95$ and $V\le V_{\rm S}+3k_{\rm B}T^\ast$, where $V_{\rm S}$ is the lowest saddle, and by a Boltzmann-weighted error. The grid solution is a numerical reference for overdamped diffusion, not an exact finite-step Monte Carlo committor. Its absorbing discs also differ from the core rules used by shooting, so their comparison includes both dynamical and boundary-definition differences. For molecules, the Results comparisons use independently refined saddle structures on the same oracle, checked for exactly one imaginary mode, and compare their energies with $E_{\rm up}$. Potential-energy gradients and finite-difference Hessians are used only in this reference calculation, not in the search. Published barriers are compared only when the level of theory matches.

\subsection{Barrier estimates from single coupled-walker runs}\label{sm:barrier}
For comparison with the estimator of MCTSS\cite{tikhonov2026two}, a single exploration run can be analysed without labelling, chain growth or committor learning. Either the pooled samples of both walkers or the weighted midpoints of equation~\MainEq{eq:midweight} are binned into 15 equal-width bins of the progress $p$. The forward and backward barriers are the highest bin mean minus the mean of the reactant and product end bins, with a within-run spread from 200 resamples of individual stored frames. This framewise resampling ignores serial dependence and is not calibrated independent-search uncertainty. Walker samples belong to no single temperature, because each walker samples $\tau_{\rm b}$ while apart, the midpoint $2\tau_{\rm b}$ when together, and the approach is not stationary. The midpoint weighting identity applies to stationary sampling of $\Pi$ at $\tau_{\rm mid}=2\tau_{\rm b}$ (Supplementary Section~\ref{prop:pairframe}). An exploration run stopped at a join does not establish stationarity. Bin means lie above the zero-temperature profile by roughly $\tfrac12(d-1)k_{\rm B}T$, and a weighted linear fit in $1/\tau$ over at least three runs extrapolates the estimate to zero effective temperature. These estimators are heuristics and are reported separately from $E_{\rm up}$.

\subsection{Software, random numbers and reproducibility}\label{sm:repro}
The implementation is written in Python 3.12 with NumPy, SciPy, ASE\cite{larsen2017ase}, PyTorch\cite{paszke2019pytorch} and tblite. The reported hardware is specified in Methods Section~\MainSec{sec:m-systems}. The complete benchmark uses the $SE(3)$-corrected implementation, as stated in Results Section~\MainSec{sec:r-systems}. Exact package versions and the generating commit identifier were not retained in the supplied archive.
Monte Carlo moves draw from the Mersenne Twister generator of the Python standard library, seeded per task from the base seed $s$ ($s=1$ by default). Run $k$ of the exploration ladder uses seed $1000s+k$, counted across rungs. Labelled path $k$ uses $2000s+k$, chain $k$ of a state pair uses $3000s+k$, and local straddling-pair ensemble job $j$ uses $3000s+j$. Committor learning seeds PyTorch with $s$ and draws the seeds of its walks and shots from a NumPy generator seeded with $s$. Tasks run in a pool of spawned worker processes, four by default, each with one thread and double-precision tensors. Every task carries its own seed and results are gathered in submission order. This design reduces scheduling dependence, but does not establish reproducibility across software revisions or numerical libraries. The fixed-oracle proofs assume a consistent converged electronic solution. Reusing self-consistent-field restart guesses can introduce history dependence if different electronic solutions are reached; its effect was not quantified in the retained benchmark. The implementation can write settings, state records, chain energies, costs and structures, with checkpoints and ensemble samples for optional stages.

\begin{SuppTable}
\SuppTableCaption{\textbf{Energy surfaces and state definitions used in the benchmark.} Each row specifies the energy oracle, system size, charge and spin multiplicity, physical-temperature parameter and fingerprint used for quench-based state assignment. Radii denote the two-dimensional basin-matching tolerance; the alanine tolerance applies separately to both wrapped backbone dihedrals. Citations beside system names identify model or input-structure sources. All gas-phase results use GFN2-xTB\cite{bannwarth2019gfn2}, including $\mathrm{S_N2}$. The search used the same molecular defaults, with the stated two-dimensional exceptions.}{tab:s-systems}
\begin{ruledtabular}
\begin{tabular}{@{}llllll@{}}
\TabCell{0.25\textwidth}{System} & \TabCell{0.13\textwidth}{Energy} & \TabCell{0.12\textwidth}{Size} & \TabCell{0.08\textwidth}{Charge, mult.} & \TabCell{0.11\textwidth}{$T_{\rm phys}$ ($\tau^\ast$)} & \TabCell{0.12\textwidth}{State fingerprint} \\
\hline
\TabCell{0.25\textwidth}{Proton transfer with bath\cite{tikhonov2026two}} & \TabCell{0.13\textwidth}{analytic (cm$^{-1}$)} & \TabCell{0.12\textwidth}{1 particle, 1~amu} & \TabCell{0.08\textwidth}{--} & \TabCell{0.11\textwidth}{$\tau^\ast=10$~fs} & \TabCell{0.12\textwidth}{region, 0.22~\AA} \\
\TabCell{0.25\textwidth}{M\"uller--Brown, scaled\cite{muller1979location,trizio2025everything}} & \TabCell{0.13\textwidth}{analytic (reduced)} & \TabCell{0.12\textwidth}{1 particle, 4~amu} & \TabCell{0.08\textwidth}{--} & \TabCell{0.11\textwidth}{300~K (25.46~fs)} & \TabCell{0.12\textwidth}{region, 0.15~\AA} \\
\TabCell{0.25\textwidth}{Two-path\cite{trizio2025everything}} & \TabCell{0.13\textwidth}{analytic (reduced)} & \TabCell{0.12\textwidth}{1 particle, 4~amu} & \TabCell{0.08\textwidth}{--} & \TabCell{0.11\textwidth}{300~K (25.46~fs)} & \TabCell{0.12\textwidth}{region, 0.20~\AA} \\
\TabCell{0.25\textwidth}{$\mathrm{Cl^-+CH_3Br}$ ($\mathrm{S_N2}$)\cite{tikhonov2026two}} & \TabCell{0.13\textwidth}{GFN2-xTB} & \TabCell{0.12\textwidth}{6 atoms} & \TabCell{0.08\textwidth}{$-1$, 1} & \TabCell{0.11\textwidth}{298.15~K} & \TabCell{0.12\textwidth}{bond graph} \\
\TabCell{0.25\textwidth}{$\mathrm{HCN\to HNC}$\cite{suresh2026sacts}} & \TabCell{0.13\textwidth}{GFN2-xTB} & \TabCell{0.12\textwidth}{3} & \TabCell{0.08\textwidth}{0, 1} & \TabCell{0.11\textwidth}{298.15~K} & \TabCell{0.12\textwidth}{bond graph} \\
\TabCell{0.25\textwidth}{$\mathrm{HCCH\to H_2CC}$\cite{suresh2026sacts}} & \TabCell{0.13\textwidth}{GFN2-xTB} & \TabCell{0.12\textwidth}{4} & \TabCell{0.08\textwidth}{0, 1} & \TabCell{0.11\textwidth}{298.15~K} & \TabCell{0.12\textwidth}{bond graph} \\
\TabCell{0.25\textwidth}{$\mathrm{HCOH\to H_2CO}$\cite{suresh2026sacts}} & \TabCell{0.13\textwidth}{GFN2-xTB} & \TabCell{0.12\textwidth}{4} & \TabCell{0.08\textwidth}{0, 1} & \TabCell{0.11\textwidth}{298.15~K} & \TabCell{0.12\textwidth}{bond graph} \\
\TabCell{0.25\textwidth}{$\mathrm{H_2CO\to H_2+CO}$\cite{suresh2026sacts}} & \TabCell{0.13\textwidth}{GFN2-xTB} & \TabCell{0.12\textwidth}{4} & \TabCell{0.08\textwidth}{0, 1} & \TabCell{0.11\textwidth}{298.15~K} & \TabCell{0.12\textwidth}{bond graph} \\
\TabCell{0.25\textwidth}{$\mathrm{CH_3O^\bullet\to{}^\bullet CH_2OH}$\cite{suresh2026sacts}} & \TabCell{0.13\textwidth}{GFN2-xTB} & \TabCell{0.12\textwidth}{5} & \TabCell{0.08\textwidth}{0, 2} & \TabCell{0.11\textwidth}{298.15~K} & \TabCell{0.12\textwidth}{bond graph} \\
\TabCell{0.25\textwidth}{$\mathrm{CH_3CHO\to CH_2CHOH}$\cite{suresh2026sacts}} & \TabCell{0.13\textwidth}{GFN2-xTB} & \TabCell{0.12\textwidth}{7} & \TabCell{0.08\textwidth}{0, 1} & \TabCell{0.11\textwidth}{298.15~K} & \TabCell{0.12\textwidth}{bond graph} \\
\TabCell{0.25\textwidth}{$\mathrm{C_2H_4+HF\to CH_3CH_2F}$\cite{suresh2026sacts}} & \TabCell{0.13\textwidth}{GFN2-xTB} & \TabCell{0.12\textwidth}{8} & \TabCell{0.08\textwidth}{0, 1} & \TabCell{0.11\textwidth}{298.15~K} & \TabCell{0.12\textwidth}{bond graph} \\
\TabCell{0.25\textwidth}{$\mathrm{C_3H_5O_2^\bullet}$ ring opening\cite{suresh2026sacts}} & \TabCell{0.13\textwidth}{GFN2-xTB} & \TabCell{0.12\textwidth}{10} & \TabCell{0.08\textwidth}{0, 2} & \TabCell{0.11\textwidth}{298.15~K} & \TabCell{0.12\textwidth}{bond graph} \\
\TabCell{0.25\textwidth}{Alanine dipeptide, $\mathrm{C7_{eq}\to C7_{ax}}$\cite{trizio2025everything}} & \TabCell{0.13\textwidth}{ff99SB, vacuum\cite{hornak2006ff99sb}} & \TabCell{0.12\textwidth}{22} & \TabCell{0.08\textwidth}{0, 1} & \TabCell{0.11\textwidth}{300~K} & \TabCell{0.12\textwidth}{$(\phi,\psi)$, $60^\circ$} \\
\end{tabular}
\end{ruledtabular}
\par\vspace{10pt}
\SuppTableCaption{\textbf{System-dependent neural descriptors and shooting budgets.} System citations identify model or input-structure sources; the network and sampling settings are those used in this work. Molecular networks receive all $N(N-1)/2$ interatomic distances; the two-dimensional models receive $(x,y)$. Network sizes list the input, hidden and output dimensions. $\Gamma$ controls the binomial shooting-loss contribution, and minibatch denotes configurations per optimization batch. A shot stops on reaching a core or exhausting $2000\max(1,3N/18)$ steps, rounded for molecules. Non-committing shots are recorded separately and excluded from the reported product fraction.}{tab:s-sysnc}
\begin{ruledtabular}
\begin{tabular}{@{}llllll@{}}
\TabCell{0.27\textwidth}{System} & \TabCell{0.1\textwidth}{Atoms} & \TabCell{0.12\textwidth}{Descriptors} & \TabCell{0.13\textwidth}{Network} & \TabCell{0.12\textwidth}{$\Gamma$, minibatch} & \TabCell{0.1\textwidth}{Shot budget} \\
\hline
\TabCell{0.27\textwidth}{Two-dimensional models (3)\cite{tikhonov2026two,muller1979location,trizio2025everything}} & \TabCell{0.1\textwidth}{--} & \TabCell{0.12\textwidth}{2} & \TabCell{0.13\textwidth}{2--32--32--1} & \TabCell{0.12\textwidth}{0, 512} & \TabCell{0.1\textwidth}{2000} \\
\TabCell{0.27\textwidth}{$\mathrm{HCN\to HNC}$\cite{suresh2026sacts}} & \TabCell{0.1\textwidth}{3} & \TabCell{0.12\textwidth}{3} & \TabCell{0.13\textwidth}{3--32--32--1} & \TabCell{0.12\textwidth}{1, 256} & \TabCell{0.1\textwidth}{2000} \\
\TabCell{0.27\textwidth}{$\mathrm{HCCH}$, $\mathrm{HCOH}$, $\mathrm{H_2CO}$ reactions\cite{suresh2026sacts}} & \TabCell{0.1\textwidth}{4} & \TabCell{0.12\textwidth}{6} & \TabCell{0.13\textwidth}{6--32--32--1} & \TabCell{0.12\textwidth}{1, 256} & \TabCell{0.1\textwidth}{2000} \\
\TabCell{0.27\textwidth}{$\mathrm{CH_3O^\bullet\to{}^\bullet CH_2OH}$\cite{suresh2026sacts}} & \TabCell{0.1\textwidth}{5} & \TabCell{0.12\textwidth}{10} & \TabCell{0.13\textwidth}{10--32--32--1} & \TabCell{0.12\textwidth}{1, 256} & \TabCell{0.1\textwidth}{2000} \\
\TabCell{0.27\textwidth}{$\mathrm{Cl^-+CH_3Br}$ ($\mathrm{S_N2}$)\cite{tikhonov2026two}} & \TabCell{0.1\textwidth}{6} & \TabCell{0.12\textwidth}{15} & \TabCell{0.13\textwidth}{15--32--32--1} & \TabCell{0.12\textwidth}{1, 256} & \TabCell{0.1\textwidth}{2000} \\
\TabCell{0.27\textwidth}{$\mathrm{CH_3CHO\to CH_2CHOH}$\cite{suresh2026sacts}} & \TabCell{0.1\textwidth}{7} & \TabCell{0.12\textwidth}{21} & \TabCell{0.13\textwidth}{21--32--32--1} & \TabCell{0.12\textwidth}{1, 256} & \TabCell{0.1\textwidth}{2333} \\
\TabCell{0.27\textwidth}{$\mathrm{C_2H_4+HF\to CH_3CH_2F}$\cite{suresh2026sacts}} & \TabCell{0.1\textwidth}{8} & \TabCell{0.12\textwidth}{28} & \TabCell{0.13\textwidth}{28--32--32--1} & \TabCell{0.12\textwidth}{1, 256} & \TabCell{0.1\textwidth}{2667} \\
\TabCell{0.27\textwidth}{$\mathrm{C_3H_5O_2^\bullet}$ ring opening\cite{suresh2026sacts}} & \TabCell{0.1\textwidth}{10} & \TabCell{0.12\textwidth}{45} & \TabCell{0.13\textwidth}{45--32--32--1} & \TabCell{0.12\textwidth}{1, 256} & \TabCell{0.1\textwidth}{3333} \\
\TabCell{0.27\textwidth}{Alanine dipeptide\cite{trizio2025everything}} & \TabCell{0.1\textwidth}{22} & \TabCell{0.12\textwidth}{231} & \TabCell{0.13\textwidth}{231--32--32--1} & \TabCell{0.12\textwidth}{1, 256} & \TabCell{0.1\textwidth}{7333} \\
\end{tabular}
\end{ruledtabular}
\end{SuppTable}

\ifdefined\MainDocument\onecolumngrid\fi
\clearpage
\ifdefined\MainDocument\twocolumngrid\fi
\section{tReax\_expedition Algorithm}\label{sm:alg}
Supplementary Algorithm~\ref{alg:search} summarizes the search; Algorithms~\ref{alg:chain} and \ref{alg:committor} specify continuation and committor learning. Values are molecular defaults unless stated otherwise; Supplementary Tables~\ref{tab:s-search}--\ref{tab:s-committor} give shared settings and two-dimensional exceptions.

\begin{SuppAlgBlock}
\SuppAlgorithm{alg:search}{tReax\_expedition search between two endpoints.}
\begin{algorithmic}[1]
\Require endpoints $\mathbf R_A^0,\mathbf R_B^0$; oracle $V$; fingerprint; settings
\State relax both endpoints by quenching; register states $A$ and $B$; stop if they coincide
\State $\tau\gets\tau_0$; run four coupled-walker runs at $\tau$; halve $\tau$ (at most three times) until at least half join
\State stop if no rung joined during the initial tests; otherwise $\tau_{\rm x}\gets\tau$
\For{at most seven colder rungs}
  \State $\tau\gets2\tau$; run four coupled-walker runs at $\tau$
  \If{fewer than half join}
    \State \textbf{break}
  \EndIf
  \State $\tau_{\rm x}\gets\tau$
\EndFor
\State pick up to eight joined paths with $\tau_{\rm b}\le\tau_{\rm x}$; label each by bisection with quenches
\ForAll{state pairs $(S_a,S_b)$ connected by a label change}
  \State build $\xi$; choose up to four seeds by farthest-point selection
  \State grow a chain from every seed (Algorithm~\ref{alg:chain}); keep certified chains
  \If{no chain is certified and committor learning is requested}
    \State learn $q_\theta$ from the seeds (Algorithm~\ref{alg:committor}); regrow the chains along $z_\theta$
  \EndIf
  \State group chains into channels; keep the lowest chain of each
\EndFor
\State on each channel, bisect the three-vote label along the chain; refine the link four times
\State optional: learn $q_\theta$ at $\tau^\ast$ for each pair (Algorithm~\ref{alg:committor}); shoot from the crossings
\State optional: sample the straddling pair at $\tau_{\rm b}=\tau^\ast/2$ on each channel
\State report the minimax route from $A$ to $B$, all sampled $E_{\rm up}$, crossings and oracle calls per stage
\end{algorithmic}
\end{SuppAlgBlock}

\begin{SuppAlgBlock}
\SuppAlgorithm{alg:chain}{One restrained continuation chain between $S_a$ and $S_b$.}
\begin{algorithmic}[1]
\Require seed $\mathbf R_0$; coordinate $\xi$; minima of $S_a,S_b$; $\tau_{\rm x}$
\State slice grid $\xi_0,\dots,\xi_{24}$ between the minima; $\kappa\gets\kappa_0/\Delta\xi^2$; $\ell\gets\sqrt{3\hbar\tau_{\rm x}/2}$
\State seed node: restrained quench at $\mathrm{clip}(\xi(\mathbf R_0))$
\ForAll{sweep directions, slices beyond $\Delta\xi/4$ from the seed, in order}
  \State move the restraint centre to $\xi_k$ in five short quenches from the previous node
  \State quench at $\xi_k$; quadruple $\kappa$ and repeat while the miss exceeds $\Delta\xi/4$ (at most three times)
  \State record the highest energy held along these quenches as $E_k^{\rm traj}$
  \State insert nodes at half spacing while two nodes are farther apart than $\ell$ (depth at most three)
\EndFor
\State quench both end nodes without restraint; register the states they reach; record $E^{\rm desc}$
\State evaluate $V$ on every straight link at $\max(4,\lceil4L/\ell\rceil)$ interior points
\State $E_{\rm up}\gets$ equation~\MainEq{eq:eup}; certified $\gets$ ends reach $(S_a,S_b)$ and $E_{\rm up}<\infty$
\end{algorithmic}
\end{SuppAlgBlock}

\begin{SuppAlgBlock}
\SuppAlgorithm{alg:committor}{Committor learning for one state pair.}
\begin{algorithmic}[1]
\Require minima of $S_a,S_b$; highest-energy chain nodes and other chain nodes; quench-label crossings; $\tau^\ast$
\State unbiased walks from both minima; core configurations define the boundary sets; store the block with $b_0=0$
\State fit $q_\theta$ on the stored data (MBAR weights)
\For{$k=1,\dots,6$}
  \State $V_{\rm bias}\gets V_{\rm K}[q_\theta]+V_z$, with the $V_z$ grid sized from the stored latents
  \State walkers take turns depositing $V_z$; freeze $V_{\rm bias}$
  \State production walks under the frozen bias; store the block with its bias
  \State draw 16 shooting points (Lorentzian in $z_\theta$ over density); run two unbiased chains from each
  \State score the current model on the new shots; add the shots to the data
  \State refit $q_\theta$ with MBAR weights capped at the 99th percentile
\EndFor
\State shoot 20 chains from each quench-label crossing; report $p_B$ against $q_\theta$
\end{algorithmic}
\end{SuppAlgBlock}

\clearpage
\onecolumngrid
\newenvironment{SuppFinalTable}{\par\addvspace{8pt}\noindent\begin{minipage}{\textwidth}\small}{\end{minipage}\par\addvspace{8pt}}
\begin{SuppFinalTable}
\SuppTableCaption{\textbf{Energy and geometric errors for the two-dimensional continuation chains.} Each row compares one retained chain with its corresponding analytic-model saddle. $\delta E_{\rm up}=E_{\rm up}-E_{\rm saddle}$ uses all evaluated chain segments, whereas $\delta E_0=V(\mathbf R^\ddagger_0)-E_{\rm saddle}$ and the Euclidean displacement refer to the quench-label crossing. Values are derived from the retained \texttt{ts\_compare.csv} records using $1$~eV$=96.4853$~kJ\,mol$^{-1}$. The five comparisons come from one search on each of three surfaces; they are not independent search replicates. Figure~\MainFig{fig:r-2d} shows the corresponding geometry. Negative $\delta E_{\rm up}$ values are consistent with finite evaluation spacing; the continuous-path correction was not computed.}{tab:s-2d-comparison}
\begin{ruledtabular}
\begin{tabular}{@{}lrrr@{}}
Surface and channel & $\delta E_{\rm up}$ (kJ\,mol$^{-1}$) & $\delta E_0$ (kJ\,mol$^{-1}$) & Displacement (\AA) \\
\hline
Proton transfer, $A\to B$ & -0.0020 & -0.4553 & 0.0526 \\
M\"uller--Brown, $A\to I_1$ & -0.0018 & -0.2632 & 0.0548 \\
M\"uller--Brown, $I_1\to B$ & +0.0008 & +0.0005 & 0.0039 \\
Two-path, lower channel & +0.0252 & +0.0056 & 0.1012 \\
Two-path, upper channel & +0.0118 & +0.0129 & 0.0283 \\
\end{tabular}
\end{ruledtabular}
\end{SuppFinalTable}

\begin{SuppFigure}
\centering
\includegraphics[width=0.72\textwidth]{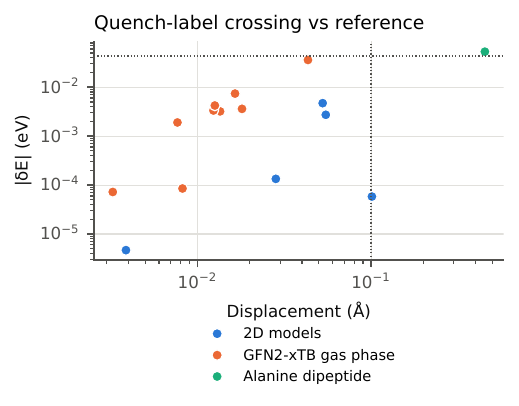}
\SuppFigureCaption{\textbf{Crossing displacement and energy deviation measure different aspects of localization.} Points compare quench-label crossings with the corresponding same-surface reference saddles. Blue denotes the five two-dimensional saddle comparisons, orange the nine primary gas-phase channels and green the lowest alanine dipeptide chain. The molecular displacement is symmetry-adjusted all-atom RMSD; two-dimensional displacements are Euclidean distances. Both axes are logarithmic; $|\delta E|=|V(\mathbf R^\ddagger_0)-E_{\rm reference}|$ is the absolute crossing-energy deviation. Dotted guides mark 0.1~\AA\ and approximately 1~kcal\,mol$^{-1}$; they are visual scales, not statistical thresholds. The 15 points derive from one search per system, with no uncertainty bars or cross-system correlation fit.}{fig:s-r-scatter}
\end{SuppFigure}

\begin{SuppFigure}
\centering
\includegraphics[width=\textwidth]{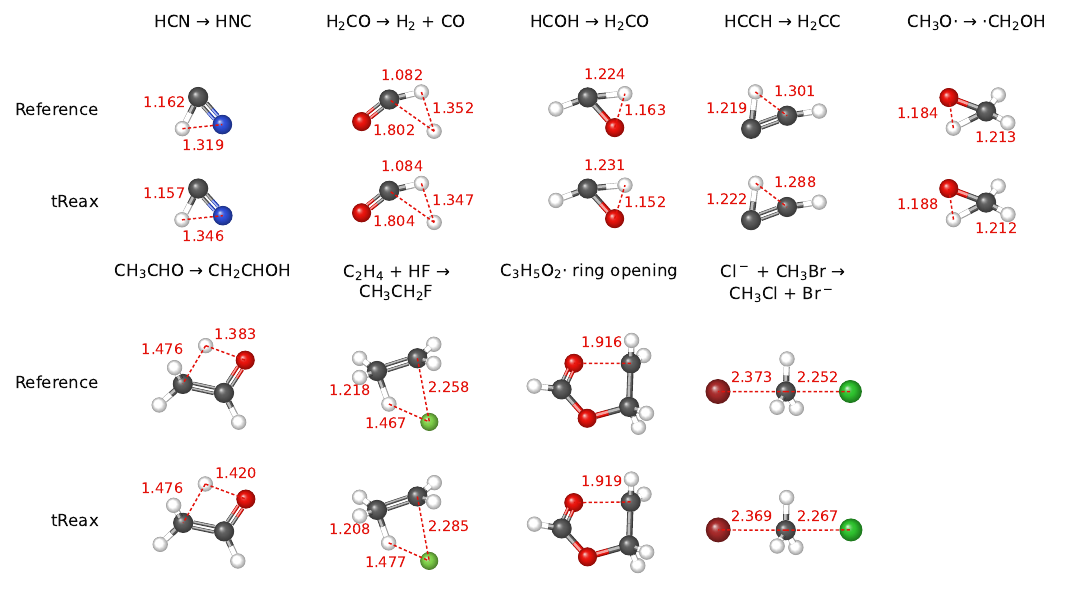}
\SuppFigureCaption{\textbf{Local geometric comparison of gas-phase reference saddles and quench-label crossings.} Each reaction is shown as a pair, with the independently refined GFN2-xTB first-order saddle (Reference) above the primary tReax\_expedition crossing (tReax). The upper group shows HCN isomerization, formaldehyde dissociation, HCOH isomerization, acetylene rearrangement and methoxy isomerization, from left to right; the lower group shows acetaldehyde tautomerization, HF addition to ethylene, radical ring opening and $\mathrm{S_N2}$ substitution. Red dashed segments identify selected interatomic distances, labelled in \AA. Carbon, hydrogen, oxygen and nitrogen are grey, white, red and blue; fluorine, chlorine and bromine are light green, green and dark red. These local distances complement the all-atom RMSDs in Table~\MainTable{tab:r-benchmark} and energy profiles in main Fig.~\MainFig{fig:r-gas}. One retained crossing is shown per reaction.}{fig:s-r-schematic}
\end{SuppFigure}

\clearpage
\begin{SuppFigure}
\centering
\includegraphics[width=\textwidth,height=0.72\textheight,keepaspectratio]{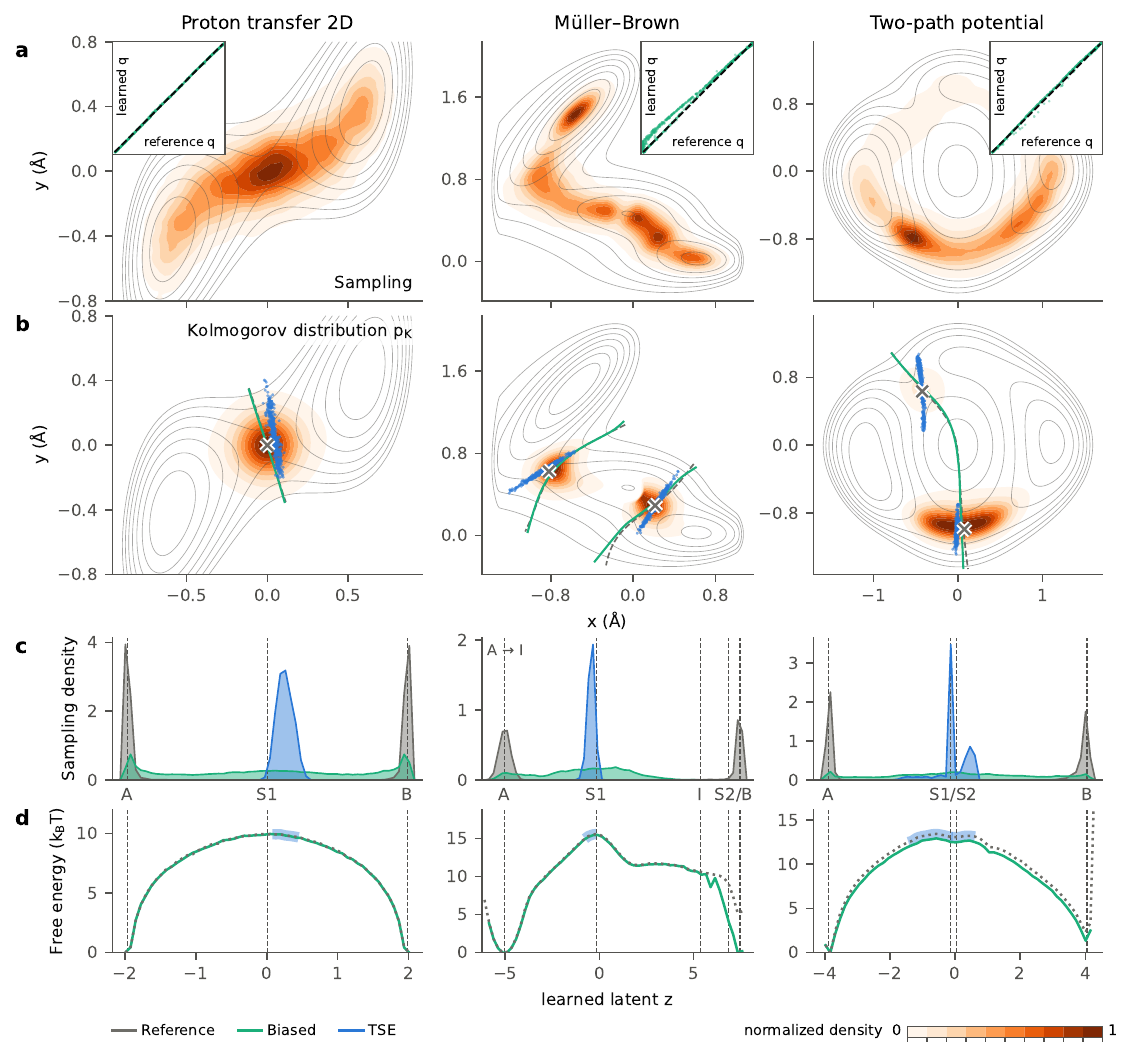}
\SuppFigureCaption{\textbf{Broad biased sampling and local transition-region sampling serve different purposes.} Rows a--d compare broad biased sampling with local midpoint sampling. Columns show proton transfer, M\"uller--Brown and the two-path potential. \textbf{a}, Two-dimensional biased sampling densities; insets compare the learned committor with the numerical diffusion reference, with dashed diagonals marking equality. \textbf{b}, Kolmogorov density $p_{\rm K}$ (orange; equation~\eqref{eq:s-pk}), midpoints from the local straddling-pair ensemble (labelled TSE, blue), learned $q_\theta=1/2$ contours (green), numerical reference $q=1/2$ contours (grey dashed) and reference saddles (crosses). Grey background contours in a,b show potential energy. Orange maps use the normalized-density colour scale shown. \textbf{c}, Distributions along the learned latent $z$: numerical equilibrium reference (grey), biased training samples (green) and midpoints from the local straddling-pair ensemble (blue). Vertical guides locate the labelled minima and saddles. Both M\"uller--Brown projections in c,d refer to the indicated $A\to I$ committor; b includes both saddle neighbourhoods. \textbf{d}, Latent free-energy projections in $k_{\rm B}T$ units from the numerical reference (grey) and MBAR-reweighted biased samples (green), up to an additive constant. Blue segments mark the latent intervals occupied by local straddling-pair midpoint samples on the reference profile; they are neither free-energy curves estimated from those midpoints nor confidence intervals. Biased walkers start from the two minima and retained highest-energy chain nodes, combining $V_{\rm K}$ with $V_z$ before frozen-bias production. The local straddling pair straddles a surface through the quench-label crossing without using $q_\theta$ in its side constraint. The curves are descriptive, without uncertainty estimates.}{fig:s-r-sampling}
\end{SuppFigure}

\clearpage


\ifdefined\MainDocument\expandafter \fi

\providecommand{\bibcommenthead}{}
\bibliographystyle{nature}
\bibliography{literature}

\end{document}